%% file: main.tex
\documentclass[12pt,draftclsnofoot,onecolumn]{IEEEtran}

\usepackage{amsmath, mathrsfs,balance}
\usepackage{bm}

\usepackage{graphicx}
\usepackage{color}
\usepackage{tikz}
\usepackage{pgfplots}
\pgfplotsset{compat=newest}
\usetikzlibrary{patterns}
\usetikzlibrary{positioning}
\usetikzlibrary{datavisualization}
\usetikzlibrary{datavisualization.formats.functions}
\usetikzlibrary{backgrounds}
\usetikzlibrary{shapes,snakes}

\makeatletter
\def\maketag@@@#1{\hbox{\m@th\normalfont\normalsize#1}}
\makeatother

\usepackage[none]{hyphenat}

\newcommand{\subparagraph}{}
\usepackage[compact]{titlesec}
\titlespacing*{\section}{2pt}{1\baselineskip}{0.9\baselineskip}

\allowdisplaybreaks

\usepackage{amsfonts}
\usepackage{amssymb}
\usepackage{color}
\usepackage{multirow}
\usepackage{graphicx}

\usepackage{caption}
\usepackage{subcaption}
\usepackage[numbers,sort&compress]{natbib}
\usepackage{balance}

\input{symbols.tex}

\input{macros.tex}

\def\snrbef{{\mathsf{SNR}_\text{BBF}}}

\usepackage{tikz}
\usepackage{pgfplots}
\pgfplotsset{compat=newest}
\usetikzlibrary{patterns}
\usetikzlibrary{positioning}
\usetikzlibrary{datavisualization}
\usetikzlibrary{datavisualization.formats.functions}
\usetikzlibrary{backgrounds}
\usetikzlibrary{shapes,snakes}
\usepackage{textcomp}
\usetikzlibrary{automata}

\allowdisplaybreaks

\usepackage{algorithm}
\usepackage{algpseudocode}
\usepackage{pifont}
\usepackage{varwidth}

\def\one{{\bf 1}}

\usepackage{multicol,lipsum}
\usepackage{array}

\usepackage{lscape}

\def\herm{{\sfH}}

\def\ptot{{P_{\ttt \tto \ttt}}}

\def\cg{{\clC\clN}}

\usepackage{color}
\newcommand{\figref}[1]{Fig.~\ref{#1}}
\usetikzlibrary{decorations.markings}

\IEEEoverridecommandlockouts

\begin{document}

\sloppy

\title{Fully-$\,$/$\,$Partially-Connected Hybrid Beamforming Architectures for mmWave MU-MIMO\thanks{X. Song is sponsored by the China Scholarship Council (201604910530). This work was funded by the European Union's Horizon 2020 research and innovation programme under grant agreement No. 779305 (SERENA).}}
\author{\IEEEauthorblockN{Xiaoshen Song, \IEEEmembership{Student Member, IEEE,} Thomas K\"uhne,   and Giuseppe Caire \IEEEmembership{Fellow, IEEE}}\\
	
}

\maketitle

\begin{abstract}
Hybrid digital analog (HDA) beamforming has attracted considerable attention in practical implementation 
of millimeter wave (mmWave) multiuser multiple-input multiple-output (MU-MIMO) systems
due to the low power consumption with respect to its fully digital baseband counterpart. 
The implementation cost, performance, and power efficiency of HDA beamforming depends on the level of connectivity
and reconfigurability of the analog beamforming network. In this paper, we investigate the performance of two typical architectures that can be regarded as
extreme cases, namely,  the fully-connected (FC) and the one-stream-per-subarray (OSPS) architectures. 
In the FC architecture each RF antenna port is connected to all antenna elements of the array, while in the OSPS architecture 
the RF antenna ports are connected to disjoint subarrays. 
We jointly consider the initial beam acquisition and data communication phases, such that the latter takes place by using the beam direction 
information obtained by the former. We use the state-of-the-art beam alignment (BA) scheme previously proposed by the authors and consider a family of 
MU-MIMO  precoding schemes well adapted to the beam information extracted from the BA phase. 
We also evaluate the power efficiency of the two HDA architectures taking into account the power dissipation at different hardware components 
as well as the power backoff under typical power amplifier constraints. 
Numerical results show that the two architectures achieve similar sum spectral efficiency, while the OSPS architecture is advantageous with respect to the FC 
case in terms of hardware complexity and power efficiency, at the sole cost of a slightly longer BA time-to-acquisition due to its 
reduced beam angle resolution. 
\end{abstract}	


\section{Introduction}\label{introduction}
Millimeter wave (mmWave) multiuser multiple-input multiple-output (MU-MIMO) communications have emerged as one of the most promising techniques for the 
second phase of 5G wireless systems, aimed at achieving broadband data communications at unprecedented high rates ($\geq 1\,$Gb/s per user), in very dense 
urban small-cell environments. The relatively underutilized mmWave spectrum ($30$-$300\,$GHz) allows to 
achieve a target $\sim1\,$Gb/s per data stream with $\sim1\,$ GHz signal bandwidth, provided that the system can support 
a  spectral efficiency of about 1 bit/s/Hz.  Such relatively low spectral efficiency per stream can be achieved with rather standard 
modulation and coding techniques (e.g., binary codes of rate $1/2$ mapped onto a QPSK constellation), when that the signal to interference plus noise ratio (SINR) at the receiver is between $0$ and $3\,$dB (depending on the gap to capacity of the 
underlying code).\footnote{With ideal single-user capacity achieving codes for the Gaussian channel, we have that $\log (1 + \SINR) = 1$ bit/s/Hz is achieved for $\SINR = 1$ (i.e., 0 dB). In practice, gaps of a fraction of a dB to $3$-$4$ dB are obtained by actual coding schemes adopted in current standards.}

Due to the severe isotropic pathloss incurred by mmWave frequencies, large antenna gains are required both at the base station (BS) 
side and the user equipment (UE) side. Fortunately, the small carrier wavelength associated with mmWave frequencies enables large antenna arrays 
to be packed in a small form factor, such that the required large antenna gain can be obtained using beamforming. 
For example, in a single-user scenario where the signal-to-noise ratio (SNR) at the receiver in isotropic propagation conditions\footnote{ Here the isotropic propagation conditions correspond to one active antenna at the transmitter (Tx) and one active antenna at the receiver (Rx), respectively.} is between $-30$ and $-20\,$dB (a quite realistic situation for outdoor mmWave channels), a combined Tx and Rx beamforming gain of $30\,$dB is needed 
such that, when the Tx and the Rx beams are well aligned, the resulting SNR {\em after beamforming} reaches the desired target (a bit above $0$ dB, as argued before). 

Realizing fast and accurate digitally steerable beamforming at mmWave, however, is not a trivial task. 
One main challenge is that the conventional full digital transceiver architecture (with one radio frequency (RF) chain per antenna element) 
is infeasible due to hardware cost, power consumption, and above all power dissipation in the small integrated array form factor. 
{ Each RF chain consists of (roughly speaking)  analog-to-digital$\,$/$\,$ digital-to-analog (A/D, D/A) converters, up$\,$/$\,$down-conversion mixers, filters, 
power amplifiers (PAs), and low-noise amplifiers (LNAs).} It follows that a design goal for mmWave transceivers is to reduce the number of RF 
chains to be significantly smaller than the number of antenna array elements. 

For this reason,  the concatenation of digital and analog beamforming, known as hybrid digital analog (HDA) beamforming architecture, 
has been widely considered. In such a context, the limited number of RF chains are used to enable the multistream baseband processing, 
while an analog processing is used to realize the antenna beamforming gain.  A primary objective of HDA beamforming is to maximize the multiuser 
sum rate, while keeping the hardware costs, complexity, and power efficiency, within some desirable targets.

\subsection{Related Work}
A large number of works have addressed HDA beamforming for mmWave communication systems. Rather than giving a complete account of
such considerable body of literature (out of scope of the present non-tutorial paper), we consider a few significant representatives and examine their proposed 
approaches in a critical manner.  A common assumption in most of existing works is that the analog part of the HDA precoder 
can only utilize phase control. This phase control can be realized through either phase shifters \cite{Time2017,Sohrabi2016HDA,AngLi2017,EldarHDA2017,JingboSubFull2018,CaoPerAntennaPower2018} or lenses \cite{OverviewHeath2016, LinglongLowComp2018}. Consequently, the problem of finding the (sub-) optimal analog and digital precoding matrices 
is transformed into a series of relatively complicated decomposition steps \cite{Sohrabi2016HDA,AngLi2017,EldarHDA2017,JingboSubFull2018,CaoPerAntennaPower2018}, since the underlying optimization problem is non-convex. 
This phase-only control assumption may somewhat reduce the hardware complexity. However, the signaling freedom is also drastically reduced and the 
corresponding optimization computational complexity is typically prohibitive for practical real-time implementations. 
These drawbacks motivate the exploration of an analog precoding architecture with both phase and amplitude controls \cite{molisch2017hybrid,MajidzadehSub2017}. In fact, it has been demonstrated in practice that simultaneous phase and amplitude control is fully feasible at mmWaves with good accuracy, 
low complexity, and low cost \cite{RaghavanPhaseAmplitude2018,JunyiPrototype2018}.

 { Another severe limitation appearing in several HDA beamforming works is the assumption of invariant instantaneous channel coefficients over a large time duration \cite{LinglongChannelEst2016,RobertSOMP2017,Time2017}.} It is known that, in order to overcome the heavy signal attenuation, communication at mmWaves requires an initial beam acquisition (which we refer as {\em beam alignment (BA)}) \cite{OverviewHeath2016,SaeidBA2016,sxsBA2017}. The goal of BA is to find a pair of narrow beams connecting each UE with the BS.\footnote{E.g., in line-of-sight (LOS) propagation, the aligned directions typically coincide with the AoA and AoD of the LoS path.} 
Thus, the nearly invariant channel assumption only makes sense {\em after BA is achieved}, since once the beams are aligned, the communication 
occurs only through a single narrow path with small effective angular spread, whose delay and Doppler shift can be easily compensated using standard synchronization 
techniques \cite{HeathVariation2017,sxs2017Time,sxs2018TimeJour}.
However, before BA is achieved, the channel delay spread and time-variation can be large due to the presence of
several mulipath components, each with its own delay and Doppler shift. In this case, the instantaneous channel coefficients change very fast. Any BA algorithms relying on an invariant instantaneous channel assumption are no-longer feasible, since for example, even a small motion of a few centimeters traverses several wavelengths, 
potentially producing multiple deep fades \cite{WeilerMeasure2014,Rappaport2017lowrank}.

In addition, a large number of works on HDA architectures investigated only the data communication phase and assume full  channel state information (CSI) \cite{subArray2013Heath,Xuehua2018Hybridly,CaoPerAntennaPower2018,MajidzadehSub2017,EldarHDA2017,Sohrabi2016HDA,AngLi2017,JingboSubFull2018}, i.e., 
that the vectors of baseband complex channel coefficients at each array element are known. These works focus on the optimization 
of the HDA precoder using the full CSI knowledge. 
{ Unfortunately, this assumption is obviously not feasible in a realistic system. In order to acquire such coefficients, one should be able to sample each antenna element, i.e., one would need an RF chain per antenna element or exhaustively measure all elements successively.}
Hence, if full CSI knowledge was possible, no HDA beamforming would be needed, since we could simply implement baseband digital 
beamforming/multiuser precoding, which is performance-wise more efficient.
As a matter of fact, it makes sense to study HDA architectures under the assumption that only a low-dimensional projection of the 
channel vectors can be measured by the limited number of RF chains.  To this end, a hybrid precoding scheme exploiting implicit CSI (i.e., the couplings of all possible pairs of analog beamforming vectors)  was proposed in \cite{FettweisHDA2018}. However, the work in \cite{FettweisHDA2018}  (as well as in \cite{subArray2013Heath,Xuehua2018Hybridly,CaoPerAntennaPower2018,MajidzadehSub2017,EldarHDA2017}) is limited to a single-user 
configuration and does not treat the MU-MIMO case. 

\begin{figure}[t]
	\centering
	\includegraphics[width=14cm]{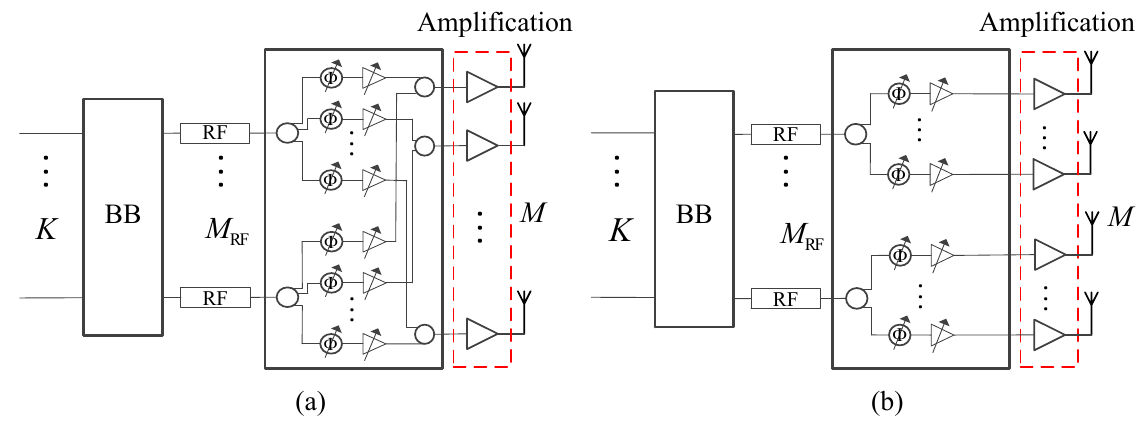}
	\caption{ \small Hybrid digital analog (HDA) transmitter architectures: (a) fully-connected (FC), (b) partially-connected with one-stream-per-subarray (OSPS).
	The ``BB'' block denotes digital baseband beamforming, $K$ is the number of data streams, $M_{\text{RF}}$ is the number of RF chains, and $M$ is the number of antennas.}
	\label{TX}
\end{figure}

It is known that MU-MIMO is superior to single-user beamforming from a network spectral efficiency perspective even under HDA, 
provided that the user density is rich enough such that the BS can schedule subsets of UEs to be served by spatial multiplexing 
with sufficient angular separation  \cite{RaghavanDirectional2016, yunyi2017scheduler}. 
Hence, this motivates us to consider the implementation of MU-MIMO schemes under realistic HDA architecture 
constraints.  
Two ``extreme'' HDA architectures are depicted in \figref{TX} \cite{sxsFC_OSPS}.  \figref{TX}$\,$(a) shows a fully-connected (FC) architecture, 
where each RF antenna port is connected to all antenna elements of the array. 
At the other extreme, \figref{TX}$\,$(b) shows what we refer to as the one-stream-per-subarray (OSPS) architecture, 
where each RF antenna port is connected to a disjoint subarray. 
A common theme that underlies most of the HDA works is that the FC architecture outperforms the OSPS architecture 
only at the cost of higher hardware complexity. However, many reference works \cite{LinglongLowComp2018,Xuehua2018Hybridly,subArray2013Heath,AngLi2017,MajidzadehSub2017} ignore hardware 
impairments \cite{CaoPerAntennaPower2018}, such as the power dissipation and the PA nonlinear distortion. 
In particular, the nonlinear PAs employed at the BS can drastically distort the transmit signal when operated close to saturation \cite{Moghadam2018}. 
To this end, a certain power backoff from the saturation power of a PA should be considered accordingly for different signaling schemes and transceiver 
architectures, such that the PAs can always work in their linear operating region. 

\subsection{Contributions}

In this paper we overcome the shortcomings of the present literature outlined before, and comprehensively evaluate the performance of 
HDA architectures (in particular, as shown in \figref{TX}),  where we assume both amplitude and phase control for each analog path.  
Our main focus is on the MU-MIMO downlink, but similar and symmetric 
conclusions can be reached for the uplink as well.  Our main contributions are summarized as follows:

{1) {\em More general and realistic mmWave channel model.} We consider a quite general  mmWave wireless channel model, taking into account the fundamental features of mmWave channels such as fast time-variation due to Doppler, frequency-selectivity, and the AoA-AoD sparsity \cite{WeilerMeasure2014,Rappaport2017lowrank,Tim2018}. The numerical results based on our proposed channel model are further verified on the 3D geometry based channel generator QuaDRiGa \cite{quadriga}, which has become a standard tool in industrial R\&D as well as in 3GPP standardization.

{2) {\em More practical hardware impairments and power efficiency analysis.} When comparing the HDA beamforming performance 
of different transmitter architectures, we take into account the practical hardware impairments, particularly, the potential power dissipation of the underlying analog network components, as well as the unavoidable power backoff for the nonlinear PAs. While the former is not difficult to be compensated, the latter is highly dependent on the peak-to-average power ratio (PAPR) of the input signal, which (as illustrated in Section \ref{simulation}) should be carefully investigated in terms of different signaling and modulating schemes. On top of the potential hardware impairments, we also evaluate the power efficiency of the most power consuming PAs with respect to different transmitter architectures. Numerical results show that the OSPS architecture with single-carrier (SC) modulation 
achieves the highest power efficiency. 
		
{3) {\em A joint evaluation of initial BA and data communication.} As mentioned before, a main limitation in most hybrid beamforming works is that they only focus on the data communication and assume full CSI. To address this issue, we consider both initial BA and consecutive data communication in this paper. We assume that the precoder in the data communication phase can only exploit a limited amount of CSI, which is obtained along the beams acquired in the 
BA phase. Hence, the signaling and communication procedure in our paper captures the fundamental features of practical mmWave communication.


{4) { {\em Low-complexity data transmit precoding.} In the BA phase, we use our previously proposed BA scheme \cite{sxsBA2017,sxs2017Time,sxs2018TimeJour}, after which each UE obtains a sparse estimate of the channel gains associated to all pairs of AoA-AoD on a finely spaced discrete grid, corresponding to the Tx and Rx beamforming codebooks. For the data communication phase, we consider three alternative precoding options 
	on top of the effective channel after the BA phase. These are referred to as beam steering (BST), 
	analog maximum ratio transmission (MRT), and joint analog maximum ratio and baseband zeroforcing (MR-ZF), respectively. 
	The proposed schemes are very suitable for practical implementations due to the low-time-overhead and low-complexity. In particular, the MR-ZF precoding scheme proposed in this 
	paper outperforms the state-of-the-art counterparts in the literature.}

{\bf Notation}: We denote vectors, matrices, and scalars by $\bfa$, $\bfA$, and $a$ ($A$), respectively. 
For an integer $K\in\intgr$, $[K]$ denotes the index set $\{1,...,K\}$.
We represent sets by calligraphic $\clA$ and their cardinality with $|\clA|$. 
We use $\bE [\cdot]$ for the expectation, $\|\cdot\|$ for $l_2$-norm, $\circledast$ for continuous-time convolution, $\otimes$ for the  Kronecker product, $\odot$ for Hadamard product.

\section{Channel and Signal Models}\label{mathModel}
\subsection{Channel Model}
	\begin{figure*}[t]
	\centering	
	\begin{subfigure}[b]{0.4\linewidth}
		\centering
		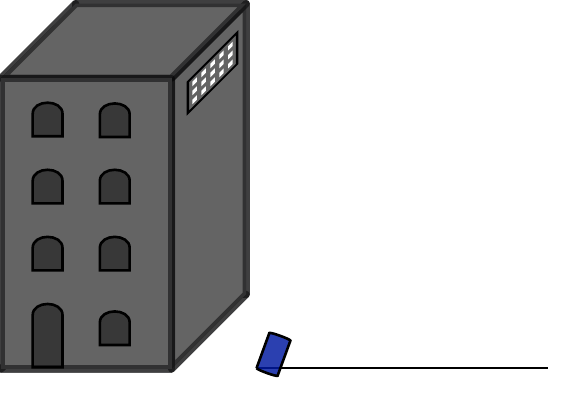
	\end{subfigure}
	\begin{subfigure}[b]{0.4\linewidth}
		\centering
		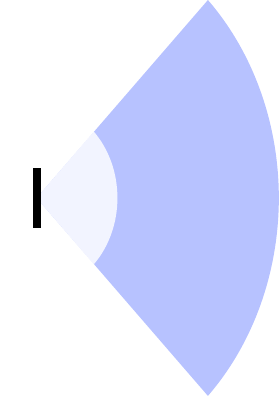
	\end{subfigure}
\vspace{0.7cm}
	\caption{{ \small Illustration of a small cell scenario with (a) 3D side view  and (b) 2D top view. In this paper, the initial beam alignment refers to the beam training/searching in the azimuth plane as shown in (b).}}
	\label{fig:scenario}
\end{figure*}
{ One of the main new features of 5G wireless networks is the densely spread small cell layer \cite{smallcell_Jha}. In small cell configurations as illustrated in \figref{fig:scenario}$\,$(a), the BS creates a fixed arc-like sectorized beam in the elevation direction. The orientation of the BS beam center in the elevation direction tends to be fixed with an elevation angle $\alpha_e$ \cite{smallcell_Saxena}. It follows that the probing area in the range direction is restricted and the intensive initial beam searching takes place mainly in the azimuth direction. For notation simplicity, in this paper we only focus on the 2D azimuth plane. Extension to the 3D geometry is conceptually 
straightforward although may lead to a rather high dimensional search for the initial beam acquisition phase. In the small cell scenario as illustrated in Fig.~\ref{fig:scenario}, where the beam shape in the elevation direction is fixed a priori in order to define the cell footprint area, the 2D azimuth geometry is fully justified.} 
We assume that the BS serving simultaneously $K$ UEs. 
The BS is equipped with a uniform linear array (ULA) of $M$ antennas and $M_{\text{RF}}$ RF chains, 
where $K \leq M_{\text{RF}} \ll M$. Each UE is equipped with a ULA of $N$ antennas and $N_{\text{RF}} \ll N$ RF chains. 
Since the focus of this paper is the BS architecture, we consider the case of $N_{\text{RF}}=1$, where the extension to $N_{\text{RF}} > 1$
is straightforward and was considered in our work on BA \cite{sxsBA2017,sxs2017Time,sxs2018TimeJour}. 
The propagation channel between the BS and the $k$-th UE, $k\in[K]$, consists of $L_k\ll \max\{M,N\}$ {\em significant} multipath components. 
As a result, the $N\times M$ baseband equivalent impulse response of the channel at time slot $s$ can be written as
\begin{align}\label{ch_mod_disc_mp}
\sfH_{s,k}(t,\tau)&=\sum_{l=1}^{L_k} \rho_{s,k,l} e^{j2\pi \nu_{k,l}t}\bfa_{\text{R}}(\phi_{k,l}) \bfa_{\text{T}}(\theta_{k,l})^\herm \delta(\tau-\tau_{k,l})\nonumber\\
&=\sum_{l=1}^{L_k}\sfH_{s,k,l}(t)\delta(\tau-\tau_{k,l}),
\end{align}
where $\sfH_{s,k,l}(t) := \rho_{s,k,l} e^{j2\pi \nu_{k,l}t} \bfa_{\text{R}}(\phi_{k,l}) \bfa_{\text{T}}(\theta_{k,l})^\herm$ and $\delta(\cdot)$ denotes the Dirac delta function. 
Each $l$-th multipath component is identified by the tuple $(\phi_{k,l}, \theta_{k,l}, \tau_{k,l}, \nu_{k,l})$ of angle of arrival (AoA), angle of departure (AoD), 
delay, and Doppler shift, respectively.
The vectors $\bfa_{\text{T}}(\theta_{k,l})\in\bC^{M}$ and $\bfa_{\text{R}}(\phi_{k,l})\in \bC^{N}$ are the array response vectors of the BS and the $k$-th UE at the AoD $\theta_{k,l}$ and the AoA $\phi_{k,l}$, respectively. With the ULA configuration and the assumption that the spacing of the ULA antennas in each array (subarray) equals to a half-wavelength $\lambda/2$, the elements of $\bfa_{\text{T}}(\theta_{k,l})$ and $\bfa_{\text{R}}(\phi_{k,l})$ are given by
\begin{subequations}  \label{array-resp}
	\begin{align}
	&[\bfa_{\text{T}}(\theta)]_{(i'-1)\cdot \hat{M}+d}=e^{j (d-1)\pi \sin(\theta)}\cdot e^{j\varPsi(i',\theta)}, d\in [\hat{M}]\label{a_resp_BS}\\
	&[\bfa_{\text{R}}(\phi)]_n=e^{j (n-1)\pi \sin(\phi)},  n\in[N],\label{a_resp_UE}
	\end{align}
\end{subequations}
where in \eqref{a_resp_BS} we assume that $(i'\equiv1,\hat{M}=M)$ for the FC architecture as shown in \figref{TX}(a), and $(i'\in[M_{\text{RF}}],\hat{M}=\frac{M}{M_{\text{RF}}})$ for the OSPS architecture as shown in \figref{TX}(b). The additional term $\varPsi(i',\theta)$ in \eqref{a_resp_BS} takes into account the phase shifts among different subarrays, given by
\begin{align}\label{subarray_phase}
\varPsi(i',\theta) = \frac{2\pi}{\lambda}(i'-1)\cdot D_x\cdot\sin(\theta),
\end{align}
where $i'$ indicates the index of the subarrays and $D_x\geq0$ denotes the subarray center-to-center spacing in the scan direction. { Hence, in the special case with $D_x=0$,  all the subarrays are co-located$\,$\footnote{ In this paper, we consider a 2D geometry w.r.t. the azimuth plane as illustrated in \figref{fig:scenario}$\,$(b). 
In practice, the co-located layout can be obtained by stacking the arrays on top of each other in the vertical dimension. Strictly speaking this yields a rectangular array configuration, but since each row forms an individually driven array, adaptive beamforming in the elevation direction is not possible, therefore the beamforming geometry is still two-dimensional.}; while with $D_x=\frac{M}{M_{\text{RF}}}\cdot\frac{\lambda}{2}$, the antenna element layout in the scan direction for the OSPS architecture is exactly the same as for the FC architecture.}

For the sake of modeling simplicity, we assume in \eqref{ch_mod_disc_mp} that each multipath component has a very narrow footprint over the AoA-AoD-delay domain. The extension to more widely spread multipath clusters is straightforward and will be applied in the numerical simulations. 
{ We adopt a block fading model, where the coefficient of the $l$-th multipath component 
$\rho_{s,k,l}$ is constant over a short interval (within one slot) and changes from slot to slot according to a wide-sense stationary process statistics characterized by its power spectral density (Doppler spectrum) \cite{john2008digital}. When the channel \textit{coherence time} (related 
to the inverse of the bandwidth of the Doppler spectrum, see \cite{john2008digital}) is significantly larger than 
the slot duration but equal or smaller than the (non-consecutive) slot separation in time, a convenient model is to consider the coefficients as  i.i.d. 
across different slots.  
Moreover, the Doppler shift $\nu_{k,l}$ as defined in \eqref{ch_mod_disc_mp} introduces a continuous  phase rotation for each channel sample.} 
Each multipath component (channel tap coefficient) is formed by the superposition of a large number of micro-scattering components (e.g., due to rough surfaces) having (approximately) the same AoA-AoD and delay.  By the central limit theorem, it is customary to model the superposition of these many small effects as Gaussian \cite{bello1963characterization,goldsmith2005wireless}. Hence, the multipath component coefficients can be modeled as Rice fading given by
\begin{align}\label{rice_fading}
\rho_{s,k,l}\sim \sqrt{\gamma_{k,l}} \left(\sqrt{\frac{\eta_{k,l}}{1+\eta_{k,l}}}+\frac{1}{\sqrt{1+\eta_{k,l}}}\check{\rho}_{s,k,l}\right),
\end{align}
where $\gamma_{k,l}$ denotes the overall multipath component strength, 
$\eta_{k,l}\in[0,\infty)$ indicates the strength ratio between the specular reflection (or LOS) and the  scattered components, 
and { $\check{\rho}_{s,k,l} \sim \cg(0, 1)$ is a zero-mean unit-variance complex Gaussian random variable whose value changes in an i.i.d. fashion across different slots}. 
In particular, $\eta_{k,l}\to \infty$ indicates a pure LOS path while  $\eta_{k,l}=0$ indicates a pure scattered path, 
affected by Rayleigh fading.

The AoA-AoDs $(\phi_{k,l}, \theta_{k,l})$ in \eqref{ch_mod_disc_mp} can take on arbitrary values in the continuous AoA-AoD domain. 
Following the widely used approach of \cite{SayeedVirtualBeam2002}, known as {\em beam-domain representation}, we obtain a finite-dimensional representation of the channel response \eqref{ch_mod_disc_mp}. More precisely, we consider the discrete set of AoA-AoDs
\begin{subequations} \label{theta-phi}
	\begin{align}\label{gridtheta}
	\Phi&:=\left \{\check{\phi}: (1+\sin(\check{\phi}))/2=\frac{n-1}{N}, \, n \in [N] \right \},\\
	\Theta&:=\left \{\check{\theta}: (1+\sin(\check{\theta}))/2=\frac{m-1}{M},  m\in[M]\right \}.
	\end{align}
\end{subequations}
It follows that the corresponding sets $\clA_{\text{R}}:=\{\bfa_{\text{R}}(\check{\phi}): \check{\phi} \in \Phi\}$ and $\clA_{\text{T}}:=\{\bfa_{\text{T}}(\check{\theta}): \check{\theta} \in \Theta\}$ form discrete dictionaries to represent the channel response. For the ULAs considered in this paper, the dictionaries $\clA_{\text{R}}$ and $\clA_{\text{T}}$, after suitable normalization, reduce to the columns of unitary {\em Discrete Fourier Transform} (DFT) matrices $\bfF_{N}\in \bC^{N\times N}$ and $\bfF_{M}\in \bC^{M\times M}$,  with elements{
\begin{subequations}
	\begin{align}
	[\bfF_{N}]_{n,n'}&=\frac{1}{\sqrt{N}}e^{j2\pi (n-1)(\frac{n'-1}{N}-\frac{1}{2})}, n,n'\in[N],\\
	[\bfF_{M}]_{m,m'}&=\frac{1}{\sqrt{M}}e^{j2\pi (m-1)(\frac{m'-1}{M}-\frac{1}{2})}, m, m'\in[M].
	\end{align}
\end{subequations}}
{ Consequently, based on a subarray basis indexed by $i'$, the beam-domain representation of the channel response \eqref{ch_mod_disc_mp} is given by \cite{SayeedVirtualBeam2002,OverviewHeath2016}}
\begin{align}\label{beamspacechannel}
\check{\sfH}_{s,k}^{i'}(t,\tau)& = \bfF_{N}^\herm\sfH_{s,k}(t,\tau)\cdot\left(\bfF_{M}\odot\one_{\{(i'-1)\hat{M}+1:i'\hat{M},1:M\}}\right) = \sum_{l=1}^{L_k}\check{\sfH}_{s,k,l}^{i'}(t)\delta(\tau-\tau_l),
\end{align}
where $(i'\equiv1,\hat{M}=M)$ for the FC architecture, and $(i'\in[M_{\text{RF}}],\hat{M}=\frac{M}{M_{\text{RF}}})$ for the OSPS architecture. Here we define $\check{\sfH}_{s,k,l}^{i'}(t) := \bfF_{N}^\herm\sfH_{s,k,l}(t)\cdot\left(\bfF_{M}\odot\one_{\{(i'-1)\hat{M}+1:i'\hat{M},1:M\}}\right)$ as the beam-domain $l$-th multipath component between the $k$-th UE and the BS, where $\one_{\{a_1:a_2,b_1:b_2\}}\in\bC^{M\times M}$ is an indicator matrix, with $1$ at the components indexed by rows from $a_1$ to $a_2$ and by columns from $b_1$ to $b_2$, otherwise zero. The indicator matrix takes into account the fact that the number of antenna elements for each subarray in the OSPS architecture is $M_{\text{RF}}$ times less than that in the FC architecture.

{ As shown in our earlier work \cite{sxsBA2017} (and the references therein), for the FC architecture, as the number of antennas $M$ at the BS and $N$ at the UE increases, the DFT basis provides a good sparsification of the propagation channel. As a result, $\check{\sfH}_{s,k}^{i'}(t,\tau)$ can be approximated as a 
sparse matrix, with non-zero elements in the locations corresponding to small clusters of discrete AoA-AoD pairs. For the OSPS architecture, note that the indices of non-zero elements in $\check{\sfH}_{s,k}^{i'}(t,\tau)$ are identical for all $i'\in[M_{\text{RF}}]$. However, the channel sparsity depends on the number of antennas in each subarray. In both cases, we may encounter a grid error in \eqref{beamspacechannel} since the AoAs-AoDs do not necessarily fall into the uniform grid $\Phi\times\Theta$. Nevertheless, as shown in \cite{sxsBA2017}, the grid error becomes negligible by increasing the number of (subarray) antennas (i.e., the grid resolution). In our simulations, we do not constrain the AoA-AoD pairs of the physical channel to take on values on the discrete grid; therefore, the grid discretization effect is fully taken into account in our numerical results. }

\subsection{Signaling Model}

Because of space limitation, in this paper we focus on SC signaling. Similar conclusions can be reached for OFDM, although the latter is generally more fragile to frame synchronization errors,  large PAPR, and, before BA is achieved, to inter-carrier interference due to the fact that the Doppler spread between the several multipath components may be large \cite{sxs2018TimeJour, MyungPAPR}. 
Let $\bfx_s(t)=[x_{s,1}(t),x_{s,2}(t),...,x_{s,K}(t)]^\transp$ denote the continuous-time baseband equivalent signal (either pilot or data signal), 
transmitted over the $s$-th slot. With HDA beamforming, the beamformed signal at the output of the transmitter over the $s$-th slot is generally given by
\begin{align}\label{TX_signal}
\hat{\bfx}_s(t)  =\sqrt{E_0}\cdot\bfU_s^{\text{RF}}\cdot \bfW_s^{\text{BB}}\cdot \bfx_s(t),
\end{align}
where for simplicity of exposition we restrict to the case of uniform power allocation, with $E_0=\frac{\ptot T_c}{K}$ indicating the per-chip energy of each signal stream, where $\ptot$ denotes the total radiated power at the BS and $T_c=\frac{1}{B}$ denotes the chip duration with $B$ indicating the signaling bandwidth. In \eqref{TX_signal}, we define $\bfW_s^{\text{BB}}\in\bC^{M_{\text{RF}}\times K}$ and $\bfU_s^{\text{RF}}\in\bC^{M\times M_{\text{RF}}}$ as the baseband (digital) and the RF analog beamforming matrices, respectively. Note that, depending on the transmitter architecture, the analog beamforming matrix $\bfU_s^{\text{RF}}$ takes on the form
\begin{align}\label{Uform}
[\tilde{\bfu}_{s,1}, \tilde{\bfu}_{s,2},\cdots, \tilde{\bfu}_{s,M_{\text{RF}}}] \quad\text{and} \quad \begin{bmatrix}
\tilde{\bfu}_{s,1} & \zerov & \cdots & \zerov \\
\zerov &  \tilde{\bfu}_{s,2} & \cdots & \zerov \\
\vdots & \vdots &\ddots & \vdots \\
\zerov & \zerov & \cdots & \tilde{\bfu}_{s,M_{\text{RF}}}
\end{bmatrix}
\end{align}
{ for the FC (left) and the OSPS (right) architectures, respectively}, where $\tilde{\bfu}_{s,i}\in \bC^{\hat{M}}$, $i\in[M_{\text{RF}}]$, with $\hat{M}=M$ for the FC architecture and $\hat{M}=\frac{M}{M_{\text{RF}}}$ for the OSPS architecture. Hence, in both cases $\bfU_s^{\text{RF}}$ has dimension $M \times M_{\text{RF}}$, but
FC has a full matrix, while OSPS has a block-diagonal matrix, due to the constrained connectivity. 
Without loss of generality, the beamforming vectors are normalized as $\sum_{i=1}^{M_{\text{RF}}}\|\bfu_{s,i}\|^2=M_{\text{RF}}$.

The beamformed signal \eqref{TX_signal} goes through the channel as defined in \eqref{ch_mod_disc_mp}. At the UE side, because of the HDA architecture, the UE does not have direct access to each antenna element. Instead, at each slot $s$, the UE obtains only a projection of the received signal by applying some beamforming vector in the analog domain. We consider a single RF chain at each UE as mentioned before. Thus, the received signal at the $k$-th UE side is given by
\begin{align}\label{data_out1}
\hat{y}_{s,k}(t)=&\bfv_{s,k}^\herm\sfH_{s,k}(t,\tau)\circledast \hat{\bfx}_s(t) +z_{s,k}(t)\nonumber\\
=&\sqrt{E_0}\bfv_{s,k}^\herm\sfH_{s,k}(t,\tau)\circledast \left(\bfU_s^{\text{RF}}\cdot \bfW_s^{\text{BB}}\cdot \bfx_s(t)\right) +z_{s,k}(t),
\end{align}
where ${\bfv_{s,k} \in \bC^N}$ denotes the normalized beamforming vector with $\|\bfv_{s,k}\|=1$ at the $k$-th UE, 
and $z_{s,k}(t)$ is the continuous-time complex {\em Additive White Gaussian Noise} (AWGN) at the output of the UE RF chain, 
with a {\em Power Spectral Density} (PSD) of $N_0$ Watt/Hz. 

In the following, we will evaluate the performance of different transmitter architectures as shown in \figref{TX}. 
For this purpose,  it is useful to first define the channel SNR before beamforming (BBF) $\snrbef$, given by
\begin{align}\label{snrBBF}
\snrbef_{\!,\,k} =\frac{\ptot \sum_{l=1}^{L_k}\gamma_{k,l}}{N_0B}.
\end{align}
{ where $k$ is the index of the UE and $\gamma_{k,l}$ denotes the strength of the $l$-th multipath component. }
The SNR in \eqref{snrBBF} indicates the ratio of the total received signal power (summing over all the multipath components)
over the total noise power at the receiver baseband processor input, assuming that the signal is isotropically transmitted 
by the BS and isotropically received at the $k$-th UE over the total bandwidth $B$. 
As mentioned before, one of the challenges of mmWaves communication is that the SNR 
before beamforming $\snrbef$ in \eqref{snrBBF} may be very low.

\section{Beam Acquisition and Data Transmission}\label{BAandData}
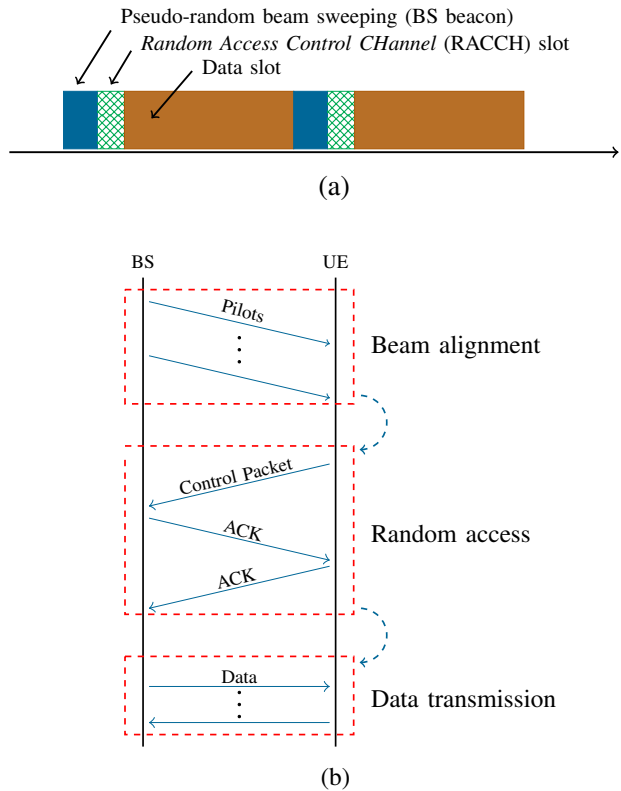
\begin{figure*}[t]
	\centering	
	\begin{subfigure}[b]{1.0\textwidth}
		\centering
		\scalebox{.9}{\input{frame_structure}}%
	\end{subfigure}
	\begin{subfigure}[b]{1.0\textwidth}
		\centering
		\vspace{0.5cm}
		\scalebox{.8}{\input{BA-timeline}}%
	\end{subfigure}
	\caption{{\small Illustration of (a) the frame structure in the underlying system and (b) the communication process between the BS and a generic UE. The initial beam alignment phase is periodically done over beacon slots, followed by a random access stage to build up the connection between the BS and the active UEs, and consecutively the data communication.}}
	\label{2_frame_timeline}
\end{figure*}
We evaluate the performance of the FC and OSPS architectures including both the BA phase and the consequent data transmission phase, where the latter uses the
beam information obtained by the former. For the BA phase we use the scheme proposed in  our previous work \cite{sxs2018TimeJour}, that compares favorably with respect to several competing schemes proposed in the literature. 
For the sake of space limitation, we provide here only 
a high-level summary of the scheme and invite the reader to consider \cite{sxs2018TimeJour} for the full details.
\figref{2_frame_timeline}$\,$(a) illustrates the considered frame structure, which consists of three parts: the beacon slot, the random access control channel (RACCH) slot, and the data slot. As shown in \figref{2_frame_timeline}$\,$(b), the BS broadcasts its pilot signals periodically over the beacon slots. 
The measurements are collected at each UE locally and independently of other UEs. 
Based on measurements accumulated over a sequence of several beacon slots, each UE can estimate a set of strongly coupled AoA-AoD pairs, corresponding to 
the directions of strong propagation paths between the UE and the BS arrays. These determine the beamforming direction for possible data transmission. 
During the RACCH slot, the BS stays in listening mode and the UEs send beamformed uplink packets. 
These packets contain basic information such as the UE ID and the beam indices of the selected BS beam directions. 
The BS responds with an acknowledgment (ACK)  data packet in the data subslot of a next frame, using the indicated beam indices for transmission. 
From this moment on, the BS and the UE are connected in the sense that, if the procedure is successful,   
they can communicate by aligning their beams along a small number of multipath components with strong average 
power transmission.

As explained in details in \cite{sxs2018TimeJour}, the BS beacon signals are formed by $M_{\text{RF}}$ different PN sequences, each of which undergoes a 
``multifinger'' beam pattern obtained by selecting a subset of the columns (or masked DFT columns as in the case of OSPS). 
The beamforming patterns send the signal energy uniformly distributed along subsets of the BS AoD grid. 
The beamforming patterns follow a pre-determined pseudo-random sequence, similar in the spirit to the primary synchronization code of a W-CDMA 3G system for BS identification.  During the beacon slot, each UE $k$ receives using its own pseudo-random sequence of multifinger beam patterns, and integrates the received signal 
energy over the multiple time segments within a beacon slot in order to obtain an estimate of the average received energy. As a result, 
this fully non-coherent energy measurement yields (approximately) the average energy sum of several multipath components. These multipath components corresponds to
the AoA-AoD pairs in the grid for which the BS transmit directions and the UE receive directions meet. 
\figref{cluster} (a) shows an example of transmit and receive multifinger beam patterns and \figref{cluster}$\,$(b) shows the corresponding masks of crossing AoA-AoD directions, superimposed with the second moments (channel gain) of the beam-domain channel matrix generated by the QuaDRiGa simulator. 
The goal of the BA algorithm run at the UE side is to identify the position of the strong components, i.e., the small dark spots in the plot of  \figref{cluster}$\,$(b).
It turns out that this problem can be cast as the reconstruction of a sparse non-negative vector from noisy linear measurements, which can be
efficiently obtained by solving a non-negative least-squares (NNLS) problem. It can be shown that NNLS naturally induce sparsity in the solution, 
and it is very efficient to solve by a plethora of well-known algorithms (e.g., projected gradient). 
The full details of the BA scheme, as well as extensive comparison with other competing state-of-the-art schemes, are provided in \cite{sxs2018TimeJour}. 
\begin{figure*}[t]
	\centering
	\scalebox{.8}{\input{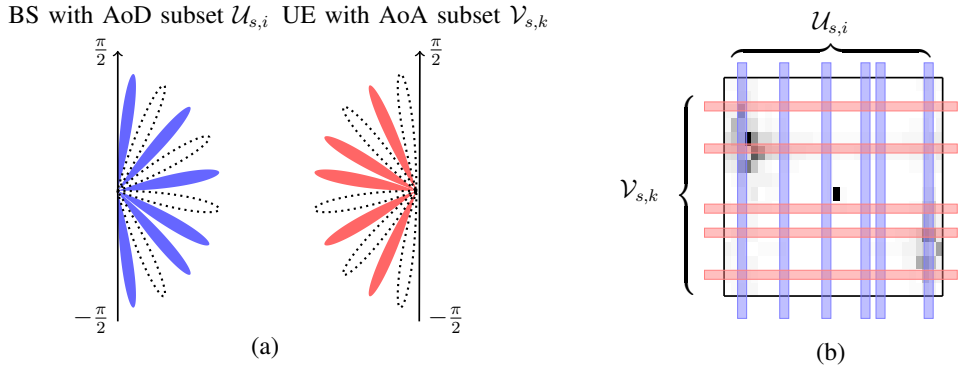}}
	\caption{{\small {(a) Illustration of the subset of AoA-AoDs at time slot $s$ probed by the $i$-th beacon stream transmitted by the BS and received by the $k$-th UE, for $\hat{M} = N = 10$. 				The AoD subset is given by  $\clU_{s,i}=\{1,3,4,6,8,10\}$ { (numbered counterclockwise)} with beamforming vector 	$\check{\bfu}_{s,i}=\frac{1}{\sqrt{6}}[1,0,1,0,1,0,1,1,0,1]^\transp$. The AoA subset is given by $\clV_{s,k}=\{2,4,5,7,9\}$ { (numbered counterclockwise)} with receive beamforming vector $\check{\bfv}_{s,k}=\frac{1}
				{\sqrt{5}}[0,1,0,1,1,0,1,0,1,0]^\transp$.
				(b) The beam-domain channel gain matrix (with one LOS component and two scattered multipath components 
				indicated by the dark spots, generated by the QuaDRiGa simulator) measured along $\clV_{s,k} \times \clU_{s,i}$ }}.}
	\label{cluster}
\end{figure*}


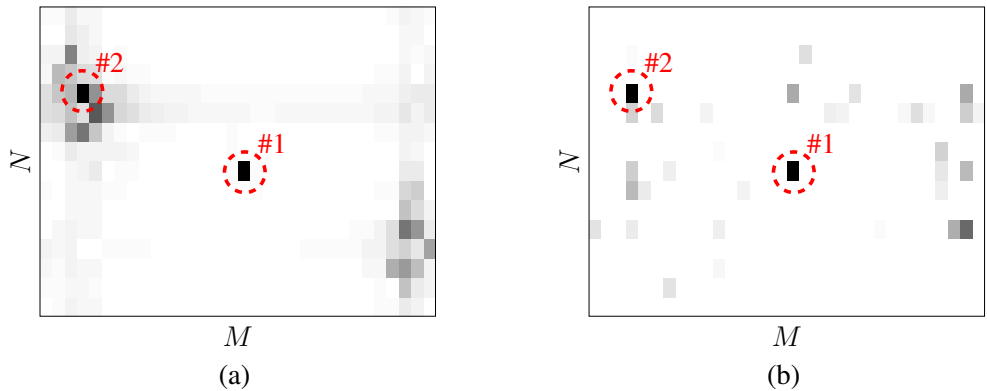
\begin{figure*}[t]
	\centering
	\begin{subfigure}[b]{0.4\textwidth}
		\centering
		\scalebox{.9}{\input{QD_bef}}%
	\end{subfigure}
	\hspace{0.4cm}
	\begin{subfigure}[b]{0.4\textwidth}
		\centering
		\scalebox{.9}{\input{QD_aft}}%
	\end{subfigure}
	\caption{{\small Illustration of the second moments of the beam-domain channel matrix $\Gammam_k$: (a) the actual QuaDRiGa generated $\Gammam_k$, (b) the NNLS estimated $\Gammam_k^\star$. The dashed circles indicate the top $p=2$ strongest components in $\Gammam_k$ and $\Gammam_k^\star$, respectively. We announce a success in the BA phase if the locations of the strongest component in $\Gammam_k$ and in $\Gammam_k^\star$ are consistent.}}
	\label{QD_BA}
\end{figure*}
We denote by $\Gammam_k\in \bC^{N\times M}$ as the matrix of second moments of the beam-domain channel coefficients 
between the BS array and the $k$-th UE array. An example of $\Gammam_k$ is illustrated in \figref{QD_BA}$\,$(a).  Also,  \figref{QD_BA}$\,$(b) shows the estimate 
$\Gammam_k^\star$ of  $\Gammam_k$ provided by the NNLS estimation at UE $k$. 
Once the BA algorithm yields $\Gammam_k^\star$,  the $k$-th UE will send a beamformed 
control packet to the BS in the RACCH. The UE chooses the beamforming direction corresponding to the strongest AoA direction obtained 
from $\Gammam_k^\star$, meanwhile the BS stays in listening mode during the RACCH, using a sectored beamforming configuration. In this way,  
full beamforming gain at the UE transmit side and a limited sector beamforming gain at the BS receive side can be achieved. Once the RACCH packet is received, 
the BS can use the transmit beam indicated by UE $k$ to communicate data. In the next section, we focus on the data communication phase
assuming that the RACCH has been correctly received, therefore, both the BS and the UE know the indices of the strong components in 
$\Gammam^\star_k$. Notice that if the NNLS estimation fails, it is likely that the RAACH will not be received or will be received in error, because the 
beamforming gain at the UE side will be poor. In this case, the UE will not receive a data packet and after a given time-out will try the BA procedure again. 
Also in the (very unlikely) case of a collision in the RACCH, the same time-out procedure can be exploited. Therefore, data communication effectively takes place
only when a) the strong multipath components in $\Gammam_k$ are correctly estimated and b) when the RACCH decoding is successful. 
In \cite{sxs2018TimeJour} we have already argued that the probability that the BA procedure fails is dominated by the error probability in the estimation
of the strong components of $\Gammam_k$. Hence, a sensible system design approach consists of allowing a sufficient 
number of beacon slots such that the probability of success in identifying the strong components of $\Gammam_k$ is close to 1, and designing the HDA beamforming scheme in the assumption that the estimation of $\Gammam_k$ is correct. As a result, we shall compare the FC and OSPS architectures in terms of 
number of beacon slots needed to achieve a BA success probability near $1$, and their achieved spectral efficiency under such condition. 
In any case, the designed HDA precoders in our simulations are always obtained from the true NNLS estimation $\Gammam_k^\star$, and not 
by the genie-aided exact knowledge of $\Gammam_k$.

\subsection{Data Communication Phase}\label{precoding}

We assume that the BS simultaneously schedules $K= M_{\text{RF}}$ UEs. { With the small cell configuration as illustrated in \figref{fig:scenario}, the distance differences between each UE and the BS are very small, implying that the received power w.r.t. the LOS path for each UE within the BS coverage are similar. Although schedulers such as random or proportionate fair scheduler are commonly used in sub 6\,GHz, the directionality of the mmWave channel instead calls for schedulers that select groups of users with good angular separation (directional scheduler) \cite{yunyi2017scheduler}. More precisely, we assume that the selected $K$ UEs have similar received power in terms of the strongest path,
and their strongest AoDs in the downlink are at least $\Delta \theta_{\text{min}}$ away from each other.}

Let $\bfx^{\text{d}}(t)=[x_{1}^{\text{d}}(t),x_{2}^{\text{d}}(t),...,x_{K}^{\text{d}}(t)]^\transp$ denote the complex baseband data signal,\footnote{From now on, we ignore the slot index $s$ for notation simplicity, also because once a successful BA is achieved, the channel statistical property, the precoding vector at the BS, and combining vector at each UE are invariant within many slots. However, note that this invariance holds only until a new updated BA takes place, implying that the underlying channel may encounter large mobility, blockage, etc.} with $x_{k}^{\text{d}}(t)$, $k\in[K]$, corresponding to the $k$-th UE, given by
\begin{align}\label{time_data_sig}
x_{k}^{\text{d}}(t)=\sum_{n=1}^{N_d}d_{k,n} p_{r}(t-nT_c),
\end{align}
where $p_{r}(t)$ is the unit-energy square-root Nyquist pulse shaping filter, $\{d_{k,n}\}$ denote the unit-energy modulation symbols belonging to a suitable modulation constellation \cite{john2008digital}, and $N_d$ indicates the number of the transmit symbols.  Accordingly, the received data signal at the $k$-th UE is given by (refer to \eqref{data_out1})
\begin{align}\label{data_out4}
\hat{y}_{k}(t)&=\sqrt{E_0}\bfv_{k}^\herm\sfH_{k}(t,\tau)\circledast \left(\bfU^{\text{RF}}\cdot \bfW^{\text{BB}}\cdot \bfx^{\text{d}}(t)\right) +z_{k}(t)\nonumber\\
&=\sum_{n=1}^{N_d}\sum_{k'=1}^{K}\sum_{l=1}^{L_k}d_{k',n}\sqrt{E_{0}}\bfv_k^\herm \sfH_{k,l}(t) \bfU^{\text{RF}} \bfw_{k'} p_{r}(t-\tau_{k',l}-nT_c)+z_{k}(t)\nonumber\\
&=\sum_{n=1}^{N_d}\sum_{k'=1}^{K}\sum_{l=1}^{L_k}C_{k,k',l,n}e^{j\Delta_{k,n,l}} p_{r}(t-\tau_{k',l}-nT_c)+z_{k}(t)
\end{align}
where $ \bfw_{k'}$ denotes the $k'$-th column of $\bfW^{\text{BB}}$, $\Delta_{k,n,l}=2\pi(\check{\nu}_{k,l}+\nu_{k,l}nT_c)$, and $C_{k,k',l,n}:=\rho_{k,l}d_{k',n}\sqrt{E_{0}}(\bfv_{k}^\herm\bfa_{\text{R}}(\phi_{k,l}) \bfa_{\text{T}}(\theta_{k,l})^\herm\bfU^{\text{RF}} \bfw_{k'}) $. We assume that each UE  uses standard timing synchronization with respect to its strongest multipath component indexed by $l^1$, which is selected by its initial BA. To decode the data signal, each UE performs matched filtering with respect to the symbol pulse $p_r(t)$, sampling at epochs $t=\hat{n}T_c+\tau_{k,l^1}$. It follows that the discrete-time baseband signal received at the $k$-th UE receiver takes on the form
\begin{align}\label{data_out5}
y_{k}[\hat{n}]&=y_k(t)|_{t=\hat{n}T_c+\tau_{k,l^1}}=\hat{y}_{k}(t)\circledast p_{r}^*(-t)\big|_{t=\hat{n}T_c+\tau_{k,l^1}}\nonumber\\
&=\sum_{n=1}^{N_d}\sum_{k'=1}^{K}\sum_{l=1}^{L_k}C_{k,k',l,n}e^{j\Delta_{k,n,l}} \varphi_{r}\left[\hat{n}^\Delta_{k,k',\hat{n},n,l}\right]+\sum_{n=1}^{N_d}z_{k}^c[\hat{n}]\nonumber\\
&=\sum_{n=1}^{N_d}\!\left(\!\sum_{l=1}^{L_k}\!C_{k,k,l,n}e^{j\Delta_{k,n,l}} \varphi_{r}\left[\hat{n}^\Delta_{k,k,\hat{n},n,l}\right]\!\!+\!\!\!\sum_{k'\neq k}\!\sum_{l=1}^{L_k}\!C_{k,k',l,n}e^{j\Delta_{k,n,l}} \varphi_{r}\left[\hat{n}^\Delta_{k,k',\hat{n},n,l}\right]\!+\!z_{k}^c[\hat{n}]\right),
\end{align}
where  $\hat{n}^\Delta_{k,k',\hat{n},n,l}:=(\hat{n}-n)T_c+\tau_{k,l^1}-\tau_{k',l}$, $\varphi_r[t^\Delta]=\varphi_r(t)|_{t=t^\Delta} := \int p_{r}(\tau)p_{r}^*(\tau-t^\Delta)d\tau$, and $z^{c}_k[\hat{n}]$ denotes the noise at the output of the matched filter with 
variance $N_0\cdot \int |p_{r}(t)|^2dt = N_0$.
As we can see, the first term in \eqref{data_out5} corresponds to the desired data symbol $d_{k,n}$ multiplied by a different complex coefficient over each path $l$.$\,$\footnote{Actually, we have shown in our precious work \cite{sxs2018TimeJour} that, the phase perturbations over several strong paths are easy to compensate by standard carrier synchronization techniques given that a successful BA is achieved and the effective channel after BA has a very small time spreading. Due to the space limit, in (\ref{data_out5})  and also in our simulations, we will keep the phase perturbations such that the numerical results coincide with the conservative worst-case scenario.} Whereas, the last two terms in \eqref{data_out5} correspond to the multiuser interference and noise, respectively. By treating the multiuser interference as noise, the asymptotic ergodic spectral efficiency of the $k$-th UE is given by
\begin{align}\label{rate_k}
R_k = \log_2\left(1+\frac{\bE\left[\big|\sum_{l=1}^{L_k}C_{k,k,l,n}e^{j\Delta_{k,n,l}} \varphi_{r}\left[\hat{n}^\Delta_{k,k,\hat{n},n,l}\right]\big|^2\right]}{\bE\left[\big|\sum_{k'\neq k}\sum_{l=1}^{L_k}C_{k,k',l,n}e^{j\Delta_{k,n,l}} \varphi_{r}\left[\hat{n}^\Delta_{k,k',\hat{n},n,l}\right]\big|^2\right]+N_0}\right),
\end{align}
and the sum rate reads $R_{\text{sum}} = \sum_{k=1}^{K}R_k$.
{ In all schemes treated here, coherent communication can be practically achieved by including per-user beamformed pilot symbols at the cost of a very small
overhead, as it is quite state of art and usual in virtually any modern wireless communication standard.} 
For simplicity, we shall not take into account this overhead or the degradation of quasi-coherent receivers, which is well known and 
not a specific feature of the systems under consideration.  

\subsubsection{Hybrid Precoding Formulation}

Now the remaining problem is how to define the precoding/combining vectors. We assume that the BS communicates with the $k$-th UE along its top-$p$ beams. We will show later that the parameter $p\geq 1$ is somehow a tradeoff between the transmitter power spreading, multiuser interference, and the system robustness to potential blockages. To simplify the practical implementation, we define the combining vector at the $k$-th UE as
\begin{align}\label{receive_vector}
\bfv_{k} = \frac{1}{\sqrt{p}}\Fm_N\cdot\sum_{p'=1}^{p} \check{\bfv}_{k,p'},
\end{align}
where $\check{\bfv}_{k,p'}\in\bC^N$ is an all-zero vector with a $1$ at the component corresponding to the $p'$-th strong AoA, i.e., the AoA index of the $p'$-th strong component in $\Gammam_k^\star$. Denoted by $\bfV\in\bC^{NK\times K}$ as the aggregated receive beamforming matrix given by $\bfV= \diag(\bfv_{1}, \bfv_{2},\, ..., \bfv_{K})$. It follows that the receive data signal vector $\bar{\bfy}(t)=[y_1(t), y_2(t), ..., y_K(t)]^{\transp}\in\bC^{K}$ corresponding to the $K$ UEs can be written as
\begin{align}\label{receive}
\bar{\bfy}(t) &= \sqrt{E_0}\bfV^\herm \cdot \overline{\sfH}(t,\tau) \circledast  \left(\bfU^{\text{RF}}\cdot \bfW^{\text{BB}}\cdot \bfx^{\text{d}}(t)\right) +\bar{\bfz}(t)\nonumber\\
& \overset{\text{(a)}}{=}  \sqrt{E_0}\left(\bfV^\herm \cdot \overline{\sfH}(t,\tau) \cdot \overline{\bfU}\cdot \bfA^{\text{RF}}\cdot\bfW^{\text{BB}}\right)\circledast\bfx^{\text{d}}(t) +\bar{\bfz}(t)\nonumber\\
&\overset{\text{(b)}}{=}  \sqrt{E_0}\left(\widetilde{\sfH}(t,\tau)\cdot \bfA^{\text{RF}}\cdot\bfW^{\text{BB}}\right)\circledast\bfx^{\text{d}}(t) +\bar{\bfz}(t)
\end{align}  
where  $\bar{\bfz}(t)\in\bC^{K}$ indicates the noise vector, $\bfU^{\text{RF}} := \overline{\bfU}\cdot \bfA^{\text{RF}}$ is the analog beamforming matrix, $\widetilde{\sfH}(t,\tau):=\bfV^\herm \cdot \overline{\sfH}(t,\tau) \cdot \overline{\bfU}$ denotes a constructed effective channel, and $\overline{\sfH}_s(t,\tau)\in\bC^{NK\times M}$ represents the aggregated instantaneous channel of all the $K$ UEs given by
\begin{align}
\overline{\sfH}(t,\tau) = \left[\sfH_{1}(t,\tau)^\transp, \sfH_{2}(t,\tau)^\transp, \cdots, \sfH_{K}(t,\tau)^\transp\right]^\transp\!,
\end{align}
where $\sfH_{k}(t,\tau)$, $k\in[K]$, is given in \eqref{ch_mod_disc_mp}. In \eqref{receive}$\,$(a), we define $\overline{\bfU}\in \bC^{M\times pK}$ as the angular support, and $\bfA^{\text{RF}} = [\bfa_1,\bfa_2, ..., \bfa_K]\in\bC^{pK\times K}$ as the coefficient tuning for the analog part. More precisely, we assume $\overline{\bfU}=[\bfU_{1}, ..., \bfU_{K}]$, where $\bfU_{k}\in\bC^{M\times p}$, $k\in[K]$, takes on the form
\begin{align}\label{RF_precoder}
\bfU_{k} =\left(\bfF_{M}\odot\one_{\{(k'-1)\hat{M}+1:k'\hat{M},1:M\}}\right)\cdot[\check{\bfu}_{k,1}, \check{\bfu}_{k,2},\, ..., \check{\bfu}_{k,p}],
\end{align}
where $(i'\equiv1, \hat{M}=M)$ for the FC architecture, and $(i'=k, \hat{M}=\frac{M}{M_{\text{RF}}})$ for the OSPS architecture. Also, we define $\check{\bfu}_{k,p'}\in\bC^{M}$, $p'\in[p]$, as an all-zero vector with a $1$ at the component corresponding to the $p'$-th strongest AoD of $\Gammam_k^\star$.

Notice that in order to construct the beamforming vector at each $k$-th UE and the precoding vectors at the BS, only the AoA-AoD indices of the $p$ strongest components in the estimated channel gain matrix $\Gammam_k^\star$ are needed. Then, once these vectors are fixed, the resulting effective channel has much lower dimensions than the original physical $N \times M$ channel (from array to array). Therefore, it can be estimated using orthogonal uplink pilots and channel reciprocity as in regular TDD MU-MIMO (e.g., see  \cite{shepard2012argos,marzetta2016fundamentals}). 
Namely, the constructed effective channel matrix $\widetilde{\sfH}(t,\tau)$ in \eqref{receive}$\,$(b) has dimension 
$K\times (p\,K)$, and can be estimated using $p K$ uplink pilot sub-slots using TDD reciprocity. 

\subsubsection{Beam Steering (BST) Scheme}
The BST scheme consists of simply steering the $K$ data streams towards the $K$ UEs along their strongest AoD. 
Hence, we have $p=1$ in \eqref{receive_vector} and in \eqref{RF_precoder}, respectively. 
In such case, the analog tuning matrix and the baseband precoding matrices under the BST precoding scheme turn to be identity, i.e., $\bfA^{\text{RF}} = \bfW^{\text{BB}}= \bfI_K$. Note that in the BST scheme, we do not need any additional uplink channel estimation of $\widetilde{\sfH}(t,\tau)$. 
Namely, once the UEs has fed back its strongest AoD control packet, the BS can immediately 
provide the BST precoder. 

\subsubsection{Analog Maximum Ratio Transmission (MRT) Scheme}\label{MR_scheme}

In this scheme, we aims to maximize the desired signal power as well as to increase the scheme blockage robustness. To this end, the baseband precoding matrix remains identity, i.e., $\bfW^{\text{BB}}= \bfI_K$, while the $k$-th analog MRT tuning vector (i.e., the $k$-th column of $\bfA^{\text{RF}}$) is given by
\begin{align}\label{analog_MR_vector}
\bfa_{k} = \left(\widetilde{\sfH}(t,\tau)_{\{k,:\}}\right)^\herm \odot \one_{\{(k'-1)\hat{p}+1:k'\hat{p}\}}\cdot \Delta^{\text{RF}},
\end{align}
where $\widetilde{\sfH}(t,\tau)_{\{k,:\}}$ indicates the $k$-th row of $\widetilde{\sfH}(t,\tau)$, and $\Delta^{\text{RF}}\in \bR_+$ denotes the normalizing factor such that $\sum_{i=1}^{M_{\text{RF}}}\|\bfu_i\|^2=M_{\text{RF}}$. The indicator vector $\one_{\{(k'-1)\hat{p}+1:k'\hat{p}\}}\in\bC^{pK}$ has components $1$ over the index $\{(k'-1)\hat{p}+1:k'\hat{p}\}$ otherwise $0$, where $(k'\equiv 1, \hat{p}=pK)$ for the FC architecture and $(k'= k, \hat{p}=p)$ for the OSPS architecture. Here the indicator vector ensures that, in the OSPS architecture, the analog beamforming matrix $\bfU^{\text{RF}} = \overline{\bfU}\cdot \bfA^{\text{RF}}$ satisfies the block diagonal structure as illustrated in \eqref{Uform}.

\subsubsection{Joint Analog Maximum Ratio and { Baseband} Zeroforcing (MR-ZF) Scheme}
On top of the previous MRT scheme, in this joint MR-ZF scheme, we propose to make use of the baseband precoding to further reduce the multiuser interference. Accordingly, the analog MRT vectors in $\bfA^{\text{RF}}$ are given by \eqref{analog_MR_vector}, while the baseband ZF matrix $\bfW^{\text{BB}}$ takes on the form
\begin{align}\label{AZF}
\bfW^{\text{BB}} = \left(\widetilde{\sfH}(t,\tau)\bfA^{\text{RF}}\right)^\herm\cdot\left(\widetilde{\sfH}(t,\tau)\bfA^{\text{RF}}\left(\widetilde{\sfH}(t,\tau)\bfA^{\text{RF}}\right)^\herm\right)^{-1}\cdot \Delta^{\text{ZF}},
\end{align}
where $\Delta^{\text{ZF}}\in \bR_+$ is the normalizing factor ensuring the total radiated power constraint, i.e., $\sum_{k=1}^{K}\|\bfw_k\|^2=K$.

\section{Hardware Impairments}

In all the above derivations, we have implicitly assumed that all the hardware components work in their ideal range without any distortion or power dissipation. However, in practical hardware systems, such assumption is not trivial to meet. For example, the implementation of HDA transceivers consists of a large number of power dividers and combiners in the analog part, particularly for the FC architecture. The power dissipation caused by these components has a severe impact on the transmit power and the power efficiency. Moreover, due to the superposition of multiple beamformed pilots$\,$/$\,$data, the input signal at the PAs may encounter a large PAPR. Also, different beamforming vectors will create different power levels for different PAs. As a result, the input power for some individual PAs may exceed their saturation limit (relevant to per-antenna power constraint) and even cause a disruption of the whole transmission. All these hardware impairment have a severe impact on the transmitter performance and should not be neglected. In this section, we will provide the mathematical model to evaluate the hardware efficiency of different transmitter architectures given in \figref{TX}.

We assume that each analog path has simultaneous amplitude and phase control as shown in \figref{TX}. Refer to \eqref{TX_signal}, let $\tilde{\bfx}\in\bC^M$ denote the pre-amplified beamformed signal\footnote{For notation simplicity, here we ignored the slot index $s$ and the time index $t$.}, given by
\begin{align}\label{beamformedX}
\tilde{\bfx} = \sqrt{\alpha_{\text{com}}}\cdot\widetilde{\bfU}^{\text{RF}}\cdot \sqrt{\alpha_{\text{div}}} \cdot \bfW^{\text{BB}}\cdot\bfx,
\end{align}
where $\bfx=[x_1,\cdots, x_{K}]\in\bC^{K}$ denotes the transmit symbol, with $\bE[|x_i|^2]=\epsilon$, $i\in[K]$. The factor $\alpha_{\text{div}}$ indicates the power splitting at the divider, with $\alpha_{\text{div}}=\frac{1}{M}$ for the FC architecture as shown in \figref{TX}$\,$(a) and $\alpha_{\text{div}}=\frac{M_\text{RF}}{M}$ for the OSPS architecture as shown in \figref{TX}$\,$(b). Moreover, the factor $\alpha_{\text{com}}$ models the power dissipation factor of the combiners, i.e., $\alpha_{\text{com}} = \frac{1}{M_\text{RF}}$ for the FC architecture, and $\alpha_{\text{com}} = 1$ for the OSPS architecture. Both $\alpha_{\text{div}}$ and $\alpha_{\text{com}}$ result from the hardware implementation and are based on the corresponding S-parameters of the dividers and combiners as in \cite{JingboSubFull2018}. We assume that the baseband beamforming matrix $\bfW^{\text{BB}}$ is of dimension $K\times K$ with $K=M_{\text{RF}}$, and the analog beamforming matrix $\widetilde{\bfU}^{\text{RF}}=[\bfu_1,..., \bfu_{M_{\text{RF}}}]\in\bC^{M\times M_{\text{RF}}}$ satisfies the specific FC$\,$/$\,$OSPS architecture as illustrated in \eqref{Uform}. 

We consider the rather simple BST precoding with $\bfW^{\text{BB}}=\bfI_K$. To first meet the total power constraint, for any $i\in[M_{\text{RF}}]$, we have $\|\bfu_i\|^2=M$ for the FC architecture and $\|\bfu_i\|^2=\frac{M}{M_{\text{RF}}}$ for the OSPS architecture, respectively. It follows that the effective pre-amplified radiated power of the beamformed signal $\tilde{\bfx}$ in \eqref{beamformedX} can be written as
\begin{align}
\tilde{P} &= \bE[\tilde{\bfx}^\herm\tilde{\bfx}]=\alpha_{\text{com}}\alpha_{\text{div}}\cdot\bE[\bfx^\herm(\widetilde{\bfU}^{\text{RF}})^\herm\widetilde{\bfU}^{\text{RF}}\bfx]=\alpha_{\text{com}}\alpha_{\text{div}}\cdot\trace\left(\bE[\bfx\bfx^\herm]\cdot(\widetilde{\bfU}^{\text{RF}})^\herm\widetilde{\bfU}^{\text{RF}}\right).
\end{align}
Accordingly, the pre-amplified radiated power for the FC and the OSPS architectures reads 
$\tilde{P}_{\text{FC}} = \epsilon M_{\text{RF}} \frac{1}{M_{\text{RF}}}$ and $\tilde{P}_{\text{OSPS}}=\epsilon M_{\text{RF}}$, respectively.  As we can see, in order to achieve the same output power, the FC transmitter should compensate for an additional combiner power dissipation. More precisely, the transmitter should either boost the input signal as $M_{\text{RF}} \bfx$ or choose PAs with 
larger gain for the amplification stage. We consider the former approach and mathematically include the potential boosting factor $M_{\text{RF}}$ as well as the factors $(\alpha_{\text{com}},\alpha_{\text{div}})$ into the beamforming matrix $\widetilde{\bfU}^{\text{RF}}$. Denoted by $\bfU^{\text{RF}}$ as the integrated analog beamforming matrix, such that the pre-amplified beamformed signal in \eqref{beamformedX} can be written as $\tilde{\bfx} = \bfU^{\text{RF}}\cdot \bfW^{\text{BB}}\cdot\bfx$, which is consistent with our assumptions and formulations in Section \ref{mathModel}.

The beamformed signal \eqref{beamformedX} then goes through the amplification stage, where at each antenna branch a PA amplifies the signal before transmission. We assume that the PAs in different antenna branches have
the same input-output relation. For any given antenna in the transmitter array, let $P_{\text{rad}}$ denote the radiated power of the antenna, and $P_{\text{cons}}$ denote the consumed power of the corresponding PA, which includes both the radiated power and the dissipated power. Following the approach in \cite{Moghadam2018}, the  power consumed by the PA takes on the form
\begin{align}
P_{\text{cons}} = \frac{\sqrt{P_{\text{max}}}}{\eta_{\text{max}}}\sqrt{P_{\text{rad}}},
\end{align}
where $P_{\text{max}}$ is the maximum output power of the PA with $P_{\text{rad}}\leq P_{\text{max}}$ and $\eta_{\text{max}}$ is the maximum efficiency of the PA. Note that this relation holds for the most common PA implementations and is therefore a good choice for the following calculation. Considering that the PAs are often the predominant power consumption part, we define $\eta_{\text{eff}}$ given by
\begin{align}
\eta_{\text{eff}} = \frac{P_{\text{rad}}}{P_{\text{cons}}}
\end{align}
as the metric to effectively compare the power efficiency of the two transmitter architectures shown in \figref{TX}. Note that due to the superposition of multiple beamforming vectors (particularly for the FC architecture) and the potentially high PAPR of the time-domain transmit waveform $\tilde{\bfx}$ in \eqref{beamformedX} (particularly with OFDM signaling), the input power for some individual PA may exceed its saturation limit. This would result in non-linear distortion and even the disruption of the whole transmission. To compare the two transmitter architectures and ensure that all the underlying $M$ PAs simultaneously work in their linear range, we generally have two options:

{\em\textbf{ Option$\,$I:}} Both the FC architecture and the OSPS architecture utilize the same PA but apply a different input back-off $\alpha_{\text{off}}\in(0,1]$, such that the peak power of the radiated signal is smaller than $P_{\text{max}}$. As a reference, we denote by $(P_{\text{rad},0},\eta_{\text{max},0})$ as the parameters of a reference PA under the reference precoding/beamforming strategy with a power backoff factor $\alpha_{\text{off},0}$ (as illustrated later in Section \ref{simulation}). For different scenarios (with certain $\alpha_{\text{off}}$) the average radiated power and the consumed power take the form $P_{\text{rad}} = \frac{\alpha_{\text{off}}}{\alpha_{\text{off},0}}P_{\text{rad},0}$, $P_{\text{cons}} = \frac{\sqrt{P_{\text{max},0}}}{\eta_{\text{max},0}}\sqrt{P_{\text{rad}}}$. The transmitter power efficiency is given by
\begin{align}\label{eff1}
\eta_{\text{eff}} = \frac{P_{\text{rad}}}{P_{\text{cons}}} = \frac{\sqrt{P_{\text{rad}}}\cdot\eta_{\text{max},0}}{\sqrt{P_{\text{max},0}}}.
\end{align}

{\em\textbf{ Option$\,$II:}} We choose to deploy different PAs for the FC architecture and the OSPS architecture. More precisely, we assume that the underlying PA has a maximum output power of $P_{\max}\! =\! \frac{\alpha_{\text{off},0}}{\alpha_{\text{off}}}P_{\text{max},0}$, where $\alpha_{\text{off}}$ has the same value as in {\em Option I}. Consequently, the average radiated power and the consumed power of the underlying PA can be written as $P_{\text{rad}} = P_{\text{rad},0}$, $P_{\text{cons}}\! = \!\frac{\sqrt{P_{\text{max},0}\cdot\alpha_{\text{off},0}/\alpha_{\text{off}}}}{\eta_{\text{max}}}\sqrt{P_{\text{rad}}}$. The transmitter power efficiency is given by
\begin{align}\label{eff2}
\eta_{\text{eff}} = \frac{P_{\text{rad}}}{P_{\text{cons}}} = \frac{\sqrt{P_{\text{rad}}}\cdot\eta_{\text{max}}}{\sqrt{P_{\text{max},0}\cdot \alpha_{\text{off},0}}}\cdot \sqrt{\alpha_{\text{off}}}.
\end{align}

Note that the characteristics $(P_{\text{max}}$ and $\eta_{\text{max}})$ of different PAs highly depend on the operation frequency, implementation, and technology. Aiming at illustrating how to apply the proposed analysis framework in practical system design, we will exemplify a set of PA parameters in Section \ref{simulation} to evaluate the efficiency $\eta_{\text{eff}}$ of the two architectures given in \figref{TX}. For the comparison of BA and data communication algorithms, we are interested in the performance of the corresponding algorithms using the different transmitter architectures but with the same channel condition as in \eqref{snrBBF}. Therefore, we assume the same total radiated power $\ptot$ constraint for both architectures in \figref{TX}. In practical systems, this assumption can be satisfied by applying a certain power backoff as in {\em Option$\,$I} or chosing different PAs  as in {\em Option$\,$II}. This in addition fulfills the per-antenna power constraint, such that all the underlying PAs work in their linear range with an identical scalar gain. However, we will show in Section \ref{simulation} that, under the same radiated power constraint, different architectures may have a different power efficiency.

\section{Numerical Evaluation}\label{simulation}
We now present the numerical results to evaluate the proposed precoding schemes and to illustrate the performance of different transmitter architectures as shown in \figref{TX}. The BA scheme was already extensively studied in \cite{sxs2017Time,sxs2018TimeJour,sxsBA2017} in terms of 
complexity, system-level scalability, and robustness to fast channel time-variations$\,$/$\,$large Doppler spread. Hence, here we focus only on the 
difference in time-to-successful BA required by the two BS architectures under comparison. 
{ We consider a system with a BS using $M=128$ antennas and $M_{\text{RF}}=4$ RF chains. The BS simultaneously schedules $K=M_{\text{RF}}=4$ UEs, each of which uses $N=16$ antennas and $N_{\text{RF}}=1$ RF chain.} We assume a short preamble structure used in IEEE 802.11ad \cite{Time2017,ParameterPerahia2010}, where the beacon slot is of duration $t_0S=1.891\,\mu$s. The system is assumed to work at $f_0=40\,$GHz with a bandwidth of $B=0.8\,$GHz, namely, each beacon slot amounts to more than $1500$ chips.

In the following simulations, otherwise stated, we will assume a fixed total radiated power constraint $\ptot$, where all the underlying 
PAs working in their linear range (w.r.t., per-antenna power constraint $P_{\max}$) with an identical scalar gain. The MU-MIMO channel is generated in two ways:

{1) In Section \ref{evalua_precoding} and Section \ref{FC_OSPS_simulation}, we use the channel model in \eqref{ch_mod_disc_mp} to generate the channel matrix between each UE $k$ and the BS. Based on the practical mmWave MIMO channel measurements in \cite{Tim2018}, we assume $L_k=3$, $k\in[K]$, multipath components for each UE, given by $(\gamma_{k,1}=1,\eta_{k,1}=100)$, $(\gamma_{k,2}=0.6,\eta_{k,2}=10)$ and $(\gamma_{k,3}=0.4,\eta_{k,3}=0)$ with respect to \eqref{rice_fading}. Thus, the first link can be roughly regarded as the LOS path, while the remaining links represent the non-LOS (NLOS) paths. 
{ We also assume that the LOS paths for the simultaneously scheduled UEs are well separated in the beam domain, while all the NLOS paths are generated in a random way}.

{2) In Section \ref{quadriga}, we use the quasi-deterministic radio channel generator (QuaDRiGa) to generate the propagation channel matrix. { The channel model is based on the 3GPP-3D urban micro-cell configuration \cite{quadriga}. In this case, the height of the BS antenna array is set to $10\,$m. The beam center of the BS orientates to the ground with an elevation angle\footnote{In QuaDRiGa, the elevation angle $90^{\circ}$ points to the zenith and  $0^{\circ}$ points to the horizon.} of $\alpha_e=-20^{\circ}$ as shown in \figref{fig:scenario}$\,$(a). The simultaneous scheduled UEs are set to $1.5\,$m in height and $18\sim25\,$m horizontally away from the BS with a downlink AoD difference of $\Delta\theta_{\text{min}}\approx8^{\circ}$ \cite{manual}. Each UE $k$ moves towards the BS at a speed of $\Delta v_{k}=1\,$m$/$s.} We will show that the numerical results based on our proposed channel model \eqref{ch_mod_disc_mp} are consistent with the results based on the QuaDRiGa generator, implying that the proposed work not only theoretically but also practically provides valuable references for mmWave system design.

\subsection{Evaluation of the Proposed Precoding Schemes}\label{evalua_precoding}

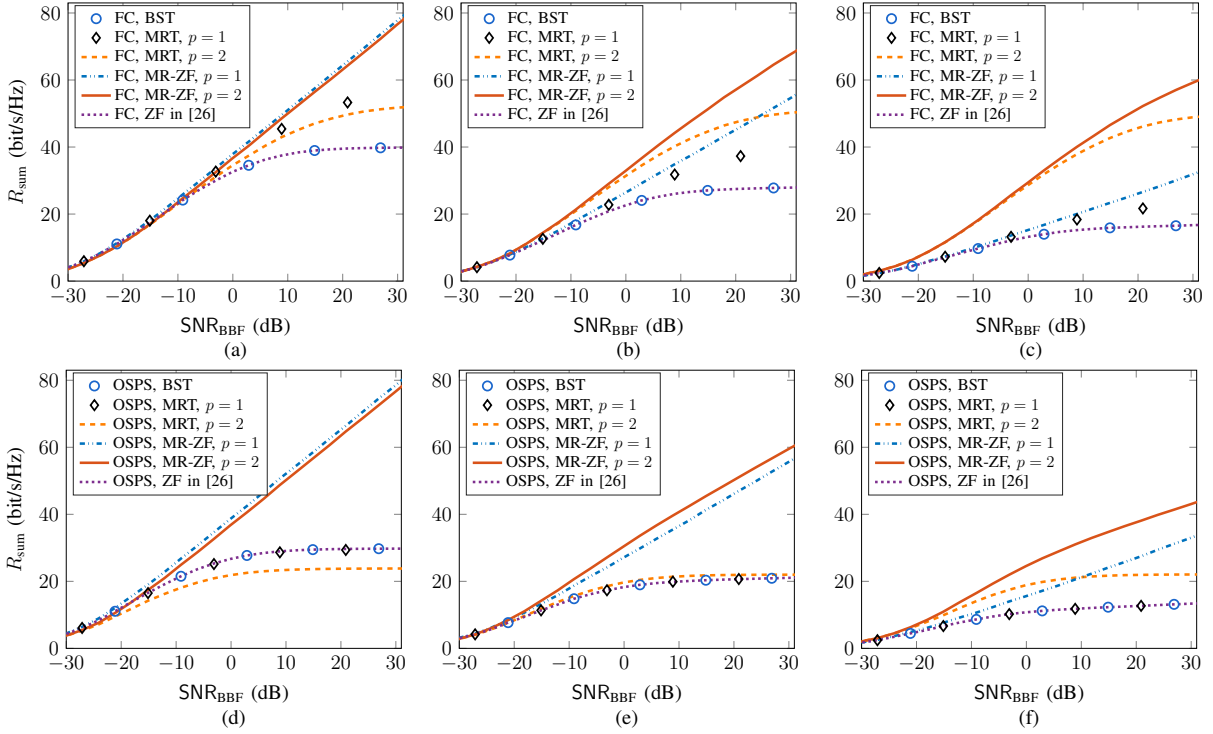
\begin{figure*}[t]
	\centering
	\hspace{-1.1cm}
	\begin{subfigure}[b]{0.3\textwidth}
		\centering
		\scalebox{.65}{\input{1_rate_fully_block0}}%
	\end{subfigure}
	\hspace{0.4cm}
	\begin{subfigure}[b]{0.3\textwidth}
		\centering
		\scalebox{.65}{\input{1_rate_fully_block1}}%
	\end{subfigure}
	\hspace{0.1cm}
	\begin{subfigure}[b]{0.3\textwidth}
		\centering
		\scalebox{.65}{\input{1_rate_fully_block2}}%
	\end{subfigure}\hspace{-1cm}

	\begin{subfigure}[b]{0.3\textwidth}
		\centering
		\scalebox{.65}{\input{1_rate_sub_block0}}%
	\end{subfigure}
	\hspace{0.4cm}
	\begin{subfigure}[b]{0.3\textwidth}
		\centering
		\scalebox{.65}{\input{1_rate_sub_block1}}%
	\end{subfigure}
	\hspace{0.1cm}
	\begin{subfigure}[b]{0.3\textwidth}
		\centering
		\scalebox{.65}{\input{1_rate_sub_block2}}%
	\end{subfigure}
	\caption{{ \small The sum spectral efficiency vs. increasing $\snrbef$. The blockage probability of the strongest path is given by (a) $0.0$, (b) $0.3$, (c) $0.6$ for the FC architecture, and (d) $0.0$, (e) $0.3$, (f) $0.6$ for the OSPS architecture.}}
	\label{FC_OSPS_SE_SNR}
\end{figure*}
The efficiency of the proposed precoding schemes are illustrated in \figref{FC_OSPS_SE_SNR}. As a comparison, we also simulate the ZF precoder proposed in \cite{yunyi2017scheduler}, where the effective channel is approximated by the initial BA vectors, and only a single path is selected between each UE and the BS. As we can see from \figref{FC_OSPS_SE_SNR}$\,$(a), for the FC architecture with no blockage, all the schemes coincide with each other in the range of $\snrbef\leq 0\,$dB. { Whereas when $\snrbef> 0\,$dB, the performance ranking of the underlying precoding schemes is as follow (MR-ZF,$\,p=2$)$\,\approx\,$(MR-ZF,$\,p=1$)$\,>\,$(MRT,$\,p=1$)$\,>\,$(MRT,$\,p=2$)$\,>\,$(BST)$\,\approx\,$(ZF in \cite{yunyi2017scheduler}). Here the MRT scheme with $p=2$ performs worse than with $p=1$ due to the fact of power spreading and the fact that with multiple receiving directions, the UE tends to have more interference. However, this effect is not observable in the MR-ZF scheme because of the further power coefficient tuning and interference cancellation, which results from the baseband zeroforcing.} Next,
by increasing the blockage probability of the strongest path while remaining unblocked for all the less strong paths between each UE and the BS, as shown in \figref{FC_OSPS_SE_SNR}$\,$(b) and \figref{FC_OSPS_SE_SNR}$\,$(c), the curves with $p=2$ drops much less than the others (equivalent to $p=1$), and the scheme of MR-ZF with $p=2$ achieves the best performance. For the OSPS architecture,  when there is no blockage as shown in \figref{FC_OSPS_SE_SNR}$\,$(d), in the low SNR range ($\snrbef\leq -10\,$dB), all the curves (roughly) coincide with each other. { Whereas, by increasing $\snrbef>-10\,$dB, the precoding schemes rank (MR-ZF,$\,p=2$)$\,\approx\,$(MR-ZF,$\,p=1$)$\,>\,$(MRT,$\,p=1$)$\,\approx\,$(BST)$\,\approx\,$(ZF in \cite{yunyi2017scheduler})$\,>\,$(MRT,$\,p=2$).}  Similar with the FC case, the MR-ZF scheme for the OSPS architecture achieves the best performance when increasing the blockage probability as shown in \figref{FC_OSPS_SE_SNR}$\,$(e) and \figref{FC_OSPS_SE_SNR}$\,$(f). As a brief summary w.r.t. the given scenario, for both architectures, when the channel SNR is weak and there is no blockage, we claim that the BST scheme is preferred since it is rather simple but adequate to achieve good performance. However, when the channel SNR is not too weak or there are potential blockages, the MR-ZF scheme with $p> 1$ outperforms the other schemes. As a side note, in practical implementation, the choice of $p$ should not be too large since it plays a trade-off between blockage robustness, power spreading and the overhead for additional channel estimation.

\subsection{Fully-Connected (FC) or One-Stream-Per-Subarray (OSPS)?}\label{FC_OSPS_simulation}
\begin{figure*}[t]
	\centering
	\begin{subfigure}[b]{0.4\textwidth}
		\centering
		\scalebox{.65}{\input{1-BA-compare}}
	\end{subfigure}
	\begin{subfigure}[b]{0.4\textwidth}
		\centering 
		\scalebox{.65}{\input{3-compare}}
	\end{subfigure}
	\begin{subfigure}[b]{0.4\textwidth}
		\centering
		\scalebox{.65}{\input{1-P-P}}
	\end{subfigure}
	\begin{subfigure}[b]{0.4\textwidth}
		\centering  
		\scalebox{.65}{\input{1-P-eff}}
	\end{subfigure}
	\caption{{ \small The performance comparison of different transmitter architectures. (a) The initial BA detection probability vs. the training overhead with $\snrbef=-19$ dB. (b) The sum spectral efficiency vs. increasing $\snrbef$, without blockage. (c) The actual radiated power under {\em Option I} vs. the radiated power of the reference scenario. (d) The power efficiency under {\em Option II} vs. the actual radiated power.}}
	\label{compare_FC_OSPS}
\end{figure*}
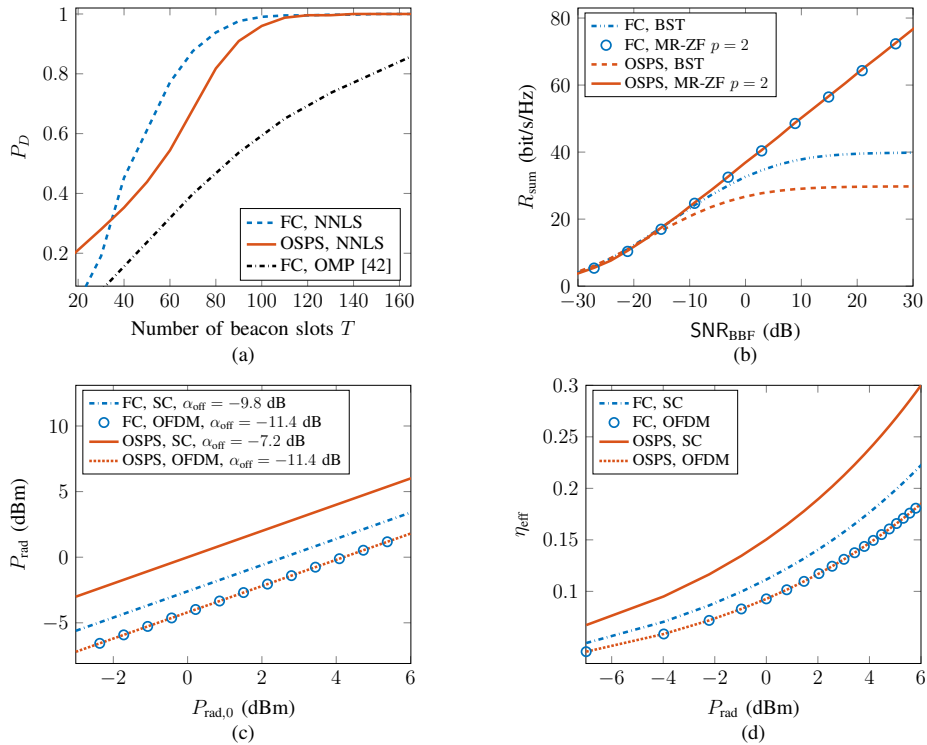
Note that the performance of different architectures highly depends on the channel condition and the underlying precoders. On top of the given scenario in this paper, we jointly evaluate the architecture performance in three aspects: 

{\bf Training efficiency for the initial BA phase.} Let $P_D$ denote the detection probability, i.e., the probability of finding the strongest AoA-AoD pair between the BS and a generic UE. The BA results are illustrated in \figref{compare_FC_OSPS}$\,$(a). As a comparison, we also simulate a recent time-domain BA algorithm proposed in \cite{AlkhateebTimeDomain2017}, which focuses on estimating the instantaneous channel coefficients with an orthogonal matching pursuit (OMP) technique. As we can see, the proposed BA scheme requires much less training overhead than that in \cite{AlkhateebTimeDomain2017}. In addition, due to the fact that the OSPS architecture has lower angular resolution and encounters larger sidelobe power leakage than the FC case, the former requires moderately $\sim 10$ more beacon slots than the latter for $P_D\geq 0.95$.

{\bf Spectral efficiency for the data communication phase.} To compare the spectral efficiency of the two transmitter architectures as shown in \figref{TX}, we consider a no-blockage scenario and focus on two precoding schemes, i.e., the simple BST scheme and the high-performance MR-ZF scheme with $p=2$. As we can see in \figref{compare_FC_OSPS}$\,$(b), in the range of $\snrbef\leq -10\,$dB, which is more relevant in mmWave channels, all the $4$ curves coincide with each other. Namely, for either the MR-ZF scheme or the BST scheme, the two architectures achieve a rather similar spectral efficiency. In contrast, when $\snrbef> -10\,$dB, the MR-ZF scheme performs better. The two architectures with the MR-ZF precoding again achieve a rather similar performance.

{\bf Hardware power efficiency.} To evaluate the architecture power efficiency, otherwise stated, we will consider the simple BST precoder. Also, since the modulation highly affects the power efficiency, we will take into account both the SC and the OFDM signaling in this section. We first assume a reference scenario as the baseline, i.e, the OSPS architecture using the BST precoder and a SC modulation. We use reference PAs with $P_{\text{max},0}=6$ dBm and $\eta_{\text{max},0}=0.3$. The backoff factor with respect to different waveforms and transmitter architectures can be written as  $\alpha_{\text{off}}=1/(P_{\text{PAPR}})$, where $P_{\text{PAPR}}$ represents the PAPR of the input signal at a PA. { The investigation for 3GPP LTE in \cite{MyungPAPR} showed that with a probability of $0.999$, the PAPR of the LTE SC waveform { (known as SC-FDMA)} is smaller than $\sim7.2\,$dB and the PAPR of the LTE OFDM waveform (with $512$ subcarriers employing QPSK) is smaller than $\sim11.4\,$dB.} We set $P_{\text{PAPR}}$ to these values for the OSPS architecture. For the FC architecture, however, the input signals of the PAs are the sum of the signals from different RF chains. Since each OFDM signal can be modeled as a Gaussian random process \cite{MyungPAPR} and the signals from different RF chains are independent,  the PAPR of the sum is the same as of one RF chain. For the case of SC signaling, there is no clear work in the literature that shows how the sum of SC signals behaves. { We simulated the sum of $M_{\text{RF}}=4$ SC signals using the same parameters as in \cite{MyungPAPR}. The result shows that with probability of $0.999$ the PAPR of the sum is smaller than $\sim9.8\,$dB. We apply these values and without loss of generality, we choose $\alpha_{\text{off},0} = -7.2\,$dB as the reference scenario.} As shown in \eqref{eff1}, by deploying the same PAs ({\em Option I}), the two architectures achieve the same efficiency for a given $P_{\text{rad}}$. However, as illustrated in \figref{compare_FC_OSPS}$\,$(c), the 
OSPS architecture with SC signaling (OSPS, SC) achieves the highest $P_{\text{rad}}$, followed by (FC, SC), (OSPS, OFDM), and (FC, OFDM). In contrast, by deploying different PAs ({\em Option II})\footnote{Since the $\eta_{\text{max}}$ of different PAs highly depends on the technology, for simplicity, we assume that different PAs working in their linear range have roughly the same maximum efficiency $\eta_{\text{max},0}$.}, \figref{compare_FC_OSPS}$\,$(d) shows that (OSPS, SC) achieves the highest power efficiency, followed by (FC, SC), (OSPS, OFDM) and (FC, OFDM).

To sum up, given the parameters in this paper, the two architectures achieve a similar sum spectral efficiency with certain precoders, but the OSPS architecture outperforms the FC case in terms of hardware complexity and power efficiency, only at the cost of a slightly longer latency for the initial BA.

\subsection{Simulations Based on QuaDRiGa}\label{quadriga}
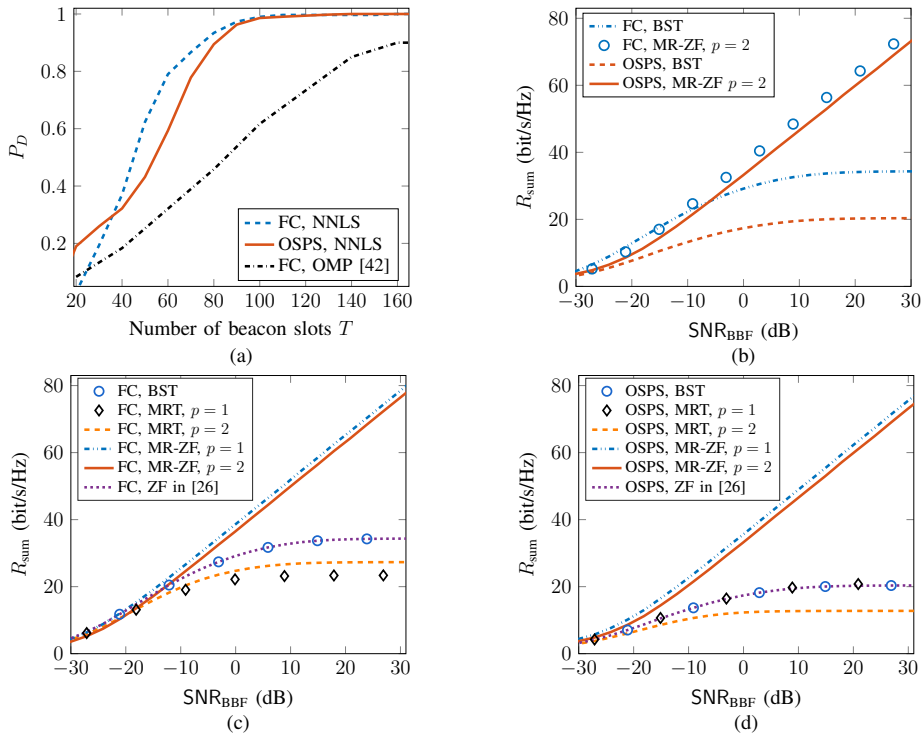
\begin{figure*}[t]
	\centering
	\begin{subfigure}[b]{0.4\textwidth}
		\centering
		\scalebox{.65}{\input{1-BA-compare-QuaDriga}}
	\end{subfigure}
	\begin{subfigure}[b]{0.4\textwidth}
		\centering 
		\scalebox{.65}{\input{3-compare-QuaDriGa}}
	\end{subfigure}
	\begin{subfigure}[b]{0.4\textwidth}
		\centering
		\scalebox{.65}{\input{1-fully-rate-snr-QuaDriGa}}
	\end{subfigure}
	\begin{subfigure}[b]{0.4\textwidth}
		\centering  
		\scalebox{.65}{\input{2-sub-rate-snr-QuaDriGa}}
	\end{subfigure}
	\caption{{ \small The simulations based on QuaDRiGa: (a) The initial BA detection probability vs. the training overhead, with $\snrbef=-19$ dB. (b) The sum spectral efficiency of different transmitter architectures vs. increasing $\snrbef$. (c) The sum spectral efficiency of the FC architecture vs. increasing $\snrbef$. (d) The sum spectral efficiency of the OSPS architecture vs. increasing $\snrbef$.}}
	\label{QuaD}
\end{figure*}
In this section, we resort to the 3D geometry based channel generator QuaDRiGa \cite{quadriga} to show that our numerical results are quite consistent with practical mmWave communication channels.\footnote{Due to the QuaDRiGa generator limits, only the no-blockage scenario is considered in this section.} More precisely, we apply our BA and precoding schemes over $\sim3\times 10^5$  channel snapshots generated by QuaDRiGa. These channel snapshots correspond to a short segment of time evolution, where the BS is stationary and the speed of each UE along its moving direction is  $1\,$m$/$s. The simulation results with respect to different transmitter architectures are shown in \figref{QuaD}. As we can see from \figref{QuaD}$\,$(a), for the initial BA with $P_D\geq 0.95$, the FC architecture requires $\sim10$ less beacon slots than the OSPS case. Whereas, for the data communication phase as shown in \figref{QuaD}$\,$(b), by using either the BST or the MR-ZF precoder in the low SNR range ($\snrbef\leq-15\,$dB), and using the MR-ZF precoder in the high SNR range ($\snrbef>-15\,$dB), the two architectures achieve a quite similar performance. In addition, for both architectures as shown in \figref{QuaD}$\,$(c) and \figref{QuaD}$\,$(d), respectively, all the curves coincides with each other in the low SNR range, whereas the MR-ZF precoder outperforms the rest in the high SNR range. As we can see, all the results based on the QuaDRiGa generator are quite consistent with the results based on our proposed channel model. This consistency implies that our models, schemes, results and statements are not only theoretically reliable but also practically applicable.

\section{Conclusion}

In this paper, we proposed an analysis framework to evaluate the performance of typical hybrid transmitters at mmWave frequencies. 
In particular, we focused on the comparison of a fully-connected (FC) architecture  and a partially-connected architecture with one-stream-per-subarray (OSPS) for a MU-MIMO base station
using HDA beamforming.   We jointly evaluated the performance of the two architectures in terms of the initial  beam alignment (BA), the data communication, and the transmitter power efficiency.  We used our recently proposed BA scheme and further proposed three simple precoding schemes on top of the effective channel after the BA. The precoding schemes are based on beam steering (BST), analog maximum ratio transmitting (MRT), and joint analog maximum ratio and baseband zero-forcing (MR-ZF), respectively. Particularly, both the BA scheme and the MR-ZF precoding scheme outperform the state-of-the-art counterparts in the literature. 
Given the parameters in this paper, our simulation results show that the two architectures achieve a similar sum spectral efficiency, 
but the OSPS architecture outperforms the FC case in terms of hardware complexity and power efficiency, 
only at the cost of a slightly longer latency for the initial BA. Therefore, the OSPS architecture emerges as a good choice
for a simple and efficient design of MU-MIMO base stations operating at mmWave.




\balance
{\footnotesize
	\bibliographystyle{IEEEtran}
	\bibliography{references}
}

\end{document}

%% file: symbols.tex
\usepackage{mathbbol}

\def\mindex#1{\index{#1}}

\def\sq{\hbox{\rlap{$\sqcap$}$\sqcup$}}
\def\qed{\ifmmode\sq\else{\unskip\nobreak\hfil
\penalty50\hskip1em\null\nobreak\hfil\sq
\parfillskip=0pt\finalhyphendemerits=0\endgraf}\fi\medskip}

\long\def\defbox#1{\framebox[.9\hsize][c]{\parbox{.85\hsize}{%
\parindent=0pt
\baselineskip=12pt plus .1pt      
\parskip=6pt plus 1.5pt minus 1pt 
 #1}}}

\long\def\beginbox#1\endbox{\subsection*{}%
\hbox{\hspace{.05\hsize}\defbox{\medskip#1\bigskip}}%
\subsection*{}}

\def\endbox{}

\def\diag{{\text{diag}}}

\newsavebox{\junk}
\savebox{\junk}[1.6mm]{\hbox{$|\!|\!|$}}

\newcommand{\field}[1]{\mathbb{#1}}

\def\intgr{\field{Z}}

\def\bC{{\mathbb C}}

\def\bE{{\mathbb E}}

\def\bR{{\mathbb R}}

\def\bfA{{\bf A}}

\def\bfF{{\bf F}}

\def\bfI{{\bf I}}

\def\bfU{{\bf U}}
\def\bfV{{\bf V}}
\def\bfW{{\bf W}}

\def\bfa{{\bf a}}

\def\bfu{{\bf u}}
\def\bfv{{\bf v}}
\def\bfw{{\bf w}}
\def\bfx{{\bf x}}
\def\bfy{{\bf y}}
\def\bfz{{\bf z}}

\def\tto{{\mathtt o}}

\def\ttt{{\mathtt t}}

\def\sfH{{\sf H}}

\def\bfmath#1{{\mathchoice{\mbox{\boldmath$#1$}}%
{\mbox{\boldmath$#1$}}%
{\mbox{\boldmath$\scriptstyle#1$}}%
{\mbox{\boldmath$\scriptscriptstyle#1$}}}}

\def\bfmY{\bfmath{Y}}

\def\bfmhhaY{\bfmath{\hhaY}} 
\def\bfmhhaY{\hbox to 0pt{$\widehat{\bfmY}$\hss}\widehat{\phantom{\raise 1.25pt\hbox{$\bfmY$}}}}

\def\til={{\widetilde =}}

\def\clA{{\cal A}}

\def\clC{{\cal C}}

\def\clN{{\cal N}}

\def\clU{{\cal U}}
\def\clV{{\cal V}}

 \def\FRAC#1#2#3{\genfrac{}{}{}{#1}{#2}{#3}}

\def\ddtp{{\mathchoice{\FRAC{1}{d^{\hbox to 2pt{\rm\tiny +\hss}}}{dt}}%
{\FRAC{1}{d^{\hbox to 2pt{\rm\tiny +\hss}}}{dt}}%
{\FRAC{3}{d^{\hbox to 2pt{\rm\tiny +\hss}}}{dt}}%
{\FRAC{3}{d^{\hbox to 2pt{\rm\tiny +\hss}}}{dt}}}}

\def\average#1,#2,{{1\over #2} \sum_{#1}^{#2}}

\def\eye(#1){{\bf(#1)}\quad}

\def\eq#1/{(\ref{e:#1})}

\newcommand{\beqn}[1]{\notes{#1}%
\begin{eqnarray} \elabel{#1}}

\newcommand{\eeqn}{\end{eqnarray} }

\newcommand{\beq}[1]{\notes{#1}%
\begin{equation}\elabel{#1}}

\newcommand{\eeq}{\end{equation}}

\def\bdes{\begin{description}}
\def\edes{\end{description}}

\newcounter{rmnum}

\newcounter{anum}

{\end{list}}

\def\ass(#1:#2){(#1\ref{#1:#2})}

\def\ritem#1{
\item[{\sf \ass(\current_model:#1)}]
}

\newenvironment{recall-ass}[1]{%
\begin{description}
\def\current_model{#1}}{
\end{description}
}

\def\ptot{{P_{\ttt \tto \ttt}}}

%% file: macros.tex
\long\def\comment#1{}

\newcommand{\zerov}{{\bf 0}}

\newcommand{\Fm}{{\bf F}}

\newcommand{\Gammam}{\boldsymbol{\Gamma}}

\newcommand{\trace}{{\hbox{tr}}}

\newcommand{\SINR}{{\sf SINR}}

\newcommand{\transp}{{\sf T}}



%% file: scenario_3d.pdf_tex
\begingroup%
  \makeatletter%
  \providecommand\color[2][]{%
    \errmessage{(Inkscape) Color is used for the text in Inkscape, but the package 'color.sty' is not loaded}%
    \renewcommand\color[2][]{}%
  }%
  \providecommand\transparent[1]{%
    \errmessage{(Inkscape) Transparency is used (non-zero) for the text in Inkscape, but the package 'transparent.sty' is not loaded}%
    \renewcommand\transparent[1]{}%
  }%
  \providecommand\rotatebox[2]{#2}%
  \ifx\svgwidth\undefined%
    \setlength{\unitlength}{163.83481844bp}%
    \ifx\svgscale\undefined%
      \relax%
    \else%
      \setlength{\unitlength}{\unitlength * \real{\svgscale}}%
    \fi%
  \else%
    \setlength{\unitlength}{\svgwidth}%
  \fi%
  \global\let\svgwidth\undefined%
  \global\let\svgscale\undefined%
  \makeatother%
  \begin{picture}(1,0.7104283)%
    \put(1.9419743,2.44177484){\color[rgb]{0,0,0}\makebox(0,0)[b]{\smash{}}}%
    \put(0,0){\includegraphics[width=\unitlength,page=1]{scenario_3d.pdf}}%
    \put(0.43931498,-0.01){\color[rgb]{0,0,0}\makebox(0,0)[lb]{\smash{UE1}}}%
    \put(0,0){\includegraphics[width=\unitlength,page=2]{scenario_3d.pdf}}%
    \put(0.36193671,0.30294479){\color[rgb]{0,0,0}\makebox(0,0)[rb]{\smash{$h$}}}%
    \put(0.48141125,0.52214293){\color[rgb]{0,0,0}\makebox(0,0)[lb]{\smash{$\alpha_2$}}}%
    \put(0.83576495,0.22010924){\color[rgb]{0,0,0}\makebox(0,0)[b]{\smash{$d_2$}}}%
    \put(0.48514933,0.16311541){\color[rgb]{0,0,0}\makebox(0,0)[b]{\smash{$d_1$}}}%
    \put(0,0){\includegraphics[width=\unitlength,page=3]{scenario_3d.pdf}}%
    \put(0.92474594,-0.01){\color[rgb]{0,0,0}\makebox(0,0)[lb]{\smash{UE2}}}%
    \put(0,0){\includegraphics[width=\unitlength,page=4]{scenario_3d.pdf}}%
    \put(0.58463336,0.24848668){\color[rgb]{0,0,0}\makebox(0,0)[lb]{\smash{$\alpha_1$}}}%
    \put(0,0){\includegraphics[width=\unitlength,page=5]{scenario_3d.pdf}}%
    \put(0.57655597,0.4796502){\color[rgb]{1,0,0}\makebox(0,0)[lb]{\smash{$\alpha_e$}}}%
    
    \put(0.45,-0.16){\color[rgb]{0,0,0}\makebox(0,0)[lb]{(a)}}%
  \end{picture}%
\endgroup%

%% file: scenario_top.pdf_tex
\begingroup%
  \makeatletter%
  \providecommand\color[2][]{%
    \errmessage{(Inkscape) Color is used for the text in Inkscape, but the package 'color.sty' is not loaded}%
    \renewcommand\color[2][]{}%
  }%
  \providecommand\transparent[1]{%
    \errmessage{(Inkscape) Transparency is used (non-zero) for the text in Inkscape, but the package 'transparent.sty' is not loaded}%
    \renewcommand\transparent[1]{}%
  }%
  \providecommand\rotatebox[2]{#2}%
  \ifx\svgwidth\undefined%
    \setlength{\unitlength}{80.30621446bp}%
    \ifx\svgscale\undefined%
      \relax%
    \else%
      \setlength{\unitlength}{\unitlength * \real{\svgscale}}%
    \fi%
  \else%
    \setlength{\unitlength}{\svgwidth}%
  \fi%
  \global\let\svgwidth\undefined%
  \global\let\svgscale\undefined%
  \makeatother%
  \begin{picture}(1,1.41978989)%
    \put(0,0){\includegraphics[width=\unitlength,page=1]{scenario_top.pdf}}%
    \put(0.10459151,0.68276124){\color[rgb]{0,0,0}\makebox(0,0)[rb]{\smash{BS}}}%
    \put(0,0){\includegraphics[width=\unitlength,page=2]{scenario_top.pdf}}%
    \put(0.43762829,0.33967064){\color[rgb]{0,0,0}\makebox(0,0)[b]{\smash{UE1}}}%
    \put(0,0){\includegraphics[width=\unitlength,page=3]{scenario_top.pdf}}%
    \put(0.90864456,0.76283872){\color[rgb]{0,0,0}\makebox(0,0)[b]{\smash{UE2}}}%
    \put(0,0){\includegraphics[width=\unitlength,page=4]{scenario_top.pdf}}%
    
    \put(0.45,-0.28){\color[rgb]{0,0,0}\makebox(0,0)[lb]{(b)}}%
  \end{picture}%
\endgroup%

%% file: frame_structure.tex
	
\begin{tikzpicture}
	\draw [->, thick] (-0.5,7)--(8.5,7);
	\draw [color={rgb:red,0;green,149;blue,221}, fill={rgb:red,0;green,149;blue,221}] (0.3,7.05) rectangle (0.8,7.9);
	\draw [color={rgb:red,0;green,128;blue,64}] (0.8,7.9)--(1.2,7.9);
	\draw [color={rgb:red,0;green,128;blue,64}] (0.8,7.05)--(1.2,7.05);
	\pattern[pattern=crosshatch, pattern color={rgb:red,0;green,128;blue,64}] (0.8,7.05)--(0.8,7.9)--(1.2,7.9)--(1.2,7.05)--cycle;
	\draw [color={rgb:red,128;green,64;blue,0}, fill= {rgb:red,128;green,64;blue,0},fill opacity=0.85] (1.2,7.05) rectangle (3.7,7.9);	
	
	\draw [color={rgb:red,0;green,149;blue,221}, fill={rgb:red,0;green,149;blue,221}] (0.3+3.4,7.05) rectangle (0.8+3.4,7.9);	
	\draw [color={rgb:red,0;green,128;blue,64}] (0.8+3.4,7.9)--(1.2+3.4,7.9);
	\draw [color={rgb:red,0;green,128;blue,64}] (0.8+3.4,7.05)--(1.2+3.4,7.05);
	\pattern[pattern=crosshatch, pattern color={rgb:red,0;green,128;blue,64}] (0.8+3.4,7.05)--(0.8+3.4,7.9)--(1.2+3.4,7.9)--(1.2+3.4,7.05)--cycle;
	\draw [color={rgb:red,128;green,64;blue,0}, fill= {rgb:red,128;green,64;blue,0},fill opacity=0.85] (1.2+3.4,7.05) rectangle (3.7+3.4,7.9);

	\draw [->, thick] (1,8.8)--(0.5,8);
	\node [above right] at (1,8.7) {\footnotesize Pseudo-random beam sweeping (BS beacon)};
	\draw [->, thick] (1.3,8.5)--(1,8);
	\node [above right] at (1.3,8.3) {\footnotesize {\em Random Access Control CHannel} (RACCH) slot};
	
	\draw [->, thick] (2.2,8.2)--(1.6,7.5);
	\node [above right] at (2.2,8) {\footnotesize Data slot};
	
	\node [below] at (4.3,6.8) {(a)};		

\end{tikzpicture}

%% file: BA-timeline.tex
%
%
	\begin{tikzpicture}
	\draw[help lines,color=white, opacity=0] (2.5,0.0) grid (9,9);
	
	\begin{scope}[transform canvas={xshift = 0cm, yshift = 0cm}]
		
		\draw [thick] (2.9,8.5)--(2.9,0.7);
		\draw [thick] (6.1,8.5)--(6.1,0.7);
		\node[above] at (2.9,8.5) {\footnotesize BS};
		\node[above] at (6.1,8.5) {\footnotesize UE};
		
		\draw [decoration={markings,mark=at position 1 with
			{\arrow[scale=1.0]{>}}},,postaction={decorate},color={rgb:red,0;green,149;blue,221}] (3.0,8.1)--(6,7.4);

		\node [above, rotate = -15] at (4.5,7.7) {\footnotesize Pilots};		
		\draw [fill] (4.5,7.1) circle [radius = 0.02];
		\draw [fill] (4.5,7.3) circle [radius = 0.02];
		\draw [fill] (4.5,7.5) circle [radius = 0.02];		
		\draw [decoration={markings,mark=at position 1 with
			{\arrow[scale=1.0]{>}}},,postaction={decorate},color={rgb:red,0;green,149;blue,221}] (3.0,7.2)--(6,6.5);
		\node [right] at (6.4,7.35) {$\,\,$Beam alignment};
		\draw [dashed, color = red, thick] (2.6,8.3)--(6.4,8.3)--(6.4,6.4)--(2.6,6.4)--(2.6,8.3);

		\draw [<-,color={rgb:red,0;green,149;blue,221},dashed,thick] (6.5,5.64) arc [radius=0.45, start angle=-90, end angle= 90];

		\begin{scope}[transform canvas={xshift = 0cm, yshift = 0.3cm}]
			\draw [decoration={markings,mark=at position 1 with
			{\arrow[scale=1.5]{>}}},,postaction={decorate},color={rgb:red,0;green,149;blue,221}] (6,5.1)--(3.0,4.4);
			\node [above, rotate = 13] at (4.5,4.65) {\footnotesize  Control Packet};

		
			\draw [decoration={markings,mark=at position 1 with
			{\arrow[scale=1.5]{>}}},,postaction={decorate},color={rgb:red,0;green,149;blue,221}] (3.0,5.0-0.8)--(6,4.3-0.8);
			\node [above, rotate = -15] at (4.5,4.5-0.75) {\footnotesize  ACK};
			
			\draw [decoration={markings,mark=at position 1 with
			{\arrow[scale=1.5]{>}}},postaction={decorate},color={rgb:red,0;green,149;blue,221}] (6.0,4.2-0.8)--(3,3.5-0.8);
			\node [above, rotate = 13] at (4.5,3.7-0.75) {\footnotesize  ACK};
		
			\draw [dashed, color = red, thick] (2.6,5.4)--(6.4,5.4)--(6.4,2.6)--(2.6,2.6)--(2.6,5.2);
			\node [right] at (6.4,3.95) {$\,\,$Random access};
		\end{scope}
		
		\draw [<-,color={rgb:red,0;green,149;blue,221},dashed,thick] (6.5,2.1) arc [radius=0.45, start angle=-90, end angle= 90];
		
		\draw [decoration={markings,mark=at position 1 with
		{\arrow[scale=1.5]{>}}},postaction={decorate},color={rgb:red,0;green,149;blue,221}] (3.0,2.5-0.8)--(6,2.5-0.8);
	
		\draw [fill] (4.5,1.6) circle [radius = 0.02];
		\draw [fill] (4.5,1.4) circle [radius = 0.02];
		\draw [fill] (4.5,1.2) circle [radius = 0.02];	
	
		\draw [decoration={markings,mark=at position 1 with
		{\arrow[scale=1.5]{>}}},postaction={decorate},color={rgb:red,0;green,149;blue,221}] (6.0,1.1)--(3,1.1);
	
		\node [above] at (4.5,2.4-0.8) {\footnotesize  Data};
		
		\draw [dashed, color = red, thick] (2.6,2.2)--(6.4,2.2)--(6.4,0.9)--(2.6,0.9)--(2.6,2.2);
		
		\node [right] at (6.4,1.5) {$\,\,$Data transmission};
	
	\end{scope}
	
	\node [below] at (6.1,0.5) {(b)};

	\end{tikzpicture}
	

%% file: QD_bef.tex
%
%
\begin{tikzpicture}

\begin{axis}[%
width=2.3in,
height=1.8in,
at={(0.758in,0.481in)},
scale only axis,
axis on top,
xmin=0.5,
xmax=32.5,
xtick={\empty},
xlabel = {$M$},
y dir=reverse,
ymin=0.5,
ymax=16.5,
ytick={\empty},
ylabel={$N$}
]
\addplot [forget plot] graphics [xmin=0.5, xmax=32.5, ymin=0.5, ymax=16.5] {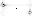};
\end{axis}
\node [above] at (4.8,0) {(a)};

\draw[dashed,red, line width=1.5pt] (4.95,3.35) circle (0.3cm);
\node [above right, red] at (5.0,3.48) {\#1};

\draw[dashed,red, line width=1.5pt] (2.55,4.55) circle (0.3cm);
\node [above right, red] at (2.6,4.68) {\#2};

\end{tikzpicture}%

%% file: QD_aft.tex
%
%
\begin{tikzpicture}

\begin{axis}[%
width=2.3in,
height=1.8in,
at={(0.758in,0.481in)},
scale only axis,
axis on top,
xmin=0.5,
xmax=32.5,
xtick={\empty},
xlabel = {$M$},
y dir=reverse,
ymin=0.5,
ymax=16.5,
ytick={\empty},
ylabel={$N$}
]
\addplot [forget plot] graphics [xmin=0.5, xmax=32.5, ymin=0.5, ymax=16.5] {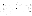};
\end{axis}

\node [above] at (4.8,0) {(b)};

\draw[dashed,red, line width=1.5pt] (4.95,3.35) circle (0.3cm);
\node [above right, red] at (5.0,3.48) {\#1};

\draw[dashed,red, line width=1.5pt] (2.55,4.55) circle (0.3cm);
\node [above right, red] at (2.6,4.68) {\#2};

\end{tikzpicture}%

%% file: 1_rate_fully_block0.tex
%
%
\definecolor{mycolor1}{rgb}{0.00000,0.44700,0.74100}%
\definecolor{mycolor3}{rgb}{0.85000,0.32500,0.09800}%
\definecolor{mycolor2}{rgb}{0.92900,0.69400,0.12500}%
\definecolor{mycolor4}{rgb}{0.49400,0.18400,0.55600}%
\definecolor{mycolor5}{rgb}{0.46600,0.67400,0.18800}%
\definecolor{mycolor6}{rgb}{0.10000,0.40100,0.8000}%
\begin{tikzpicture}

\begin{axis}[%
xmin=-30,
xmax=31,
xlabel={$\snrbef$ (dB)},
ymin=0,
ymax=83,
ylabel={$R_{\text{sum}}$ (bit/s/Hz)},
title style={at={(0.5,-0.02)},anchor=north,yshift=-1.2cm},
title = {(a)},
legend style={at={(0.02,0.55)}, nodes={scale=0.8, transform shape}, anchor=south west, legend cell align=left, align=left, draw=white!15!black}
]
\addplot [only marks, line width = 1.0pt, mark size=2.7pt, mark=o, mark options={solid, mycolor6},mark repeat=4,mark phase=3]
  table[row sep=crcr]{%
-33.1132995230379	2.60020052656018\\
-30.1132995230379	3.97053017323255\\
-27.1132995230379	5.84094069546417\\
-24.1132995230379	8.27311547797526\\
-21.1132995230379	11.0780128522433\\
-18.1132995230379	14.1943768053639\\
-15.1132995230379	17.4522329399139\\
-12.1132995230379	20.7649791758005\\
-9.11329952303793	24.1689911462114\\
-6.11329952303793	27.218246330641\\
-3.11329952303793	29.9957358341098\\
-0.113299523037931	32.5596664358839\\
2.88670047696207	34.5472721204632\\
5.88670047696207	36.2589297350353\\
8.88670047696207	37.4732559812558\\
11.8867004769621	38.3976440860537\\
14.8867004769621	38.9644366388394\\
17.8867004769621	39.3316896943072\\
20.8867004769621	39.5662451229206\\
23.8867004769621	39.6917173015248\\
26.8867004769621	39.7712961464956\\
29.8867004769621	39.8195499020598\\
32.8867004769621	39.8549708123029\\
};
\addlegendentry{FC, BST}

\addplot [only marks, line width = 1.0pt, mark size=3pt, mark=diamond, mark options={solid, black},mark repeat=4,mark phase=1]
  table[row sep=crcr]{%
-33.1132995230379	2.63092295877343\\
-30.1132995230379	4.02464463198627\\
-27.1132995230379	5.92989333506986\\
-24.1132995230379	8.42487375633187\\
-21.1132995230379	11.3304685637688\\
-18.1132995230379	14.558591640798\\
-15.1132995230379	18.0213550046953\\
-12.1132995230379	21.6228709793278\\
-9.11329952303793	25.4604661270357\\
-6.11329952303793	29.1008100201866\\
-3.11329952303793	32.6702602102845\\
-0.113299523037931	36.2605033075222\\
2.88670047696207	39.4583684746683\\
5.88670047696207	42.5969215042596\\
8.88670047696207	45.4110019919135\\
11.8867004769621	47.9515842155369\\
14.8867004769621	50.0985601488904\\
17.8867004769621	51.9001284741449\\
20.8867004769621	53.2968661846696\\
23.8867004769621	54.3186619071126\\
26.8867004769621	55.0866629418197\\
29.8867004769621	55.5835449449466\\
32.8867004769621	55.9045069368937\\
};
\addlegendentry{FC, MRT, $p=1$}

\addplot [color=orange, dashed, line width=1.5pt]
  table[row sep=crcr]{%
-33.1132995230379	2.3413335363685\\
-30.1132995230379	3.51457723910593\\
-27.1132995230379	5.39183648585944\\
-24.1132995230379	7.69062378560817\\
-21.1132995230379	10.3503106183114\\
-18.1132995230379	13.4139402448847\\
-15.1132995230379	16.8297100863796\\
-12.1132995230379	20.4004558834258\\
-9.11329952303793	24.1476711239928\\
-6.11329952303793	27.6547686050698\\
-3.11329952303793	31.1078622649327\\
-0.113299523037931	34.4796368000474\\
2.88670047696207	37.4177329634438\\
5.88670047696207	40.364155413374\\
8.88670047696207	42.9555490202523\\
11.8867004769621	45.107802278458\\
14.8867004769621	47.0240434122463\\
17.8867004769621	48.4720894238178\\
20.8867004769621	49.6900578765903\\
23.8867004769621	50.6215531454323\\
26.8867004769621	51.2665022483775\\
29.8867004769621	51.7450871690502\\
32.8867004769621	52.0713517751568\\
};
\addlegendentry{FC, MRT, $p=2$}

\addplot [color=mycolor1, dashdotdotted, line width=1.5pt]
table[row sep=crcr]{%
-33.1132995230379	2.59150430252167\\
-30.1132995230379	3.97178601600433\\
-27.1132995230379	5.87235647188098\\
-24.1132995230379	8.36761259511824\\
-21.1132995230379	11.3075246477635\\
-18.1132995230379	14.5790287426809\\
-15.1132995230379	18.1176561072966\\
-12.1132995230379	21.8569454201913\\
-9.11329952303793	25.8532293252036\\
-6.11329952303793	29.7714382726767\\
-3.11329952303793	33.694415516524\\
-0.113299523037931	37.7711436958669\\
2.88670047696207	41.5960032145201\\
5.88670047696207	45.5848169232549\\
8.88670047696207	49.524114490783\\
11.8867004769621	53.5443609098486\\
14.8867004769621	57.4711441757215\\
17.8867004769621	61.5553146101827\\
20.8867004769621	65.5352487721018\\
23.8867004769621	69.5490759118722\\
26.8867004769621	73.6434434031832\\
29.8867004769621	77.6610192092075\\
32.8867004769621	81.607050097759\\
};
\addlegendentry{FC, MR-ZF, $p=1$}

\addplot [color=mycolor3, line width=1.5pt]
table[row sep=crcr]{%
-33.1132995230379	2.31534055845656\\
-30.1132995230379	3.48734815434876\\
-27.1132995230379	5.37189534216527\\
-24.1132995230379	7.67408846351626\\
-21.1132995230379	10.3630091671397\\
-18.1132995230379	13.4912891006412\\
-15.1132995230379	16.9720627895696\\
-12.1132995230379	20.7195499684753\\
-9.11329952303793	24.714245751862\\
-6.11329952303793	28.5587580137156\\
-3.11329952303793	32.5016974179287\\
-0.113299523037931	36.5764182806629\\
2.88670047696207	40.3627639221253\\
5.88670047696207	44.4784489439384\\
8.88670047696207	48.538216298723\\
11.8867004769621	52.4760256976933\\
14.8867004769621	56.4390195687979\\
17.8867004769621	60.3071928133991\\
20.8867004769621	64.3016289300879\\
23.8867004769621	68.2906226554492\\
26.8867004769621	72.3077073122472\\
29.8867004769621	76.521473553026\\
32.8867004769621	80.4568813063385\\
};
\addlegendentry{FC, MR-ZF, $p=2$}

\addplot [color=mycolor4, dotted, line width=1.5pt]
  table[row sep=crcr]{%
-33.1132995230379	2.60020052656018\\
-30.1132995230379	3.97053017323255\\
-27.1132995230379	5.84094069546417\\
-24.1132995230379	8.27311547797526\\
-21.1132995230379	11.0780128522433\\
-18.1132995230379	14.1943768053639\\
-15.1132995230379	17.4522329399139\\
-12.1132995230379	20.7649791758005\\
-9.11329952303793	24.1689911462114\\
-6.11329952303793	27.218246330641\\
-3.11329952303793	29.9957358341098\\
-0.113299523037931	32.5596664358839\\
2.88670047696207	34.5472721204632\\
5.88670047696207	36.2589297350353\\
8.88670047696207	37.4732559812558\\
11.8867004769621	38.3976440860537\\
14.8867004769621	38.9644366388394\\
17.8867004769621	39.3316896943072\\
20.8867004769621	39.5662451229206\\
23.8867004769621	39.6917173015248\\
26.8867004769621	39.7712961464956\\
29.8867004769621	39.8195499020599\\
32.8867004769621	39.8549708123029\\
};
\addlegendentry{FC, ZF in \cite{yunyi2017scheduler}}

\end{axis}
\end{tikzpicture}%

%% file: 1_rate_fully_block1.tex
%
%
\definecolor{mycolor1}{rgb}{0.00000,0.44700,0.74100}%
\definecolor{mycolor3}{rgb}{0.85000,0.32500,0.09800}%
\definecolor{mycolor2}{rgb}{0.92900,0.69400,0.12500}%
\definecolor{mycolor4}{rgb}{0.49400,0.18400,0.55600}%
\definecolor{mycolor5}{rgb}{0.46600,0.67400,0.18800}%
\definecolor{mycolor6}{rgb}{0.10000,0.40100,0.8000}%
\begin{tikzpicture}

\begin{axis}[%
xmin=-30,
xmax=31,
xlabel={$\snrbef$ (dB)},
ymin=0,
ymax=83,
title style={at={(0.5,-0.02)},anchor=north,yshift=-1.2cm},
title = {(b)},
legend style={at={(0.02,0.55)}, nodes={scale=0.8, transform shape}, anchor=south west, legend cell align=left, align=left, draw=white!15!black}
]
\addplot [only marks, line width = 1.0pt, mark size=2.7pt, mark=o, mark options={solid, mycolor6},mark repeat=4,mark phase=3]
  table[row sep=crcr]{%
-33.1132995230379	1.75713123397638\\
-30.1132995230379	2.82939057159659\\
-27.1132995230379	4.07703881535713\\
-24.1132995230379	5.80489878254085\\
-21.1132995230379	7.75077512552668\\
-18.1132995230379	9.86543102037297\\
-15.1132995230379	12.1774068971182\\
-12.1132995230379	14.3796727073624\\
-9.11329952303793	16.7636388471282\\
-6.11329952303793	18.8533199211404\\
-3.11329952303793	20.8337740381495\\
-0.113299523037931	22.552803791329\\
2.88670047696207	24.0715287839238\\
5.88670047696207	25.1586950716826\\
8.88670047696207	26.0437481936197\\
11.8867004769621	26.6453961073393\\
14.8867004769621	27.0556299348834\\
17.8867004769621	27.3319268634938\\
20.8867004769621	27.5296714227143\\
23.8867004769621	27.6737437092988\\
26.8867004769621	27.7891537373586\\
29.8867004769621	27.8843385638696\\
32.8867004769621	27.9823992148741\\
};
\addlegendentry{FC, BST}

\addplot [only marks, line width = 1.0pt, mark size=3pt, mark=diamond, mark options={solid, black},mark repeat=4,mark phase=1]
  table[row sep=crcr]{%
-33.1132995230379	1.78852527870992\\
-30.1132995230379	2.87763618884753\\
-27.1132995230379	4.14932213390061\\
-24.1132995230379	5.92627129712153\\
-21.1132995230379	7.93290471195977\\
-18.1132995230379	10.1453295588689\\
-15.1132995230379	12.6138977917885\\
-12.1132995230379	15.0211371137692\\
-9.11329952303793	17.7139081857746\\
-6.11329952303793	20.1906506725496\\
-3.11329952303793	22.7686674311266\\
-0.113299523037931	25.1772701131929\\
2.88670047696207	27.6185746983117\\
5.88670047696207	29.7729943075652\\
8.88670047696207	31.7914289894851\\
11.8867004769621	33.5389833985023\\
14.8867004769621	35.0476643837731\\
17.8867004769621	36.2936180861326\\
20.8867004769621	37.2907721806252\\
23.8867004769621	38.0642436298719\\
26.8867004769621	38.6226047827107\\
29.8867004769621	39.0490673566064\\
32.8867004769621	39.3643202038741\\
};
\addlegendentry{FC, MRT, $p=1$}

\addplot [color=orange, dashed, line width=1.5pt]
  table[row sep=crcr]{%
-33.1132995230379	1.73131781937861\\
-30.1132995230379	2.76044681491758\\
-27.1132995230379	4.25740763698867\\
-24.1132995230379	6.003854702071\\
-21.1132995230379	8.28001636861115\\
-18.1132995230379	11.113119504082\\
-15.1132995230379	14.2916405212712\\
-12.1132995230379	17.4854111040945\\
-9.11329952303793	20.931743086773\\
-6.11329952303793	24.5084386249316\\
-3.11329952303793	28.0247181008145\\
-0.113299523037931	31.2783025247343\\
2.88670047696207	34.5297602285434\\
5.88670047696207	37.5264253602283\\
8.88670047696207	40.2093061451714\\
11.8867004769621	42.6060867321669\\
14.8867004769621	44.6194020140732\\
17.8867004769621	46.4330492380753\\
20.8867004769621	47.7974134206007\\
23.8867004769621	48.8376377410568\\
26.8867004769621	49.6599743678207\\
29.8867004769621	50.2184172912469\\
32.8867004769621	50.6081296727028\\
};
\addlegendentry{FC, MRT, $p=2$}

\addplot [color=mycolor1, dashdotdotted, line width=1.5pt]
table[row sep=crcr]{%
-33.1132995230379	1.75832388255264\\
-30.1132995230379	2.83943595376909\\
-27.1132995230379	4.10691283433749\\
-24.1132995230379	5.89448892013607\\
-21.1132995230379	7.91710232449978\\
-18.1132995230379	10.179793884579\\
-15.1132995230379	12.7167295998424\\
-12.1132995230379	15.2170241757892\\
-9.11329952303793	18.0579283230119\\
-6.11329952303793	20.749837758035\\
-3.11329952303793	23.5798764791925\\
-0.113299523037931	26.3332287802902\\
2.88670047696207	29.2235959776631\\
5.88670047696207	32.0349265167813\\
8.88670047696207	34.8745761969231\\
11.8867004769621	37.534168947596\\
14.8867004769621	40.3655975823561\\
17.8867004769621	43.2379852177147\\
20.8867004769621	45.9782404284367\\
23.8867004769621	48.7973661292196\\
26.8867004769621	51.6810857240906\\
29.8867004769621	54.518840868148\\
32.8867004769621	57.5242109256267\\
};
\addlegendentry{FC, MR-ZF, $p=1$}

\addplot [color=mycolor3, line width=1.5pt]
table[row sep=crcr]{%
-33.1132995230379	1.71340354645088\\
-30.1132995230379	2.7368404772004\\
-27.1132995230379	4.23399107356791\\
-24.1132995230379	5.98564398954447\\
-21.1132995230379	8.28165215359207\\
-18.1132995230379	11.1451314563218\\
-15.1132995230379	14.4049205863576\\
-12.1132995230379	17.7029372569376\\
-9.11329952303793	21.338420845843\\
-6.11329952303793	25.1971194046358\\
-3.11329952303793	29.0516700469838\\
-0.113299523037931	32.8001214719959\\
2.88670047696207	36.6730369388236\\
5.88670047696207	40.5039105330949\\
8.88670047696207	44.2153906346981\\
11.8867004769621	47.7765552089923\\
14.8867004769621	51.2386310160265\\
17.8867004769621	54.8363326164198\\
20.8867004769621	58.0338797465356\\
23.8867004769621	61.2365977598985\\
26.8867004769621	64.5724252019895\\
29.8867004769621	67.5836285337284\\
32.8867004769621	70.5245829537421\\
};
\addlegendentry{FC, MR-ZF, $p=2$}

\addplot [color=mycolor4, dotted, line width=1.5pt]
  table[row sep=crcr]{%
-33.1132995230379	1.75713123397638\\
-30.1132995230379	2.82939057159659\\
-27.1132995230379	4.07703881535713\\
-24.1132995230379	5.80489878254085\\
-21.1132995230379	7.75077512552668\\
-18.1132995230379	9.86543102037297\\
-15.1132995230379	12.1774068971182\\
-12.1132995230379	14.3796727073624\\
-9.11329952303793	16.7636388471282\\
-6.11329952303793	18.8533199211404\\
-3.11329952303793	20.8337740381495\\
-0.113299523037931	22.552803791329\\
2.88670047696207	24.0715287839238\\
5.88670047696207	25.1586950716826\\
8.88670047696207	26.0437481936197\\
11.8867004769621	26.6453961073393\\
14.8867004769621	27.0556299348834\\
17.8867004769621	27.3319268634938\\
20.8867004769621	27.5296714227143\\
23.8867004769621	27.6737437092988\\
26.8867004769621	27.7891537373586\\
29.8867004769621	27.8843385638696\\
32.8867004769621	27.9823992148741\\
};
\addlegendentry{FC, ZF in \cite{yunyi2017scheduler}}

\end{axis}
\end{tikzpicture}%

%% file: 1_rate_fully_block2.tex
%
%
\definecolor{mycolor1}{rgb}{0.00000,0.44700,0.74100}%
\definecolor{mycolor3}{rgb}{0.85000,0.32500,0.09800}%
\definecolor{mycolor2}{rgb}{0.92900,0.69400,0.12500}%
\definecolor{mycolor4}{rgb}{0.49400,0.18400,0.55600}%
\definecolor{mycolor5}{rgb}{0.46600,0.67400,0.18800}%
\definecolor{mycolor6}{rgb}{0.10000,0.40100,0.8000}%
\begin{tikzpicture}

\begin{axis}[%
xmin=-30,
xmax=31,
xlabel={$\snrbef$ (dB)},
ymin=0,
ymax=83,
title style={at={(0.5,-0.02)},anchor=north,yshift=-1.2cm},
title = {(c)},
legend style={at={(0.02,0.55)}, nodes={scale=0.8, transform shape}, anchor=south west, legend cell align=left, align=left, draw=white!15!black}
]
\addplot [only marks, line width = 1.0pt, mark size=2.7pt, mark=o, mark options={solid, mycolor6},mark repeat=4,mark phase=3]
  table[row sep=crcr]{%
-33.1132995230379	1.05020467751044\\
-30.1132995230379	1.59822046759192\\
-27.1132995230379	2.3866791558885\\
-24.1132995230379	3.3064478260124\\
-21.1132995230379	4.4340166666122\\
-18.1132995230379	5.80827313088842\\
-15.1132995230379	7.04812351904351\\
-12.1132995230379	8.40443923204939\\
-9.11329952303793	9.70804136054449\\
-6.11329952303793	10.9409758471513\\
-3.11329952303793	12.1320781198803\\
-0.113299523037931	13.1688528868522\\
2.88670047696207	13.983585764913\\
5.88670047696207	14.713474732493\\
8.88670047696207	15.205211352519\\
11.8867004769621	15.5891172083562\\
14.8867004769621	15.8851410646242\\
17.8867004769621	16.0505864685514\\
20.8867004769621	16.2389999720081\\
23.8867004769621	16.379421476119\\
26.8867004769621	16.513624785456\\
29.8867004769621	16.6856107013343\\
32.8867004769621	16.82799888358\\
};
\addlegendentry{FC, BST}

\addplot [only marks, line width = 1.0pt, mark size=3pt, mark=diamond, mark options={solid, black},mark repeat=4,mark phase=1]
  table[row sep=crcr]{%
-33.1132995230379	1.06284615593239\\
-30.1132995230379	1.61500270506045\\
-27.1132995230379	2.42021609032348\\
-24.1132995230379	3.3574845759448\\
-21.1132995230379	4.52380245759071\\
-18.1132995230379	5.95721801203473\\
-15.1132995230379	7.26720465343049\\
-12.1132995230379	8.73021139453384\\
-9.11329952303793	10.1961280179744\\
-6.11329952303793	11.6625365961684\\
-3.11329952303793	13.2007175109297\\
-0.113299523037931	14.6074184688572\\
2.88670047696207	15.9327424708862\\
5.88670047696207	17.2581394302958\\
8.88670047696207	18.3797112438348\\
11.8867004769621	19.3895798614854\\
14.8867004769621	20.3240126137523\\
17.8867004769621	21.0502794314585\\
20.8867004769621	21.6783385009581\\
23.8867004769621	22.1968347623088\\
26.8867004769621	22.60735436273\\
29.8867004769621	22.9755922762967\\
32.8867004769621	23.250554133176\\
};
\addlegendentry{FC, MRT, $p=1$}

\addplot [color=orange, dashed, line width=1.5pt]
  table[row sep=crcr]{%
-33.1132995230379	1.26082207649648\\
-30.1132995230379	1.97436222997089\\
-27.1132995230379	3.07369001796114\\
-24.1132995230379	4.49600570087576\\
-21.1132995230379	6.35565119393648\\
-18.1132995230379	8.79740024813136\\
-15.1132995230379	11.570235543402\\
-12.1132995230379	14.7002927774774\\
-9.11329952303793	18.0195399894936\\
-6.11329952303793	21.5030741514483\\
-3.11329952303793	25.1305636475259\\
-0.113299523037931	28.5248278072451\\
2.88670047696207	31.9095612656831\\
5.88670047696207	35.0057983825649\\
8.88670047696207	37.8779547219314\\
11.8867004769621	40.3989755663696\\
14.8867004769621	42.6860964045024\\
17.8867004769621	44.5708632115676\\
20.8867004769621	46.1448431901414\\
23.8867004769621	47.3283377601919\\
26.8867004769621	48.2216589965416\\
29.8867004769621	48.860236632822\\
32.8867004769621	49.2826201752793\\
};
\addlegendentry{FC, MRT, $p=2$}

\addplot [color=mycolor1, dashdotdotted, line width=1.5pt]
table[row sep=crcr]{%
-33.1132995230379	1.04983914977258\\
-30.1132995230379	1.59694799434629\\
-27.1132995230379	2.40364988214988\\
-24.1132995230379	3.33693149246975\\
-21.1132995230379	4.52023702931301\\
-18.1132995230379	5.98076973394625\\
-15.1132995230379	7.31525767735616\\
-12.1132995230379	8.81940098641019\\
-9.11329952303793	10.3452133536615\\
-6.11329952303793	11.9198156089061\\
-3.11329952303793	13.6066820838115\\
-0.113299523037931	15.1888540394347\\
2.88670047696207	16.7698512766995\\
5.88670047696207	18.4569774956512\\
8.88670047696207	20.0262505395762\\
11.8867004769621	21.6974668073802\\
14.8867004769621	23.337682214217\\
17.8867004769621	24.9282696865503\\
20.8867004769621	26.6090029149771\\
23.8867004769621	28.2663207314136\\
26.8867004769621	30.0375008332728\\
29.8867004769621	31.8197644800284\\
32.8867004769621	33.6110460701805\\
};
\addlegendentry{FC, MR-ZF, $p=1$}

\addplot [color=mycolor3, line width=1.5pt]
table[row sep=crcr]{%
-33.1132995230379	1.25263471371531\\
-30.1132995230379	1.96144590721555\\
-27.1132995230379	3.05249886200641\\
-24.1132995230379	4.48396691464992\\
-21.1132995230379	6.34299703390677\\
-18.1132995230379	8.81751123424751\\
-15.1132995230379	11.6038178726524\\
-12.1132995230379	14.7813953580348\\
-9.11329952303793	18.1920839325268\\
-6.11329952303793	21.846395982818\\
-3.11329952303793	25.6520113550306\\
-0.113299523037931	29.2873666904668\\
2.88670047696207	33.039938241312\\
5.88670047696207	36.6133038470191\\
8.88670047696207	40.0649789532511\\
11.8867004769621	43.2819337585822\\
14.8867004769621	46.5478409929661\\
17.8867004769621	49.4086291894611\\
20.8867004769621	52.233108465344\\
23.8867004769621	54.6697285249359\\
26.8867004769621	57.0119049799818\\
29.8867004769621	59.1387068537698\\
32.8867004769621	61.1925748931176\\
};
\addlegendentry{FC, MR-ZF, $p=2$}

\addplot [color=mycolor4, dotted, line width=1.5pt]
  table[row sep=crcr]{%
-33.1132995230379	1.05020467751044\\
-30.1132995230379	1.59822046759192\\
-27.1132995230379	2.3866791558885\\
-24.1132995230379	3.3064478260124\\
-21.1132995230379	4.4340166666122\\
-18.1132995230379	5.80827313088842\\
-15.1132995230379	7.04812351904351\\
-12.1132995230379	8.40443923204939\\
-9.11329952303793	9.70804136054449\\
-6.11329952303793	10.9409758471513\\
-3.11329952303793	12.1320781198803\\
-0.113299523037931	13.1688528868522\\
2.88670047696207	13.983585764913\\
5.88670047696207	14.713474732493\\
8.88670047696207	15.205211352519\\
11.8867004769621	15.5891172083562\\
14.8867004769621	15.8851410646242\\
17.8867004769621	16.0505864685514\\
20.8867004769621	16.2389999720081\\
23.8867004769621	16.379421476119\\
26.8867004769621	16.513624785456\\
29.8867004769621	16.6856107013343\\
32.8867004769621	16.82799888358\\
};
\addlegendentry{FC, ZF in \cite{yunyi2017scheduler}}

\end{axis}
\end{tikzpicture}%

%% file: 1_rate_sub_block0.tex
%
%
\definecolor{mycolor1}{rgb}{0.00000,0.44700,0.74100}%
\definecolor{mycolor3}{rgb}{0.85000,0.32500,0.09800}%
\definecolor{mycolor2}{rgb}{0.92900,0.69400,0.12500}%
\definecolor{mycolor4}{rgb}{0.49400,0.18400,0.55600}%
\definecolor{mycolor5}{rgb}{0.46600,0.67400,0.18800}%
\definecolor{mycolor6}{rgb}{0.10000,0.40100,0.8000}%
\begin{tikzpicture}

\begin{axis}[%
xmin=-30,
xmax=31,
xlabel={$\snrbef$ (dB)},
ymin=0,
ymax=83,
ylabel={$R_{\text{sum}}$ (bit/s/Hz)},
title style={at={(0.5,-0.02)},anchor=north,yshift=-1.2cm},
title = {(d)},
legend style={at={(0.02,0.55)}, nodes={scale=0.8, transform shape}, anchor=south west, legend cell align=left, align=left, draw=white!15!black}
]
\addplot [only marks, line width = 1.0pt, mark size=2.7pt, mark=o, mark options={solid, mycolor6},mark repeat=4,mark phase=3]
  table[row sep=crcr]{%
-33.1132995230379	2.84164383483401\\
-30.1132995230379	4.23342916990659\\
-27.1132995230379	6.09952943519654\\
-24.1132995230379	8.36977510162132\\
-21.1132995230379	11.06210101789\\
-18.1132995230379	13.8792936953459\\
-15.1132995230379	16.5491015978343\\
-12.1132995230379	19.1185344473451\\
-9.11329952303793	21.5140397382615\\
-6.11329952303793	23.6268995224936\\
-3.11329952303793	25.3339035216193\\
-0.113299523037931	26.6795551206176\\
2.88670047696207	27.7228322878319\\
5.88670047696207	28.4460932708011\\
8.88670047696207	28.9489617278806\\
11.8867004769621	29.2654755148093\\
14.8867004769621	29.4653984418618\\
17.8867004769621	29.5901184224516\\
20.8867004769621	29.6686038474536\\
23.8867004769621	29.7127776559511\\
26.8867004769621	29.7392127341132\\
29.8867004769621	29.7542701740576\\
32.8867004769621	29.7638051500946\\
};
\addlegendentry{OSPS, BST}

\addplot [only marks, line width = 1.0pt, mark size=3pt, mark=diamond, mark options={solid, black},mark repeat=4,mark phase=1]
  table[row sep=crcr]{%
-33.1132995230379	2.84364863236256\\
-30.1132995230379	4.22611581632069\\
-27.1132995230379	6.09165199316264\\
-24.1132995230379	8.36327364242653\\
-21.1132995230379	11.0186108045178\\
-18.1132995230379	13.8402056248008\\
-15.1132995230379	16.4984135819177\\
-12.1132995230379	19.0415250334708\\
-9.11329952303793	21.4003502065529\\
-6.11329952303793	23.4725346922626\\
-3.11329952303793	25.1566429383802\\
-0.113299523037931	26.4886192992557\\
2.88670047696207	27.4581077726152\\
5.88670047696207	28.1754654247066\\
8.88670047696207	28.652830888439\\
11.8867004769621	28.972023240616\\
14.8867004769621	29.1611906216601\\
17.8867004769621	29.2900594540556\\
20.8867004769621	29.3631952799142\\
23.8867004769621	29.4117789000691\\
26.8867004769621	29.43571987066\\
29.8867004769621	29.4482032435046\\
32.8867004769621	29.4555320827443\\
};
\addlegendentry{OSPS, MRT, $p=1$}

\addplot [color=orange, dashed, line width=1.5pt]
  table[row sep=crcr]{%
-33.1132995230379	2.43584020068874\\
-30.1132995230379	3.70201034299428\\
-27.1132995230379	5.287330268927\\
-24.1132995230379	7.13081000914\\
-21.1132995230379	9.48196768548684\\
-18.1132995230379	11.8384441537577\\
-15.1132995230379	14.2083845359924\\
-12.1132995230379	16.2455599166994\\
-9.11329952303793	18.0958560995383\\
-6.11329952303793	19.6394294802627\\
-3.11329952303793	20.9118896340406\\
-0.113299523037931	21.8398101130637\\
2.88670047696207	22.5360957653801\\
5.88670047696207	22.9957045533417\\
8.88670047696207	23.3175341741324\\
11.8867004769621	23.5164724869855\\
14.8867004769621	23.6460833904143\\
17.8867004769621	23.719468464909\\
20.8867004769621	23.7711955809285\\
23.8867004769621	23.7948653456815\\
26.8867004769621	23.8157772755191\\
29.8867004769621	23.8252670256905\\
32.8867004769621	23.8301023319453\\
};
\addlegendentry{OSPS, MRT, $p=2$}

\addplot [color=mycolor1, dashdotdotted, line width=1.5pt]
table[row sep=crcr]{%
-33.1132995230379	2.92079951885032\\
-30.1132995230379	4.36817617275265\\
-27.1132995230379	6.40175033965057\\
-24.1132995230379	8.92564980813054\\
-21.1132995230379	12.0681690096396\\
-18.1132995230379	15.5336148038121\\
-15.1132995230379	19.0292422349504\\
-12.1132995230379	22.7751002465405\\
-9.11329952303793	26.7980233290681\\
-6.11329952303793	30.6504794228787\\
-3.11329952303793	34.6476013593092\\
-0.113299523037931	38.6296137372225\\
2.88670047696207	42.483175595695\\
5.88670047696207	46.5762394105452\\
8.88670047696207	50.6926959003088\\
11.8867004769621	54.5337685073509\\
14.8867004769621	58.4675597118393\\
17.8867004769621	62.5048381549908\\
20.8867004769621	66.4132866690601\\
23.8867004769621	70.466557927506\\
26.8867004769621	74.5610304491985\\
29.8867004769621	78.3852027940746\\
32.8867004769621	82.5429242633342\\
};
\addlegendentry{OSPS, MR-ZF, $p=1$}

\addplot [color=mycolor3, line width=1.5pt]
table[row sep=crcr]{%
-33.1132995230379	2.37078537167793\\
-30.1132995230379	3.72943518234258\\
-27.1132995230379	5.47455957775466\\
-24.1132995230379	7.6258119796927\\
-21.1132995230379	10.5669316683391\\
-18.1132995230379	13.7948454309746\\
-15.1132995230379	17.4266410779038\\
-12.1132995230379	20.9678264772795\\
-9.11329952303793	24.9381688892527\\
-6.11329952303793	28.6951922263877\\
-3.11329952303793	32.6643081018105\\
-0.113299523037931	36.7496074444842\\
2.88670047696207	40.5780956149554\\
5.88670047696207	44.5510028975334\\
8.88670047696207	48.7483517257448\\
11.8867004769621	52.6743848603638\\
14.8867004769621	56.6396353938565\\
17.8867004769621	60.584612101616\\
20.8867004769621	64.7201024703684\\
23.8867004769621	68.5829169070326\\
26.8867004769621	72.5622739389744\\
29.8867004769621	76.5447285548545\\
32.8867004769621	80.5962010814112\\
};
\addlegendentry{OSPS, MR-ZF, $p=2$}

\addplot [color=mycolor4, dotted, line width=1.5pt]
  table[row sep=crcr]{%
-33.1132995230379	2.841643834834\\
-30.1132995230379	4.23342916990659\\
-27.1132995230379	6.09952943519654\\
-24.1132995230379	8.36977510162132\\
-21.1132995230379	11.06210101789\\
-18.1132995230379	13.8792936953459\\
-15.1132995230379	16.5491015978343\\
-12.1132995230379	19.1185344473451\\
-9.11329952303793	21.5140397382615\\
-6.11329952303793	23.6268995224936\\
-3.11329952303793	25.3339035216193\\
-0.113299523037931	26.6795551206176\\
2.88670047696207	27.7228322878319\\
5.88670047696207	28.4460932708012\\
8.88670047696207	28.9489617278806\\
11.8867004769621	29.2654755148093\\
14.8867004769621	29.4653984418618\\
17.8867004769621	29.5901184224516\\
20.8867004769621	29.6686038474536\\
23.8867004769621	29.7127776559511\\
26.8867004769621	29.7392127341132\\
29.8867004769621	29.7542701740576\\
32.8867004769621	29.7638051500946\\
};
\addlegendentry{OSPS, ZF in \cite{yunyi2017scheduler}}

\end{axis}
\end{tikzpicture}%

%% file: 1_rate_sub_block1.tex
%
%
\definecolor{mycolor1}{rgb}{0.00000,0.44700,0.74100}%
\definecolor{mycolor3}{rgb}{0.85000,0.32500,0.09800}%
\definecolor{mycolor2}{rgb}{0.92900,0.69400,0.12500}%
\definecolor{mycolor4}{rgb}{0.49400,0.18400,0.55600}%
\definecolor{mycolor5}{rgb}{0.46600,0.67400,0.18800}%
\definecolor{mycolor6}{rgb}{0.10000,0.40100,0.8000}%
\begin{tikzpicture}

\begin{axis}[%
xmin=-30,
xmax=31,
xlabel={$\snrbef$ (dB)},
ymin=0,
ymax=83,
title style={at={(0.5,-0.02)},anchor=north,yshift=-1.2cm},
title = {(e)},
legend style={at={(0.02,0.55)}, nodes={scale=0.8, transform shape}, anchor=south west, legend cell align=left, align=left, draw=white!15!black}
]
\addplot [only marks, line width = 1.0pt, mark size=2.7pt, mark=o, mark options={solid, mycolor6},mark repeat=4,mark phase=3]
  table[row sep=crcr]{%
-33.1132995230379	1.95100270855724\\
-30.1132995230379	2.97467406898009\\
-27.1132995230379	4.18193682652305\\
-24.1132995230379	5.73145037221511\\
-21.1132995230379	7.67618479679157\\
-18.1132995230379	9.42603449643021\\
-15.1132995230379	11.327894223739\\
-12.1132995230379	13.0537111683824\\
-9.11329952303793	14.8128792555541\\
-6.11329952303793	16.1776715737916\\
-3.11329952303793	17.298563521864\\
-0.113299523037931	18.2732520556232\\
2.88670047696207	18.9603769785943\\
5.88670047696207	19.475249538011\\
8.88670047696207	19.8472154874778\\
11.8867004769621	20.1044879873883\\
14.8867004769621	20.3345832590299\\
17.8867004769621	20.5028328960551\\
20.8867004769621	20.6441893059181\\
23.8867004769621	20.7857116211856\\
26.8867004769621	20.914966530691\\
29.8867004769621	21.0402612662028\\
32.8867004769621	21.149411132915\\
};
\addlegendentry{OSPS, BST}

\addplot [only marks, line width = 1.0pt, mark size=3pt, mark=diamond, mark options={solid, black},mark repeat=4,mark phase=1]
  table[row sep=crcr]{%
-33.1132995230379	1.95098523543073\\
-30.1132995230379	2.97408351852383\\
-27.1132995230379	4.17956387423602\\
-24.1132995230379	5.73269265140467\\
-21.1132995230379	7.6654964129269\\
-18.1132995230379	9.43029212550081\\
-15.1132995230379	11.3406993668892\\
-12.1132995230379	13.0556950702157\\
-9.11329952303793	14.8190675501155\\
-6.11329952303793	16.2065187544687\\
-3.11329952303793	17.3379837487132\\
-0.113299523037931	18.2945726820945\\
2.88670047696207	18.9864125509369\\
5.88670047696207	19.508996185229\\
8.88670047696207	19.8926895166117\\
11.8867004769621	20.1415560911216\\
14.8867004769621	20.3737552009082\\
17.8867004769621	20.54214838473\\
20.8867004769621	20.6829505786705\\
23.8867004769621	20.8212087106245\\
26.8867004769621	20.9358619977894\\
29.8867004769621	21.0602679155377\\
32.8867004769621	21.1390141043347\\
};
\addlegendentry{OSPS, MRT, $p=1$}

\addplot [color=orange, dashed, line width=1.5pt]
  table[row sep=crcr]{%
-33.1132995230379	1.82524093280735\\
-30.1132995230379	2.78178912635715\\
-27.1132995230379	4.03141841896776\\
-24.1132995230379	5.69625469615547\\
-21.1132995230379	7.61999062633371\\
-18.1132995230379	9.68172078630016\\
-15.1132995230379	11.8025025244737\\
-12.1132995230379	13.7952243392582\\
-9.11329952303793	15.6528697324481\\
-6.11329952303793	17.2159358897842\\
-3.11329952303793	18.5563529563768\\
-0.113299523037931	19.6522684390125\\
2.88670047696207	20.4331691077866\\
5.88670047696207	20.9983511087609\\
8.88670047696207	21.3656772970802\\
11.8867004769621	21.6249708515577\\
14.8867004769621	21.7832524202683\\
17.8867004769621	21.871898197183\\
20.8867004769621	21.9264690839686\\
23.8867004769621	21.9574356832954\\
26.8867004769621	21.9753223421777\\
29.8867004769621	21.9830516445285\\
32.8867004769621	21.9883736262313\\
};
\addlegendentry{OSPS, MRT, $p=2$}

\addplot [color=mycolor1, dashdotdotted, line width=1.5pt]
table[row sep=crcr]{%
-33.1132995230379	2.01676761645495\\
-30.1132995230379	3.10226119146831\\
-27.1132995230379	4.42339941345444\\
-24.1132995230379	6.20699348629818\\
-21.1132995230379	8.49699992742384\\
-18.1132995230379	10.6841246065114\\
-15.1132995230379	13.2829286562846\\
-12.1132995230379	15.8291554559958\\
-9.11329952303793	18.7158511160061\\
-6.11329952303793	21.512551976536\\
-3.11329952303793	24.1610252141313\\
-0.113299523037931	27.074905117799\\
2.88670047696207	29.81606928139\\
5.88670047696207	32.6901618938375\\
8.88670047696207	35.4641706407247\\
11.8867004769621	38.2263360333936\\
14.8867004769621	41.3134718845081\\
17.8867004769621	44.0440824962507\\
20.8867004769621	46.9941874104526\\
23.8867004769621	49.7693161670542\\
26.8867004769621	52.7033527573207\\
29.8867004769621	55.5571545548269\\
32.8867004769621	58.4253290675232\\
};
\addlegendentry{OSPS, MR-ZF, $p=1$}

\addplot [color=mycolor3, line width=1.5pt]
table[row sep=crcr]{%
-33.1132995230379	1.80087287687749\\
-30.1132995230379	2.79706608443168\\
-27.1132995230379	4.16839900830043\\
-24.1132995230379	6.10054089618687\\
-21.1132995230379	8.45477188193376\\
-18.1132995230379	11.103791606153\\
-15.1132995230379	14.1878936372496\\
-12.1132995230379	17.2731915872066\\
-9.11329952303793	20.6609169454591\\
-6.11329952303793	23.838321963161\\
-3.11329952303793	27.1246942270306\\
-0.113299523037931	30.377833465796\\
2.88670047696207	33.6091599902355\\
5.88670047696207	36.6795057771947\\
8.88670047696207	39.6323718705384\\
11.8867004769621	42.4934175287379\\
14.8867004769621	45.4152910213681\\
17.8867004769621	48.2292630424431\\
20.8867004769621	51.0721029191373\\
23.8867004769621	53.9013264184699\\
26.8867004769621	56.6623094734414\\
29.8867004769621	59.4266147253989\\
32.8867004769621	62.1257566954723\\
};
\addlegendentry{OSPS, MR-ZF, $p=2$}

\addplot [color=mycolor4, dotted, line width=1.5pt]
  table[row sep=crcr]{%
-33.1132995230379	1.95100270855724\\
-30.1132995230379	2.97467406898009\\
-27.1132995230379	4.18193682652305\\
-24.1132995230379	5.73145037221511\\
-21.1132995230379	7.67618479679157\\
-18.1132995230379	9.42603449643021\\
-15.1132995230379	11.327894223739\\
-12.1132995230379	13.0537111683824\\
-9.11329952303793	14.8128792555541\\
-6.11329952303793	16.1776715737915\\
-3.11329952303793	17.298563521864\\
-0.113299523037931	18.2732520556232\\
2.88670047696207	18.9603769785943\\
5.88670047696207	19.475249538011\\
8.88670047696207	19.8472154874778\\
11.8867004769621	20.1044879873883\\
14.8867004769621	20.3345832590299\\
17.8867004769621	20.5028328960551\\
20.8867004769621	20.6441893059181\\
23.8867004769621	20.7857116211856\\
26.8867004769621	20.914966530691\\
29.8867004769621	21.0402612662028\\
32.8867004769621	21.149411132915\\
};
\addlegendentry{OSPS, ZF in \cite{yunyi2017scheduler}}

\end{axis}
\end{tikzpicture}%

%% file: 1_rate_sub_block2.tex
%
%
\definecolor{mycolor1}{rgb}{0.00000,0.44700,0.74100}%
\definecolor{mycolor3}{rgb}{0.85000,0.32500,0.09800}%
\definecolor{mycolor2}{rgb}{0.92900,0.69400,0.12500}%
\definecolor{mycolor4}{rgb}{0.49400,0.18400,0.55600}%
\definecolor{mycolor5}{rgb}{0.46600,0.67400,0.18800}%
\definecolor{mycolor6}{rgb}{0.10000,0.40100,0.8000}%
\begin{tikzpicture}

\begin{axis}[%
xmin=-30,
xmax=31,
xlabel={$\snrbef$ (dB)},
ymin=0,
ymax=83,
title style={at={(0.5,-0.02)},anchor=north,yshift=-1.2cm},
title = {(f)},
legend style={at={(0.02,0.55)}, nodes={scale=0.8, transform shape}, anchor=south west, legend cell align=left, align=left, draw=white!15!black}
]
\addplot [only marks, line width = 1.0pt, mark size=2.7pt, mark=o, mark options={solid, mycolor6},mark repeat=4,mark phase=3]
  table[row sep=crcr]{%
-33.1132995230379	1.13289787978065\\
-30.1132995230379	1.70112312836334\\
-27.1132995230379	2.43654431534884\\
-24.1132995230379	3.42811402518689\\
-21.1132995230379	4.46328337590256\\
-18.1132995230379	5.48457682494105\\
-15.1132995230379	6.67307956487158\\
-12.1132995230379	7.720796274938\\
-9.11329952303793	8.64027966106657\\
-6.11329952303793	9.50590555515065\\
-3.11329952303793	10.2458748173451\\
-0.113299523037931	10.752493987879\\
2.88670047696207	11.1850618674139\\
5.88670047696207	11.5439664672393\\
8.88670047696207	11.801973380857\\
11.8867004769621	12.0633472869314\\
14.8867004769621	12.2863195167729\\
17.8867004769621	12.4826918067557\\
20.8867004769621	12.6921257046216\\
23.8867004769621	12.9100852583629\\
26.8867004769621	13.1339857047083\\
29.8867004769621	13.3407744193636\\
32.8867004769621	13.5400782463835\\
};
\addlegendentry{OSPS, BST}

\addplot [only marks, line width = 1.0pt, mark size=3pt, mark=diamond, mark options={solid, black},mark repeat=4,mark phase=1]
  table[row sep=crcr]{%
-33.1132995230379	1.13247943292403\\
-30.1132995230379	1.70063067686861\\
-27.1132995230379	2.43677209940265\\
-24.1132995230379	3.42783499151875\\
-21.1132995230379	4.45983893811817\\
-18.1132995230379	5.47611198663289\\
-15.1132995230379	6.65603450548442\\
-12.1132995230379	7.71353593655646\\
-9.11329952303793	8.62498671093168\\
-6.11329952303793	9.4723345127201\\
-3.11329952303793	10.2192647397396\\
-0.113299523037931	10.7139386712802\\
2.88670047696207	11.1703094137616\\
5.88670047696207	11.5185564149585\\
8.88670047696207	11.7894526841172\\
11.8867004769621	12.0514716419512\\
14.8867004769621	12.2722586738704\\
17.8867004769621	12.4753860216462\\
20.8867004769621	12.6860515390994\\
23.8867004769621	12.9094856414065\\
26.8867004769621	13.1332996828628\\
29.8867004769621	13.3499381292654\\
32.8867004769621	13.555779354862\\
};
\addlegendentry{OSPS, MRT, $p=1$}

\addplot [color=orange, dashed, line width=1.5pt]
  table[row sep=crcr]{%
-33.1132995230379	1.32541641865287\\
-30.1132995230379	2.09535180326303\\
-27.1132995230379	3.06591278175589\\
-24.1132995230379	4.42072728215444\\
-21.1132995230379	6.02287817511373\\
-18.1132995230379	7.90929157822904\\
-15.1132995230379	9.88740764372519\\
-12.1132995230379	12.0503026751732\\
-9.11329952303793	14.0859523164038\\
-6.11329952303793	15.9327403647843\\
-3.11329952303793	17.4904850239852\\
-0.113299523037931	18.8477245917203\\
2.88670047696207	19.8131920831175\\
5.88670047696207	20.5589611063727\\
8.88670047696207	21.0922115950702\\
11.8867004769621	21.459506985627\\
14.8867004769621	21.6890664029307\\
17.8867004769621	21.8522941664065\\
20.8867004769621	21.9308418159368\\
23.8867004769621	21.995309939631\\
26.8867004769621	22.0259889165565\\
29.8867004769621	22.0413706339444\\
32.8867004769621	22.0526453783103\\
};
\addlegendentry{OSPS, MRT, $p=2$}

\addplot [color=mycolor1, dashdotdotted, line width=1.5pt]
table[row sep=crcr]{%
-33.1132995230379	1.15611833809478\\
-30.1132995230379	1.75792102775738\\
-27.1132995230379	2.5313933481793\\
-24.1132995230379	3.64111666356361\\
-21.1132995230379	4.86973527646058\\
-18.1132995230379	6.09780972646388\\
-15.1132995230379	7.6330161109292\\
-12.1132995230379	9.20372008105353\\
-9.11329952303793	10.6643779383531\\
-6.11329952303793	12.2993773608427\\
-3.11329952303793	14.002977112745\\
-0.113299523037931	15.5488251875119\\
2.88670047696207	17.1665940562183\\
5.88670047696207	18.7971802493274\\
8.88670047696207	20.5675311115222\\
11.8867004769621	22.3012859414171\\
14.8867004769621	24.0937393424\\
17.8867004769621	25.7553866856632\\
20.8867004769621	27.4785562305358\\
23.8867004769621	29.4012542688066\\
26.8867004769621	31.1876084985859\\
29.8867004769621	32.9518555874062\\
32.8867004769621	34.7593048070183\\
};
\addlegendentry{OSPS, MR-ZF, $p=1$}

\addplot [color=mycolor3, line width=1.5pt]
table[row sep=crcr]{%
-33.1132995230379	1.25335702254228\\
-30.1132995230379	2.03686737755956\\
-27.1132995230379	3.04670144491983\\
-24.1132995230379	4.47253755221816\\
-21.1132995230379	6.29152678230609\\
-18.1132995230379	8.46389545556446\\
-15.1132995230379	10.9400901244667\\
-12.1132995230379	13.7851735102767\\
-9.11329952303793	16.4986124617399\\
-6.11329952303793	19.3098744196212\\
-3.11329952303793	21.9809285127967\\
-0.113299523037931	24.5113550853597\\
2.88670047696207	26.8507917972241\\
5.88670047696207	28.9448231503672\\
8.88670047696207	31.0167653009944\\
11.8867004769621	32.9467591341042\\
14.8867004769621	34.7170287313999\\
17.8867004769621	36.4612587321601\\
20.8867004769621	38.1018269474192\\
23.8867004769621	39.8133375021856\\
26.8867004769621	41.3890292923764\\
29.8867004769621	43.0047052819453\\
32.8867004769621	44.6735198077878\\
};
\addlegendentry{OSPS, MR-ZF, $p=2$}

\addplot [color=mycolor4, dotted, line width=1.5pt]
  table[row sep=crcr]{%
-33.1132995230379	1.13289787978065\\
-30.1132995230379	1.70112312836334\\
-27.1132995230379	2.43654431534884\\
-24.1132995230379	3.42811402518689\\
-21.1132995230379	4.46328337590256\\
-18.1132995230379	5.48457682494105\\
-15.1132995230379	6.67307956487158\\
-12.1132995230379	7.720796274938\\
-9.11329952303793	8.64027966106657\\
-6.11329952303793	9.50590555515065\\
-3.11329952303793	10.2458748173451\\
-0.113299523037931	10.752493987879\\
2.88670047696207	11.1850618674139\\
5.88670047696207	11.5439664672393\\
8.88670047696207	11.801973380857\\
11.8867004769621	12.0633472869314\\
14.8867004769621	12.2863195167729\\
17.8867004769621	12.4826918067557\\
20.8867004769621	12.6921257046216\\
23.8867004769621	12.9100852583629\\
26.8867004769621	13.1339857047083\\
29.8867004769621	13.3407744193636\\
32.8867004769621	13.5400782463835\\
};
\addlegendentry{OSPS, ZF in \cite{yunyi2017scheduler}}

\end{axis}
\end{tikzpicture}%

%% file: 1-BA-compare.tex
%
%
\definecolor{mycolor1}{rgb}{0.00000,0.44700,0.74100}%
\definecolor{mycolor2}{rgb}{0.85000,0.32500,0.09800}%
%


\begin{tikzpicture}

\begin{axis}[%
xmin=19,
xmax=165,
xlabel={Number of beacon slots $T$},
ymin=0.09,
ymax=1.02,
ytick={  0, 0.2, 0.4, 0.6, 0.8,   1},
yminorticks=true,
ylabel={$P_D$},
title style={at={(0.5,-0.02)},anchor=north,yshift=-1.2cm},
title = {(a)},
legend style={at={(0.49,0.02)},nodes={scale=0.9, transform shape}, anchor=south west, legend cell align=left, align=left, draw=black}
]
\addplot [color=mycolor1, dashed, line width=1.5pt]
  table[row sep=crcr]{%
10	0.00952380952380952\\
20	0.0380952380952381\\
30	0.19047619047619\\
40	0.452380952380952\\
50	0.611904761904762\\
60	0.771428571428571\\
70	0.876190476190476\\
80	0.938095238095238\\
90	0.976190476190476\\
100	0.990476190476191\\
110	0.995238095238095\\
120	0.995238095238095\\
130	0.997617084237755\\
140	0.997617084181514\\
150	1\\
160	1\\
170	1\\
180	1\\
190	1\\
200	1\\
};
\addlegendentry{FC, NNLS}

\addplot [color=mycolor2, line width=1.5pt]
  table[row sep=crcr]{%
10	0.141666666666667\\
20	0.210881736788365\\
30	0.280096806910062\\
40	0.352685401733105\\
50	0.438325701366231\\
60	0.543966000999357\\
70	0.680906272832092\\
80	0.817846544664826\\
90	0.909513211331855\\
100	0.96\\
110	0.9875\\
120	0.995833333333333\\
130	0.995833333328471\\
140	1\\
150	1\\
160	1\\
170	1\\
180	1\\
190	1\\
200	1\\
};
\addlegendentry{OSPS, NNLS}

\addplot [color=black, dashdotted, line width=1.5pt]
table[row sep=crcr]{%
10	0\\
30	0.075\\
50	0.236749946502372\\
70	0.398499893004745\\
90	0.536426016415187\\
110	0.650000001805474\\
130	0.735981990189607\\
150	0.805490995094803\\
170	0.875\\
190	0.962464382336804\\
};
\addlegendentry{FC, OMP \cite{AlkhateebTimeDomain2017}}

\end{axis}
\end{tikzpicture}%

%% file: 3-compare.tex
%
%
\definecolor{mycolor1}{rgb}{0.00000,0.44700,0.74100}%
\definecolor{mycolor2}{rgb}{0.85000,0.32500,0.09800}%
\definecolor{mycolor3}{rgb}{0.92900,0.69400,0.12500}%
\definecolor{mycolor4}{rgb}{0.49400,0.18400,0.55600}%
\definecolor{mycolor5}{rgb}{0.46600,0.67400,0.18800}%
\definecolor{mycolor6}{rgb}{0.30100,0.74500,0.93300}%
\definecolor{mycolor7}{rgb}{0.10000,0.40100,0.8000}%
\begin{tikzpicture}

\begin{axis}[%
xmin=-30,
xmax=30,
xlabel={$\snrbef$ (dB)},
ymin=0,
ymax=83,
ylabel={$R_{\text{sum}}$ (bit/s/Hz)},
title style={at={(0.5,-0.02)},anchor=north,yshift=-1.2cm},
title = {(b)},
legend style={at={(0.02,0.68)},nodes={scale=0.8, transform shape}, anchor=south west, legend cell align=left, align=left, draw=black}
]
\addplot [color=mycolor1, dashdotdotted, line width=1.5pt]
table[row sep=crcr]{%
-33.1132995230379	2.60020052656018\\
-30.1132995230379	3.97053017323255\\
-27.1132995230379	5.84094069546417\\
-24.1132995230379	8.27311547797526\\
-21.1132995230379	11.0780128522433\\
-18.1132995230379	14.1943768053639\\
-15.1132995230379	17.4522329399139\\
-12.1132995230379	20.7649791758005\\
-9.11329952303793	24.1689911462114\\
-6.11329952303793	27.218246330641\\
-3.11329952303793	29.9957358341098\\
-0.113299523037931	32.5596664358839\\
2.88670047696207	34.5472721204632\\
5.88670047696207	36.2589297350353\\
8.88670047696207	37.4732559812558\\
11.8867004769621	38.3976440860537\\
14.8867004769621	38.9644366388394\\
17.8867004769621	39.3316896943072\\
20.8867004769621	39.5662451229206\\
23.8867004769621	39.6917173015248\\
26.8867004769621	39.7712961464956\\
29.8867004769621	39.8195499020598\\
32.8867004769621	39.8549708123029\\
};
\addlegendentry{FC, BST}


\addplot [only marks, line width = 1.0pt, mark size=2.8pt, mark=o, mark options={solid, mycolor1},mark repeat=2,mark phase=1]
table[row sep=crcr]{%
-33.1132995230379	2.31534055845656\\
-30.1132995230379	3.48734815434876\\
-27.1132995230379	5.37189534216527\\
-24.1132995230379	7.67408846351626\\
-21.1132995230379	10.3630091671397\\
-18.1132995230379	13.4912891006412\\
-15.1132995230379	16.9720627895696\\
-12.1132995230379	20.7195499684753\\
-9.11329952303793	24.714245751862\\
-6.11329952303793	28.5587580137156\\
-3.11329952303793	32.5016974179287\\
-0.113299523037931	36.5764182806629\\
2.88670047696207	40.3627639221253\\
5.88670047696207	44.4784489439384\\
8.88670047696207	48.538216298723\\
11.8867004769621	52.4760256976933\\
14.8867004769621	56.4390195687979\\
17.8867004769621	60.3071928133991\\
20.8867004769621	64.3016289300879\\
23.8867004769621	68.2906226554492\\
26.8867004769621	72.3077073122472\\
29.8867004769621	76.521473553026\\
32.8867004769621	80.4568813063385\\
};
\addlegendentry{FC, MR-ZF $p=2$}

\addplot [color=mycolor2, dashed, line width = 1.5pt]
table[row sep=crcr]{%
-33.1132995230379	2.84164383483401\\
-30.1132995230379	4.23342916990659\\
-27.1132995230379	6.09952943519654\\
-24.1132995230379	8.36977510162132\\
-21.1132995230379	11.06210101789\\
-18.1132995230379	13.8792936953459\\
-15.1132995230379	16.5491015978343\\
-12.1132995230379	19.1185344473451\\
-9.11329952303793	21.5140397382615\\
-6.11329952303793	23.6268995224936\\
-3.11329952303793	25.3339035216193\\
-0.113299523037931	26.6795551206176\\
2.88670047696207	27.7228322878319\\
5.88670047696207	28.4460932708011\\
8.88670047696207	28.9489617278806\\
11.8867004769621	29.2654755148093\\
14.8867004769621	29.4653984418618\\
17.8867004769621	29.5901184224516\\
20.8867004769621	29.6686038474536\\
23.8867004769621	29.7127776559511\\
26.8867004769621	29.7392127341132\\
29.8867004769621	29.7542701740576\\
32.8867004769621	29.7638051500946\\
};
\addlegendentry{OSPS, BST}
%

\addplot [color=mycolor2, line width=1.5pt]
table[row sep=crcr]{%
-33.1132995230379	2.37078537167793\\
-30.1132995230379	3.72943518234258\\
-27.1132995230379	5.47455957775466\\
-24.1132995230379	7.6258119796927\\
-21.1132995230379	10.5669316683391\\
-18.1132995230379	13.7948454309746\\
-15.1132995230379	17.4266410779038\\
-12.1132995230379	20.9678264772795\\
-9.11329952303793	24.9381688892527\\
-6.11329952303793	28.6951922263877\\
-3.11329952303793	32.6643081018105\\
-0.113299523037931	36.7496074444842\\
2.88670047696207	40.5780956149554\\
5.88670047696207	44.5510028975334\\
8.88670047696207	48.7483517257448\\
11.8867004769621	52.6743848603638\\
14.8867004769621	56.6396353938565\\
17.8867004769621	60.584612101616\\
20.8867004769621	64.7201024703684\\
23.8867004769621	68.5829169070326\\
26.8867004769621	72.5622739389744\\
29.8867004769621	76.5447285548545\\
32.8867004769621	80.5962010814112\\
};
\addlegendentry{OSPS, MR-ZF $p=2$}


\end{axis}
\end{tikzpicture}%

%% file: 1-P-P.tex
%
%
\definecolor{mycolor1}{rgb}{0.00000,0.44700,0.74100}%
\definecolor{mycolor2}{rgb}{0.85000,0.32500,0.09800}%
\definecolor{mycolor3}{rgb}{0.85000,0.32500,0.09800}%
\definecolor{mycolor4}{rgb}{0.49400,0.18400,0.55600}%
\begin{tikzpicture}

\begin{axis}[%
xmin=-3.01,
xmax=6.01,
xlabel={$P_{\text{rad},0}$ (dBm)},
ymax=13.1,
ymin=-8.1,
yminorticks=true,
ylabel={$P_{\text{rad}}$ (dBm)},
title style={at={(0.5,-0.02)},anchor=north,yshift=-1.2cm},
title = {(c)},
legend style={at={(0.02,0.68)},nodes={scale=0.8, transform shape}, anchor=south west, legend cell align=left, align=left, draw=black}
]
\addplot [color=mycolor1, dashdotted, line width=1.5pt]
table[row sep=crcr]{%
-3.01029995663981	-5.61029995663981\\
0	-2.6\\
1.76091259055681	-0.839087409443189\\
3.01029995663981	0.410299956639811\\
3.97940008672038	1.37940008672037\\
4.77121254719662	2.17121254719662\\
5.44068044350276	2.84068044350276\\
6.02059991327962	3.42059991327962\\
};
\addlegendentry{FC, SC, $\alpha_{\text{off}}=-9.8$ dB}

\addplot [only marks, color=mycolor1, line width=1.0pt, mark=o, mark size =2.7pt, mark options={solid, mycolor1}]
table[row sep=crcr]{%
-3.01029995663981	-7.21029995663982\\
-2.365235680217	-6.565235680217\\
-1.72017140379418	-5.92017140379418\\
-1.07510712737136	-5.27510712737136\\
-0.430042850948545	-4.63004285094855\\
0.215021425474272	-3.98497857452573\\
0.860085701897089	-3.33991429810291\\
1.50514997831991	-2.6948500216801\\
2.15021425474272	-2.04978574525728\\
2.79527853116554	-1.40472146883446\\
3.44034280758836	-0.759657192411646\\
4.08540708401117	-0.11459291598883\\
4.73047136043399	0.530471360433987\\
5.37553563685681	1.1755356368568\\
6.02059991327962	1.82059991327962\\
};
\addlegendentry{FC, OFDM, $\alpha_{\text{off}}=-11.4$ dB}

\addplot [color=mycolor2, line width=1.5pt]
  table[row sep=crcr]{%
-3.01029995663981	-3.01029995663981\\
0	0\\
1.76091259055681	1.76091259055681\\
3.01029995663981	3.01029995663981\\
3.97940008672038	3.97940008672038\\
4.77121254719662	4.77121254719662\\
5.44068044350276	5.44068044350276\\
6.02059991327962	6.02059991327962\\
};
\addlegendentry{OSPS, SC, $\alpha_{\text{off}}=-7.2$ dB}

\addplot [color=mycolor2, densely dotted,line width=1.5pt]
  table[row sep=crcr]{%
-3.01029995663981	-7.21029995663981\\
0	-4.2\\
1.76091259055681	-2.43908740944319\\
3.01029995663981	-1.18970004336019\\
3.97940008672038	-0.220599913279626\\
4.77121254719662	0.571212547196623\\
5.44068044350276	1.24068044350275\\
6.02059991327962	1.82059991327962\\
};
\addlegendentry{OSPS, OFDM, $\alpha_{\text{off}}=-11.4$ dB}

\end{axis}
\end{tikzpicture}%

%% file: 1-P-eff.tex
%
%
\definecolor{mycolor1}{rgb}{0.00000,0.44700,0.74100}%
\definecolor{mycolor2}{rgb}{0.85000,0.32500,0.09800}%
\definecolor{mycolor3}{rgb}{0.92900,0.69400,0.12500}%
\definecolor{mycolor4}{rgb}{0.49400,0.18400,0.55600}%
\begin{tikzpicture}

\begin{axis}[%
xmin=-7,
xmax=6,
xlabel={$P_{\text{rad}}$ (dBm)},
ymax=0.3,
ymin=0.03,
ytick={  0, 0.1,0.15,0.2,0.25,0.3},
yminorticks=true,
ylabel={$\eta_{\text{eff}}$},
title style={at={(0.5,-0.02)},anchor=north,yshift=-1.2cm},
title = {(d)},
legend style={at={(0.02,0.68)},nodes={scale=0.8, transform shape}, anchor=south west, legend cell align=left, align=left, draw=black}
]
\addplot [color=mycolor1, dashdotted, line width=1.5pt]
table[row sep=crcr]{%
-6.98970004336019	0.0498466816978341\\
-3.97940008672038	0.0704938532963718\\
-2.21848749616356	0.0863369852893624\\
-0.969100130080564	0.0996933633956682\\
0	0.111460568729152\\
0.791812460476249	0.122098935530623\\
1.46128035678238	0.131881923454264\\
2.04119982655925	0.140987706592744\\
2.55272505103306	0.149540045093502\\
3.01029995663981	0.157629047966585\\
3.42422680822206	0.165322740235991\\
3.80211241711606	0.172673970578725\\
4.14973347970818	0.179724766773273\\
4.47158031342219	0.18650920478087\\
4.77121254719662	0.193055368079414\\
5.05149978319906	0.199386726791336\\
5.31478917042255	0.205523133726713\\
5.56302500767287	0.211481559889115\\
5.7978359661681	0.217276648191697\\
6.02059991327962	0.222921137458304\\
6.232492903979	0.228426192022695\\
6.43452676486187	0.233801661410423\\
6.62757831681574	0.239056287419008\\
6.81241237375587	0.244197871061245\\
6.98970004336019	0.249233408489171\\
7.16003343634799	0.254169202665105\\
7.32393759822969	0.259010955868087\\
7.48188027006201	0.263763846908528\\
7.63427993562937	0.268432596031612\\
7.78151250383644	0.273021519826837\\
7.92391689498254	0.277534577962532\\
8.06179973983887	0.281975413185487\\
8.19543935541869	0.286347385735249\\
8.32508912706236	0.290653603097736\\
8.45098040014257	0.294896945847158\\
8.57332496431269	0.299080090187005\\
8.69231719730976	0.303205527691287\\
8.80813592280791	0.307275582659666\\
8.9209460269048	0.311292427429776\\
9.03089986991944	0.31525809593317\\
9.13813852383717	0.319174495735002\\
9.24279286061882	0.323043418759736\\
9.34498451243568	0.326866550873997\\
9.44482672150169	0.330645480471983\\
9.54242509439325	0.334381706187455\\
9.63787827345555	0.338076643838522\\
9.73127853599699	0.341731632696483\\
9.82271233039569	0.345347941157449\\
9.91226075692495	0.348926771884839\\
10	0.352469266481859\\
10.0860017176192	0.355976509745439\\
};
\addlegendentry{FC, SC}

\addplot [only marks, color=mycolor1, line width=1.0pt, draw=none, mark=o, mark size =2.7pt, mark options={solid, mycolor1}]
table[row sep=crcr]{%
-6.98970004336019	0.041460663945945\\
-3.97940008672038	0.0586342332573486\\
-2.21848749616356	0.0718119764699159\\
-0.969100130080564	0.08292132789189\\
0	0.0927088629754077\\
0.791812460476249	0.101557471064573\\
1.46128035678238	0.109694605992592\\
2.04119982655925	0.117268466514697\\
2.55272505103306	0.124381991837835\\
3.01029995663981	0.13111013137201\\
3.42422680822206	0.137509465867716\\
3.80211241711606	0.143623952939832\\
4.14973347970818	0.149488549771886\\
4.47158031342219	0.155131599513897\\
4.77121254719662	0.160576460985347\\
5.05149978319906	0.16584265578378\\
5.31478917042255	0.170946696757369\\
5.56302500767287	0.175902699772046\\
5.7978359661681	0.180722844272475\\
6.02059991327962	0.185417725950815\\
6.232492903979	0.18999663089542\\
6.43452676486187	0.194467751584804\\
6.62757831681574	0.198838359129438\\
6.81241237375587	0.203114942129145\\
6.98970004336019	0.207303319729725\\
7.16003343634799	0.211408734506886\\
7.32393759822969	0.215435929409748\\
7.48188027006201	0.219389211985185\\
7.63427993562937	0.223272508362133\\
7.78151250383644	0.227089408923352\\
7.92391689498254	0.230843207177504\\
8.06179973983887	0.234536933029395\\
8.19543935541869	0.238173381404542\\
8.32508912706236	0.241755136997152\\
8.45098040014257	0.245284595764492\\
8.57332496431269	0.24876398367567\\
8.69231719730976	0.252195373131681\\
8.80813592280791	0.255580697400774\\
8.9209460269048	0.258921763354695\\
9.03089986991944	0.262220262744021\\
9.13813852383717	0.265477782212339\\
9.24279286061882	0.268695812217497\\
9.34498451243568	0.271875755002279\\
9.44482672150169	0.275018931735431\\
9.54242509439325	0.278126588926223\\
9.63787827345555	0.281199904200864\\
9.73127853599699	0.2842399915167\\
9.82271233039569	0.287247905879664\\
9.91226075692495	0.290224647621615\\
10	0.293171166286743\\
10.0860017176192	0.296088364169833\\
};
\addlegendentry{FC, OFDM}

\addplot [color=mycolor2, line width=1.5pt]
  table[row sep=crcr]{%
-6.98970004336019	0.067241323430739\\
-3.97940008672038	0.0950935915476668\\
-2.21848749616356	0.116465388550211\\
-0.969100130080564	0.134482646861478\\
0	0.150356170088182\\
0.791812460476249	0.164706932034761\\
1.46128035678238	0.177903819624596\\
2.04119982655925	0.190187183095334\\
2.55272505103306	0.201723970292217\\
3.01029995663981	0.212635734925182\\
3.42422680822206	0.223014240226694\\
3.80211241711606	0.232930777100423\\
4.14973347970818	0.242442039459588\\
4.47158031342219	0.25159399451108\\
4.77121254719662	0.260424525824199\\
5.05149978319906	0.268965293722956\\
5.31478917042255	0.277243078911256\\
5.56302500767287	0.285280774643\\
5.7978359661681	0.293098133664525\\
6.02059991327962	0.300712340176363\\
6.232492903979	0.308138454450369\\
6.43452676486187	0.315389763130923\\
6.62757831681574	0.322478058578404\\
6.81241237375587	0.329413864069523\\
6.98970004336019	0.336206617153695\\
7.16003343634799	0.342864820293142\\
7.32393759822969	0.349396165650634\\
7.48188027006201	0.355807639249192\\
7.63427993562937	0.362105608524364\\
7.78151250383644	0.368295896395164\\
7.92391689498254	0.374383844307196\\
8.06179973983887	0.380374366190667\\
8.19543935541869	0.386271994884005\\
8.32508912706236	0.392080922270373\\
8.45098040014257	0.397805034137457\\
8.57332496431269	0.403447940584434\\
8.69231719730976	0.409013002652163\\
8.80813592280791	0.414503355734613\\
8.9209460269048	0.419921930234624\\
9.03089986991944	0.425271469850365\\
9.13813852383717	0.430554547816426\\
9.24279286061882	0.435773581372396\\
9.34498451243568	0.440930844689779\\
9.44482672150169	0.446028480453389\\
9.54242509439325	0.451068510264545\\
9.63787827345555	0.456052844009325\\
9.73127853599699	0.460983288315\\
9.82271233039569	0.465861554200846\\
9.91226075692495	0.470689264015173\\
10	0.475467957738334\\
10.0860017176192	0.480199098721124\\
};
\addlegendentry{OSPS, SC}

\addplot [color=mycolor2, densely dotted, line width=1.5pt]
  table[row sep=crcr]{%
-6.98970004336019	0.041460663945945\\
-3.97940008672038	0.0586342332573486\\
-2.21848749616356	0.0718119764699159\\
-0.969100130080564	0.08292132789189\\
0	0.0927088629754077\\
0.791812460476249	0.101557471064573\\
1.46128035678238	0.109694605992592\\
2.04119982655925	0.117268466514697\\
2.55272505103306	0.124381991837835\\
3.01029995663981	0.13111013137201\\
3.42422680822206	0.137509465867716\\
3.80211241711606	0.143623952939832\\
4.14973347970818	0.149488549771886\\
4.47158031342219	0.155131599513897\\
4.77121254719662	0.160576460985347\\
5.05149978319906	0.16584265578378\\
5.31478917042255	0.170946696757369\\
5.56302500767287	0.175902699772046\\
5.7978359661681	0.180722844272475\\
6.02059991327962	0.185417725950815\\
6.232492903979	0.18999663089542\\
6.43452676486187	0.194467751584804\\
6.62757831681574	0.198838359129438\\
6.81241237375587	0.203114942129145\\
6.98970004336019	0.207303319729725\\
7.16003343634799	0.211408734506886\\
7.32393759822969	0.215435929409748\\
7.48188027006201	0.219389211985185\\
7.63427993562937	0.223272508362133\\
7.78151250383644	0.227089408923352\\
7.92391689498254	0.230843207177504\\
8.06179973983887	0.234536933029395\\
8.19543935541869	0.238173381404542\\
8.32508912706236	0.241755136997152\\
8.45098040014257	0.245284595764492\\
8.57332496431269	0.24876398367567\\
8.69231719730976	0.252195373131681\\
8.80813592280791	0.255580697400774\\
8.9209460269048	0.258921763354695\\
9.03089986991944	0.262220262744021\\
9.13813852383717	0.265477782212339\\
9.24279286061882	0.268695812217497\\
9.34498451243568	0.271875755002279\\
9.44482672150169	0.275018931735431\\
9.54242509439325	0.278126588926223\\
9.63787827345555	0.281199904200864\\
9.73127853599699	0.2842399915167\\
9.82271233039569	0.287247905879664\\
9.91226075692495	0.290224647621615\\
10	0.293171166286743\\
10.0860017176192	0.296088364169833\\
};
\addlegendentry{OSPS, OFDM}

\end{axis}
\end{tikzpicture}%

%% file: 1-BA-compare-QuaDriga.tex
%
%
\definecolor{mycolor1}{rgb}{0.00000,0.44700,0.74100}%
\definecolor{mycolor2}{rgb}{0.85000,0.32500,0.09800}%
%


\begin{tikzpicture}

\begin{axis}[%
xmin=19,
xmax=165,
xlabel={Number of beacon slots $T$},
ymin=0.05,
ymax=1.02,
ytick={  0, 0.2, 0.4, 0.6, 0.8,   1},
yminorticks=true,
ylabel={$P_D$},
title style={at={(0.5,-0.02)},anchor=north,yshift=-1.2cm},
title = {(a)},
legend style={at={(0.49,0.02)},nodes={scale=0.9, transform shape}, anchor=south west, legend cell align=left, align=left, draw=black}
]
\addplot [color=mycolor1, dashed, line width=1.5pt]
  table[row sep=crcr]{%
10	0.0261810625513119\\
20	0.0295143957687653\\
30	0.193333333333333\\
40	0.37\\
50	0.623333333333333\\
60	0.79\\
70	0.866666666666667\\
80	0.934000000000628\\
90	0.973333333333333\\
100	0.99\\
110	0.996666666666667\\
120	0.996666666666667\\
130	0.996666666666667\\
140	0.996666666666667\\
150	0.996666666666667\\
160	1\\
170	1\\
180	1\\
190	1\\
200	1\\
};
\addlegendentry{FC, NNLS}

\addplot [color=mycolor2, line width=1.5pt]
  table[row sep=crcr]{%
10	-0.00666666666666667\\
20	0.189778777182509\\
30	0.259415423360055\\
40	0.321734117836882\\
50	0.431653027233396\\
60	0.593333333333333\\
70	0.7783333333333333\\
80	0.894444444444445\\
90	0.963046637211498\\
100	0.985970088639662\\
110	0.990414533095611\\
120	0.99415774831688\\
130	0.997777777777779\\
140	1\\
150	1\\
160	1\\
170	1\\
180	1\\
190	1\\
200	1\\
};
\addlegendentry{OSPS, NNLS}

\addplot [color=black, dashdotted, line width=1.5pt]
table[row sep=crcr]{%
20	0.0833333333333333\\
40	0.183333333333333\\
60	0.320833333333333\\
80	0.458333333333333\\
100	0.616666666666667\\
120	0.733333333333333\\
140	0.85\\
160	0.9\\
180	0.9\\
200	0.95\\
};
\addlegendentry{FC, OMP \cite{AlkhateebTimeDomain2017}}

\end{axis}
\end{tikzpicture}%

%% file: 3-compare-QuaDriGa.tex
%
%
\definecolor{mycolor1}{rgb}{0.00000,0.44700,0.74100}%
\definecolor{mycolor2}{rgb}{0.85000,0.32500,0.09800}%
\definecolor{mycolor3}{rgb}{0.92900,0.69400,0.12500}%
\definecolor{mycolor4}{rgb}{0.49400,0.18400,0.55600}%
\definecolor{mycolor5}{rgb}{0.46600,0.67400,0.18800}%
\definecolor{mycolor6}{rgb}{0.30100,0.74500,0.93300}%
\definecolor{mycolor7}{rgb}{0.10000,0.40100,0.8000}%
\begin{tikzpicture}

\begin{axis}[%
xmin=-30,
xmax=30,
xlabel={$\snrbef$ (dB)},
ymin=0,
ymax=83,
ylabel={$R_{\text{sum}}$ (bit/s/Hz)},
title style={at={(0.5,-0.02)},anchor=north,yshift=-1.2cm},
title = {(b)},
legend style={at={(0.02,0.68)},nodes={scale=0.8, transform shape}, anchor=south west, legend cell align=left, align=left, draw=black}
]
\addplot [color=mycolor1, dashdotdotted, line width=1.5pt]
table[row sep=crcr]{%
-33.1132995230379	2.97663000563727\\
-30.1132995230379	4.49706759289952\\
-27.1132995230379	6.52107346809193\\
-24.1132995230379	8.91044009003098\\
-21.1132995230379	11.7668256359569\\
-18.1132995230379	14.713814949159\\
-15.1132995230379	17.6346679639157\\
-12.1132995230379	20.4767152221275\\
-9.11329952303793	23.0659109030633\\
-6.11329952303793	25.3944258068446\\
-3.11329952303793	27.3986732783706\\
-0.113299523037931	29.1124266070899\\
2.88670047696207	30.5500109080231\\
5.88670047696207	31.7109641539882\\
8.88670047696207	32.5959066007882\\
11.8867004769621	33.2614284809162\\
14.8867004769621	33.7008088513297\\
17.8867004769621	33.9833303232997\\
20.8867004769621	34.1535846220497\\
23.8867004769621	34.2498166857598\\
26.8867004769621	34.3009167596837\\
29.8867004769621	34.3272096713522\\
32.8867004769621	34.3408787615728\\
};
\addlegendentry{FC, BST}
%

\addplot [only marks, line width = 1.0pt, mark size=2.8pt, mark=o, mark options={solid, mycolor1},mark repeat=2,mark phase=1]
table[row sep=crcr]{%
-33.1132995230379	2.25978491793321\\
-30.1132995230379	3.52724400475574\\
-27.1132995230379	5.22444743005392\\
-24.1132995230379	7.46780763528107\\
-21.1132995230379	10.3066697406668\\
-18.1132995230379	13.5407348360264\\
-15.1132995230379	17.0275620298443\\
-12.1132995230379	20.7526172593876\\
-9.11329952303793	24.6535160570179\\
-6.11329952303793	28.5300657324415\\
-3.11329952303793	32.5167774336226\\
-0.113299523037931	36.3994123582613\\
2.88670047696207	40.4325292452613\\
5.88670047696207	44.371452329208\\
8.88670047696207	48.4362836988691\\
11.8867004769621	52.4348262253666\\
14.8867004769621	56.3645991285735\\
17.8867004769621	60.4536290236896\\
20.8867004769621	64.2837333523909\\
23.8867004769621	68.3515658630604\\
26.8867004769621	72.3466405565174\\
29.8867004769621	76.3026841887942\\
32.8867004769621	80.2912884385513\\
};
\addlegendentry{FC, MR-ZF, $p=2$}

\addplot [color=mycolor2, dashed, line width = 1.5pt]
table[row sep=crcr]{%
-33.1132995230379	2.3953457757092\\
-30.1132995230379	3.15609511965213\\
-27.1132995230379	4.15773762463014\\
-24.1132995230379	5.51094110559075\\
-21.1132995230379	7.05132291253994\\
-18.1132995230379	8.72231173897743\\
-15.1132995230379	10.5177842830423\\
-12.1132995230379	12.1498603534737\\
-9.11329952303793	13.6801094432067\\
-6.11329952303793	15.1037685636452\\
-3.11329952303793	16.3279080375142\\
-0.113299523037931	17.3508650369805\\
2.88670047696207	18.2092925948635\\
5.88670047696207	18.8841214590506\\
8.88670047696207	19.4273401342882\\
11.8867004769621	19.7909431038174\\
14.8867004769621	20.0364173058024\\
17.8867004769621	20.1830991348691\\
20.8867004769621	20.2674175171761\\
23.8867004769621	20.3125646890608\\
26.8867004769621	20.3371961726485\\
29.8867004769621	20.3488577089123\\
32.8867004769621	20.3549893884583\\
};
\addlegendentry{OSPS, BST}
%

\addplot [color=mycolor2, line width=1.5pt]
table[row sep=crcr]{%
-33.1132995230379	2.92699886256337\\
-30.1132995230379	3.69602145330416\\
-27.1132995230379	4.81463729663961\\
-24.1132995230379	6.47071539358145\\
-21.1132995230379	8.60343648421443\\
-18.1132995230379	11.2138483695698\\
-15.1132995230379	14.3999482798274\\
-12.1132995230379	17.8134618606173\\
-9.11329952303793	21.5185067060469\\
-6.11329952303793	25.2936774728715\\
-3.11329952303793	29.2482161392375\\
-0.113299523037931	33.1057799655344\\
2.88670047696207	37.1168112908494\\
5.88670047696207	41.1501487166243\\
8.88670047696207	45.0817193826425\\
11.8867004769621	49.0585506122491\\
14.8867004769621	52.9711914283705\\
17.8867004769621	57.1241404749628\\
20.8867004769621	61.052283871567\\
23.8867004769621	64.9477733925773\\
26.8867004769621	69.0229740953324\\
29.8867004769621	73.0263498507282\\
32.8867004769621	76.9146871932821\\
};
\addlegendentry{OSPS, MR-ZF $p=2$}


\end{axis}
\end{tikzpicture}%

%% file: 1-fully-rate-snr-QuaDriGa.tex
%
%
\definecolor{mycolor1}{rgb}{0.00000,0.44700,0.74100}%
\definecolor{mycolor3}{rgb}{0.85000,0.32500,0.09800}%
\definecolor{mycolor2}{rgb}{0.92900,0.69400,0.12500}%
\definecolor{mycolor4}{rgb}{0.49400,0.18400,0.55600}%
\definecolor{mycolor5}{rgb}{0.46600,0.67400,0.18800}%
\definecolor{mycolor6}{rgb}{0.10000,0.40100,0.8000}%
\begin{tikzpicture}

\begin{axis}[%
xmin=-30,
xmax=31,
xlabel={$\snrbef$ (dB)},
ymin=0,
ymax=83,
ylabel={$R_{\text{sum}}$ (bit/s/Hz)},
title style={at={(0.5,-0.02)},anchor=north,yshift=-1.2cm},
title = {(c)},
legend style={at={(0.02,0.55)}, nodes={scale=0.8, transform shape}, anchor=south west, legend cell align=left, align=left, draw=white!15!black}
]
\addplot [only marks, line width = 1.0pt, mark size=2.7pt, mark=o, mark options={solid, mycolor6},mark repeat=3,mark phase=3]
  table[row sep=crcr]{%
-33.1132995230379	2.97663000563727\\
-30.1132995230379	4.49706759289952\\
-27.1132995230379	6.52107346809193\\
-24.1132995230379	8.91044009003098\\
-21.1132995230379	11.7668256359569\\
-18.1132995230379	14.713814949159\\
-15.1132995230379	17.6346679639157\\
-12.1132995230379	20.4767152221275\\
-9.11329952303793	23.0659109030633\\
-6.11329952303793	25.3944258068446\\
-3.11329952303793	27.3986732783706\\
-0.113299523037931	29.1124266070899\\
2.88670047696207	30.5500109080231\\
5.88670047696207	31.7109641539882\\
8.88670047696207	32.5959066007882\\
11.8867004769621	33.2614284809162\\
14.8867004769621	33.7008088513297\\
17.8867004769621	33.9833303232997\\
20.8867004769621	34.1535846220497\\
23.8867004769621	34.2498166857598\\
26.8867004769621	34.3009167596837\\
29.8867004769621	34.3272096713522\\
32.8867004769621	34.3408787615728\\
};
\addlegendentry{FC, BST}

\addplot [only marks, line width = 1.0pt, mark size=3pt, mark=diamond, mark options={solid, black},mark repeat=3,mark phase=1]
  table[row sep=crcr]{%
-33.1132995230379	2.87434307973996\\
-30.1132995230379	4.29572435472451\\
-27.1132995230379	6.14924870542098\\
-24.1132995230379	8.27076921039407\\
-21.1132995230379	10.7158781719723\\
-18.1132995230379	13.1175981769829\\
-15.1132995230379	15.3577204790546\\
-12.1132995230379	17.3760854827383\\
-9.11329952303793	19.0508847265118\\
-6.11329952303793	20.4216183390154\\
-3.11329952303793	21.4362321544163\\
-0.113299523037931	22.1789793298826\\
2.88670047696207	22.6715835196329\\
5.88670047696207	22.981849563475\\
8.88670047696207	23.1576750405902\\
11.8867004769621	23.2602972083366\\
14.8867004769621	23.3115540911564\\
17.8867004769621	23.3388498257367\\
20.8867004769621	23.352518505189\\
23.8867004769621	23.359523669442\\
26.8867004769621	23.3630448204913\\
29.8867004769621	23.3647936854193\\
32.8867004769621	23.3657071291124\\
};
\addlegendentry{FC, MRT, $p=1$}

\addplot [color=orange, dashed, line width=1.5pt]
  table[row sep=crcr]{%
-33.1132995230379	2.73479269941313\\
-30.1132995230379	4.11793571170781\\
-27.1132995230379	5.86994394141132\\
-24.1132995230379	8.04357233207078\\
-21.1132995230379	10.5680167094079\\
-18.1132995230379	13.2419073347967\\
-15.1132995230379	15.8226215836768\\
-12.1132995230379	18.2221921345566\\
-9.11329952303793	20.3653846355547\\
-6.11329952303793	22.1544507505404\\
-3.11329952303793	23.5843949786112\\
-0.113299523037931	24.6919601598776\\
2.88670047696207	25.5717927948907\\
5.88670047696207	26.2080487450117\\
8.88670047696207	26.6507969317834\\
11.8867004769621	26.9423609312812\\
14.8867004769621	27.1101722417259\\
17.8867004769621	27.2098737203433\\
20.8867004769621	27.2628545647095\\
23.8867004769621	27.2910723904427\\
26.8867004769621	27.3054182255361\\
29.8867004769621	27.3126403934962\\
32.8867004769621	27.3163803050834\\
};
\addlegendentry{FC, MRT, $p=2$}

\addplot [color=mycolor1,dashdotdotted, line width=1.5pt]
table[row sep=crcr]{%
-33.1132995230379	2.82526506546149\\
-30.1132995230379	4.32875465953706\\
-27.1132995230379	6.36330452171592\\
-24.1132995230379	8.86482994523982\\
-21.1132995230379	11.956253715001\\
-18.1132995230379	15.3365662972378\\
-15.1132995230379	18.915537272407\\
-12.1132995230379	22.7342030735443\\
-9.11329952303793	26.6040010766699\\
-6.11329952303793	30.5539591529772\\
-3.11329952303793	34.5148465328055\\
-0.113299523037931	38.4922100100014\\
2.88670047696207	42.4697023295091\\
5.88670047696207	46.4429034658113\\
8.88670047696207	50.3644349127524\\
11.8867004769621	54.497653291565\\
14.8867004769621	58.4199524411582\\
17.8867004769621	62.487776461672\\
20.8867004769621	66.3876793298709\\
23.8867004769621	70.3272698832777\\
26.8867004769621	74.3311660038169\\
29.8867004769621	78.3266117693479\\
32.8867004769621	82.3511750042321\\
};
\addlegendentry{FC, MR-ZF, $p=1$}

\addplot [color=mycolor3, line width=1.5pt]
table[row sep=crcr]{%
-33.1132995230379	2.25978491793321\\
-30.1132995230379	3.52724400475574\\
-27.1132995230379	5.22444743005392\\
-24.1132995230379	7.46780763528107\\
-21.1132995230379	10.3066697406668\\
-18.1132995230379	13.5407348360264\\
-15.1132995230379	17.0275620298443\\
-12.1132995230379	20.7526172593876\\
-9.11329952303793	24.6535160570179\\
-6.11329952303793	28.5300657324415\\
-3.11329952303793	32.5167774336226\\
-0.113299523037931	36.3994123582613\\
2.88670047696207	40.4325292452613\\
5.88670047696207	44.371452329208\\
8.88670047696207	48.4362836988691\\
11.8867004769621	52.4348262253666\\
14.8867004769621	56.3645991285735\\
17.8867004769621	60.4536290236896\\
20.8867004769621	64.2837333523909\\
23.8867004769621	68.3515658630604\\
26.8867004769621	72.3466405565174\\
29.8867004769621	76.3026841887942\\
32.8867004769621	80.2912884385513\\
};
\addlegendentry{FC, MR-ZF, $p=2$}

\addplot [color=mycolor4, dotted, line width=1.5pt]
  table[row sep=crcr]{%
-33.1132995230379	2.97663000563727\\
-30.1132995230379	4.49706759289952\\
-27.1132995230379	6.52107346809193\\
-24.1132995230379	8.91044009003097\\
-21.1132995230379	11.7668256359569\\
-18.1132995230379	14.7138149491589\\
-15.1132995230379	17.6346679639157\\
-12.1132995230379	20.4767152221274\\
-9.11329952303793	23.0659109030632\\
-6.11329952303793	25.3944258068445\\
-3.11329952303793	27.3986732783705\\
-0.113299523037931	29.1124266070898\\
2.88670047696207	30.5500109080229\\
5.88670047696207	31.710964153988\\
8.88670047696207	32.5959066007879\\
11.8867004769621	33.2614284809159\\
14.8867004769621	33.7008088513294\\
17.8867004769621	33.9833303232994\\
20.8867004769621	34.1535846220493\\
23.8867004769621	34.2498166857594\\
26.8867004769621	34.3009167596832\\
29.8867004769621	34.3272096713518\\
32.8867004769621	34.3408787615724\\
};
\addlegendentry{FC, ZF in \cite{yunyi2017scheduler}}

\end{axis}
\end{tikzpicture}%

%% file: 2-sub-rate-snr-QuaDriGa.tex
%
%
\definecolor{mycolor1}{rgb}{0.00000,0.44700,0.74100}%
\definecolor{mycolor3}{rgb}{0.85000,0.32500,0.09800}%
\definecolor{mycolor2}{rgb}{0.92900,0.69400,0.12500}%
\definecolor{mycolor4}{rgb}{0.49400,0.18400,0.55600}%
\definecolor{mycolor5}{rgb}{0.46600,0.67400,0.18800}%
\definecolor{mycolor6}{rgb}{0.10000,0.40100,0.8000}%
\begin{tikzpicture}

\begin{axis}[%
xmin=-30,
xmax=31,
xlabel={$\snrbef$ (dB)},
ymin=0,
ymax=83,
ylabel={$R_{\text{sum}}$ (bit/s/Hz)},
title style={at={(0.5,-0.02)},anchor=north,yshift=-1.2cm},
title = {(d)},
legend style={at={(0.02,0.55)}, nodes={scale=0.8, transform shape}, anchor=south west, legend cell align=left, align=left, draw=white!15!black}
]
\addplot [only marks, line width = 1.0pt, mark size=2.7pt, mark=o, mark options={solid, mycolor6},mark repeat=4,mark phase=3]
  table[row sep=crcr]{%
-33.1132995230379	2.3953457757092\\
-30.1132995230379	3.15609511965213\\
-27.1132995230379	4.15773762463014\\
-24.1132995230379	5.51094110559075\\
-21.1132995230379	7.05132291253994\\
-18.1132995230379	8.72231173897743\\
-15.1132995230379	10.5177842830423\\
-12.1132995230379	12.1498603534737\\
-9.11329952303793	13.6801094432067\\
-6.11329952303793	15.1037685636452\\
-3.11329952303793	16.3279080375142\\
-0.113299523037931	17.3508650369805\\
2.88670047696207	18.2092925948635\\
5.88670047696207	18.8841214590506\\
8.88670047696207	19.4273401342882\\
11.8867004769621	19.7909431038174\\
14.8867004769621	20.0364173058024\\
17.8867004769621	20.1830991348691\\
20.8867004769621	20.2674175171761\\
23.8867004769621	20.3125646890608\\
26.8867004769621	20.3371961726485\\
29.8867004769621	20.3488577089123\\
32.8867004769621	20.3549893884583\\
};
\addlegendentry{OSPS, BST}

\addplot [only marks, line width = 1.0pt, mark size=3pt, mark=diamond, mark options={solid, black},mark repeat=4,mark phase=1]
  table[row sep=crcr]{%
-33.1132995230379	2.43236874866146\\
-30.1132995230379	3.21052095274965\\
-27.1132995230379	4.23114822969026\\
-24.1132995230379	5.60735658668017\\
-21.1132995230379	7.16817279352076\\
-18.1132995230379	8.85271821864349\\
-15.1132995230379	10.654886180565\\
-12.1132995230379	12.2846989733051\\
-9.11329952303793	13.8051606937046\\
-6.11329952303793	15.2201751424244\\
-3.11329952303793	16.4463893392044\\
-0.113299523037931	17.4850897898037\\
2.88670047696207	18.3836793319595\\
5.88670047696207	19.1133130468583\\
8.88670047696207	19.7298873995306\\
11.8867004769621	20.1626930131329\\
14.8867004769621	20.4708659920737\\
17.8867004769621	20.6639159244618\\
20.8867004769621	20.7792334036815\\
23.8867004769621	20.8427941154423\\
26.8867004769621	20.878063052226\\
29.8867004769621	20.8949197195963\\
32.8867004769621	20.9038396918862\\
};
\addlegendentry{OSPS, MRT, $p=1$}

\addplot [color=orange, dashed, line width=1.5pt]
  table[row sep=crcr]{%
-33.1132995230379	2.27038976117759\\
-30.1132995230379	2.92744921230633\\
-27.1132995230379	3.77149325524164\\
-24.1132995230379	4.86786994448379\\
-21.1132995230379	6.06289899551548\\
-18.1132995230379	7.31578955993261\\
-15.1132995230379	8.57108047824925\\
-12.1132995230379	9.69700694500016\\
-9.11329952303793	10.6421295769366\\
-6.11329952303793	11.4006184998703\\
-3.11329952303793	11.9318592360603\\
-0.113299523037931	12.2878811762988\\
2.88670047696207	12.5040486964874\\
5.88670047696207	12.6286373094968\\
8.88670047696207	12.6963777562936\\
11.8867004769621	12.7310832358271\\
14.8867004769621	12.7490122674728\\
17.8867004769621	12.7582143966997\\
20.8867004769621	12.7627716021155\\
23.8867004769621	12.7651181364602\\
26.8867004769621	12.7663044260453\\
29.8867004769621	12.7668851350373\\
32.8867004769621	12.7671750881678\\
};
\addlegendentry{OSPS, MRT, $p=2$}

\addplot [color=mycolor1,dashdotdotted, line width=1.5pt]
table[row sep=crcr]{%
-33.1132995230379	3.35440714629127\\
-30.1132995230379	4.34247787218702\\
-27.1132995230379	5.71635403267605\\
-24.1132995230379	7.68411236897352\\
-21.1132995230379	10.1511798001011\\
-18.1132995230379	13.015819698758\\
-15.1132995230379	16.4708264025832\\
-12.1132995230379	19.9949393139956\\
-9.11329952303793	23.7623619355105\\
-6.11329952303793	27.621557181227\\
-3.11329952303793	31.6309794807874\\
-0.113299523037931	35.4805556601019\\
2.88670047696207	39.4841159331761\\
5.88670047696207	43.4763287824737\\
8.88670047696207	47.4706961310428\\
11.8867004769621	51.444598648578\\
14.8867004769621	55.371256172774\\
17.8867004769621	59.4743796244396\\
20.8867004769621	63.4472992547517\\
23.8867004769621	67.3568269671441\\
26.8867004769621	71.4362265175607\\
29.8867004769621	75.3827296293445\\
32.8867004769621	79.2695796625678\\
};
\addlegendentry{OSPS, MR-ZF, $p=1$}

\addplot [color=mycolor3, line width=1.5pt]
table[row sep=crcr]{%
-33.1132995230379	2.92699886256337\\
-30.1132995230379	3.69602145330416\\
-27.1132995230379	4.81463729663961\\
-24.1132995230379	6.47071539358145\\
-21.1132995230379	8.60343648421443\\
-18.1132995230379	11.2138483695698\\
-15.1132995230379	14.3999482798274\\
-12.1132995230379	17.8134618606173\\
-9.11329952303793	21.5185067060469\\
-6.11329952303793	25.2936774728715\\
-3.11329952303793	29.2482161392375\\
-0.113299523037931	33.1057799655344\\
2.88670047696207	37.1168112908494\\
5.88670047696207	41.1501487166243\\
8.88670047696207	45.0817193826425\\
11.8867004769621	49.0585506122491\\
14.8867004769621	52.9711914283705\\
17.8867004769621	57.1241404749628\\
20.8867004769621	61.052283871567\\
23.8867004769621	64.9477733925773\\
26.8867004769621	69.0229740953324\\
29.8867004769621	73.0263498507282\\
32.8867004769621	76.9146871932821\\
};
\addlegendentry{OSPS, MR-ZF, $p=2$}

\addplot [color=mycolor4, dotted, line width=1.5pt]
  table[row sep=crcr]{%
-33.1132995230379	2.3953457757092\\
-30.1132995230379	3.15609511965213\\
-27.1132995230379	4.15773762463014\\
-24.1132995230379	5.51094110559075\\
-21.1132995230379	7.05132291253994\\
-18.1132995230379	8.72231173897743\\
-15.1132995230379	10.5177842830423\\
-12.1132995230379	12.1498603534737\\
-9.11329952303793	13.6801094432067\\
-6.11329952303793	15.1037685636452\\
-3.11329952303793	16.3279080375142\\
-0.113299523037931	17.3508650369805\\
2.88670047696207	18.2092925948635\\
5.88670047696207	18.8841214590506\\
8.88670047696207	19.4273401342882\\
11.8867004769621	19.7909431038175\\
14.8867004769621	20.0364173058024\\
17.8867004769621	20.1830991348691\\
20.8867004769621	20.2674175171761\\
23.8867004769621	20.3125646890608\\
26.8867004769621	20.3371961726485\\
29.8867004769621	20.3488577089123\\
32.8867004769621	20.3549893884583\\
};
\addlegendentry{OSPS, ZF in \cite{yunyi2017scheduler}}

\end{axis}
\end{tikzpicture}%